\documentclass[a4paper,fleqn]{cas-dc}
\usepackage{lipsum,lineno,hyperref,suffix}

\usepackage[authoryear]{natbib}
\usepackage{caption}
\usepackage[normalem]{ulem}
\usepackage{afterpage}
\usepackage{float}
\usepackage{stfloats}
\def\floatpagepagefraction{.7}
\def\textpagefraction{.1}
\usepackage{url}
\usepackage{hyperref}
\usepackage{bm}
\usepackage{amsmath}
\usepackage{cuted}

\usepackage[authoryear]{natbib}

\graphicspath{{./}{figures}}

\def\tsc#1{\csdef{#1}{\textsc{\lowercase{#1}}\xspace}}
\tsc{WGM}
\tsc{QE}

\begin{document}
\let\WriteBookmarks\relax
\def\floatpagepagefraction{1}
\def\textpagefraction{.001}
\modulolinenumbers[1]
\renewcommand{\topfraction}{.9}
\renewcommand{\bottomfraction}{.8}
\renewcommand{\floatpagefraction}{.8}
\shorttitle{Dynamical Behavior \& Surface Features of 2011~UW$_{158}$}
\shortauthors{C. Gomes et~al. (2026)}

\title [mode = title]{Dynamical Behavior and Surface Features of Large Super-Fast Rotator Near-Earth Asteroid (436724) 2011~UW\texorpdfstring{$_{158}$}{158}}



\author[1,2]{Caio {Gomes}}[type=editor,
                        auid=000,bioid=1,
                        prefix=,
                        role=,
                        orcid=0009-0004-0408-5918]
\cormark[1]
\ead{caio.gomes-oliveira@unesp.br}
\credit{Conceptualization, Formal analysis, Investigation, Methodology, Software, Visualization, Writing - original draft}

\author[1,2]{Andre {Amarante}}[orcid=0000-0002-9448-141X]
\ead{andre.amarante@unesp.br}
\credit{Conceptualization, Formal analysis, Investigation, Methodology, Software, Project administration, Software, Supervision, Validation, Writing - review and editing}

\author[1,3]{Filipe Monteiro}[type=author, orcid=0000-0003-1549-4587]
\ead{filipeastro@on.br}
\credit{Data curation, Formal analysis, Investigation, Writing - original draft, Visualization, Writing - review and editing}

\author[1,2]{Alessandra {Ferreira}}[orcid=0000-0002-6162-9078]
\ead{alessandra.ferraz@unesp.br}
\credit{Formal analysis, Investigation, Methodology, Writing - original draft, Visualization, Writing - review and editing}

\author[1,2]{Leonardo {Braga}}[orcid=0009-0008-5757-3121]
\ead{lb.braga@unesp.br}
\credit{Formal analysis, Validation, Writing - original draft}

\author[1,4,7]{A.K. {de Almeida Jr}}[orcid=0000-0002-9488-4462]
\ead{allan.junior@uc.pt}
\credit{Formal analysis, Investigation, Methodology, Validation, Visualization, Writing - original draft, Writing - review and editing}

\author[1,5]{Duncan Lyster}[orcid=0009-0006-3295-7685] 
\ead{duncan.lyster@physics.ox.ac.uk}
\credit{Investigation, Methodology, Software, Writing - original draft, Writing - review and editing}

\author[1,2]{Nicoli {Rocha}}[orcid=0000-0001-7259-4934]
\ead{nicoli.rocha@unesp.br}
\credit{Formal analysis, Investigation, Methodology, Writing - original draft}

\author[1,2]{Gabriel {Damilano}}[orcid=0009-0008-9586-4317]
\ead{gabriel.damilano@unesp.br}
\credit{Investigation}

\author[1,6]{Leonardo {Barbosa}}[orcid=0000-0001-7754-2324]
\ead{leonardo.torres@upe.br}
\credit{Formal analysis, Validation, Investigation, Methodology, Visualization, Writing - review and editing}

\affiliation[1]{organization={Group of Orbital \& Rotational Research in Irregular Objects \& Observational astroNomy (ORION)},
                addressline={UNESP},
                city={Guaratinguetá},
                state={São Paulo},
                country={Brazil}}

\affiliation[2]{organization={São Paulo State University (UNESP), School of Engineering and Sciences  },
                city={Guaratinguetá},
                postcode={12516-410},
                state={São Paulo},
                country={Brazil}}

\affiliation[3]{organization={Observatório Nacional (ON/MCTI)},
            addressline={R. Gen. José Cristino, 77 - São Cristóvão},
            city={Rio de Janeiro},
            postcode={20921-400},
            state={Rio de Janeiro},
            country={Brazil}}
            
\affiliation[4]{organization={Faculty of Sciences and Technology, University of the Azores},
                addressline={Rua da Mãe de Deus, 9500-321 Ponta Delgada},
                city={Açores},
                country={Portugal}}

\affiliation[7]{organization={CFisUC, Departamento de Física, Universidade de Coimbra},
                addressline={3004-516},
                city={Coimbra},
                country={Portugal}}

\affiliation[5]{organization={Atmospheric, Oceanic \& Planetary Physics},
                addressline={Clarendon Laboratory, Parks Rd},
                city={Oxford},
                postcode={OX1 3PU},
                country={UK}}

\affiliation[6]{organization={Polytechnic School of Pernambuco, University of Pernambuco (UPE)},
                city={Recife},
                postcode={50100-010},
                state={Pernambuco},
                country={Brazil}}

\cortext[cor1]{Corresponding author}


\begin{abstract}
The sub-kilometer Super-Fast Rotator and Potentially Hazardous Asteroid (436724) 2011~UW$_{158}$ exhibits an extreme dynamical environment driven by its rotational period of $0.61072$~h. Using a 3-D polyhedral shape model, we investigate the surface dynamic stability under both cohesionless and cohesion-modified conditions, while also addressing families of periodic orbits, station-keeping maneuvers, and energy states in this rapidly rotating, irregular-body regime.
Our results show that, with the inclusion of cohesion, the effective slopes decrease to ${\sim}45^{\circ}$, permitting the retention of boulders with maximum sizes of ${\sim}50$~m near the equatorial region.
The dominance of the centrifugal term in the effective potential shifts the exterior equilibrium points inward, creating highly unstable geopotential regions where loose regolith is susceptible to ejection, preventing long-term surface retention outside the polar regions.
To quantify this instability, we introduce the equivalent instability speed ($v_{\mathrm{ei}}$), extending conventional escape speed analyses to the SFR regime. These results are consistent with significant internal cohesion in 2011~UW$_{158}$, which enables the body to maintain structural integrity under rotational stresses exceeding the classical spin barrier.
Additionally, thermophysical modeling reveals that the super-fast rotation produces a nearly uniform surface temperature distribution.
Finally, we apply a spherical harmonics method to compute families of periodic orbits around the asteroid to assess the fuel cost of station-keeping maneuvers at reduced computational cost.
Overall, this study characterizes the dynamical environment of 2011~UW$_{158}$, contributing to the understanding of Super-Fast Rotator asteroids recently identified in LSST survey images.
\end{abstract}


\begin{highlights}
\item Cohesion reduces surface slopes by up to 45$^\circ$ on 2011~UW$_{158}$;
\item Cohesion may retain $\sim50$\,m boulders on the surface despite super-fast rotation;
\item Centrifugal potential shifts equilibrium points inward;
\item Rapid rotation produces nearly uniform surface thermal patterns;
\item Periodic orbit families support mission design around 2011~UW$_{158}$.
\end{highlights}

\begin{keywords}
Near-Earth objects \sep Asteroids, dynamics \sep Asteroids, rotation
\end{keywords}

\ExplSyntaxOn
\keys_set:nn { stm / mktitle } { nologo }
\ExplSyntaxOff

\maketitle

\section{Introduction}
The exploration of Small Solar System Bodies (SSSBs), particularly Near-Earth Objects (NEOs), is motivated both by their relevance to understanding Solar System formation and by the practical challenges they pose for planetary defense \citep{2023Natur.616..443D} and In-Situ Resource Utilization (ISRU) \citep{Lewis_1996}. In this context, sub-kilometer Super-Fast Rotators remain an understudied class, and their formation and survival mechanisms remain poorly understood. The asteroid (436724) 2011~UW$_{158}$ is a prime example of this category. Classified as both a Potentially Hazardous Asteroid (PHA) and a Super-Fast Rotator (SFR), it possesses an absolute magnitude of $H = 19.98$ and an Earth Minimum Orbit Intersection Distance (MOID) of $0.003026$~au ($\sim1.2\times$ Lunar distance \citep{UAT}). Its trajectory includes several nominal close approaches to Earth, such as $0.01644$~au in 2015, $0.04316$~au in 2048, $0.08166$~au in 2079, and $0.01094$~au in 2108 (passing at just $0.00961$~au from the Moon) \citep{JPL_2011}. These orbital characteristics make 2011~UW$_{158}$ a highly compelling target for dynamical and structural analyses. Beyond this, NEOs are increasingly viewed as promising ISRU targets, as approximately one in five NEOs require a lower accessibility fuel cost than the lunar surface \citep{jedicke2018availability}. Furthermore, these asteroids contain diverse resources, including minerals, water, and organic compounds, which are fundamentally important for sustainable space exploration and space-based industries \citep{jiang2026emerging}. Given its low MOID and favorable accessibility relative to other NEOs, 2011~UW$_{158}$ stands out as a particularly attractive candidate for such applications.

Among the currently $\sim$41,000 known NEOs, the shapes range from highly irregular, such as (4179) Toutatis \citep{Ostro1999}, to top-shaped bodies like (162173) Ryugu \citep{Icarus_2008, asteroidsiv, Muller_2017}. Radar observations of 2011~UW$_{158}$ \citep{naidu2015radar} showed that the asteroid has an equivalent semimajor axis of $300 \,\mathrm{m}$ and an exceptionally short rotational period of about 37 min. Light curve observations provided a rotational period of $0.61072\,\mathrm{h}$ \citep{Gary_2016, Ipatov2016, Monteiro_2020}. Objects in this size range are generally expected to be gravitationally bound aggregates with a ``rubble pile'' structure \citep{richardson_2002rubblepiles, Kevin2018}. Such a short rotational period suggests that the body maintains its structural integrity with no signs of disintegration or material ejection \citep{Monteiro_2020}. Due to rapid rotation, 2011~UW$_{158}$ has a more cohesive structure, possibly being a coherent monolithic fragment if its composition is consistent with an E-type asteroid, or a high-cohesion aggregate if it is closer to a C-type classification \citep{Monteiro_2020}. \citet{chapman1975surface} historically consolidated asteroid classification into major compositional groups by synthesizing albedo, polarimetry, radiometry, and spectrophotometry. This work remains a fundamental reference for modern taxonomic classifications, including those proposed by \citet{roh2022new} and recent approaches utilizing neural networks \citep{luo2024taxonomic}. This anomaly challenges traditional classifications and highlights the need for further research into asteroids' dynamic and structural properties to understand their formation and behavior better.

The observed ``spin barrier'' indicates that cohesionless asteroids with diameters larger than $\sim 200 \, \text{m} $ have a minimum rotational period of $\sim 2.2 \, \text{h} $ without disintegration or the formation of a binary system \citep{harris1996rotation, pravec_2002, Pravec2007}. The initial understanding assumed that most asteroids up to 100 km in size are fragile, gravitationally bound aggregates with low tensile strength \citep{Pravec_Harris_2000, richardson_2002rubblepiles}. However, \citet{harris1996rotation} hypothesized the existence of an asteroid that exceeds this barrier, possibly not being a rubble pile, but rather a monolithic object that remains intact even at high rotation rates. This was confirmed by the discovery of the first asteroid with non-zero tensile strength and a diameter greater than 0.2\,km, the 2001 OE$_{84}$ \citep{Pravec2002b}. 

As of early 2026, there were 436 asteroids from the NASA Small-Body Database \citep{nasa_sbdb} (8 of which are NEOs) with a rotational period $\le$ 2.2 h and a diameter $\ge$ 200 m. Furthermore, the recent discovery of 19 new SFRs in Legacy Survey of Space and Time (LSST) images \citep{lsst_2026} reinforces the current paradigm that internal cohesion, rather than a strictly monolithic structure, is the primary mechanism preventing rotational disruption \citep{Holsapple_2007, Scheeres_2010, Sanchez_2014}. Consequently, cohesive forces have become a fundamental assumption in contemporary dynamical studies of fast-rotating asteroids \citep{Chang_2014, Polishook_2016, Monteiro_2020, Chang_2022, Pan_2022}.

Because of its rapid rotation, this object offers an opportunity to examine surface processes under conditions that differ significantly from those typically expected for slow rotators and cohesionless asteroids. In particular, we analyze how rapid rotation and cohesive forces influence surface dynamics, slope stability, and material retention in sub-kilometer SFRs. 

Researchers have extensively explored asteroid structural and near-surface dynamics using various approaches. Continuum theories and discrete element methods are frequently employed to assess structural limits and evolution \citep{holsapple2006tidal, walsh2012spin}. In contrast, semi-analytical methods provide deep insights into nonlinear dynamic regimes, orbit families, and their potential orbital and rotational evolution \citep{ershkov2025particular, ershkov2026semi}. Furthermore, long-term continuous granular flow and the secular evolution of surface regolith, while outside the scope of this study, remain fundamental research topics. Although these robust methods provide excellent dynamical analyses, investigating the highly perturbed surface dynamics of a rapidly rotating asteroid requires a precise mapping of its topographic irregularities. For this reason, we adopt the polyhedron method as our primary approach.

The specific implementations in this work are novel generalizations of concepts well established in the literature. Regarding boulder retention, while previous studies such as \citet{polishook2017_boulder} defined analytical conditions using a point-mass gravity approximation restricted to the equator, we extend this mathematical framework across all latitudes and facets for 3-D polyhedral models. Furthermore, although perturbed effective acceleration is a known concept \citep{Amarante2021bennu}, formulating macroscopic cohesion as an equivalent perturbing acceleration and mapping it globally across a polyhedral mesh introduces a distinct approach to the dynamical analysis of fast-rotating asteroids. In addition, we generalize the classical escape-speed formulation \citep{scheeres2016orbital} to map structurally unstable regions characterized by imaginary velocity solutions (unbound states), providing a qualitative demonstration of dynamical asymmetries between the asteroid's leading and trailing edges. Thermophysical modeling of asteroids is standard practice, and applying the Isothermal Latitude Model (ILM) to objects in instantaneous rotation has already been done assuming infinite thermal inertia and uniform longitudinal temperatures \citep{cruikshank2005high}. Explicit surface mapping for this dynamic environment employs a methodology similar to the study by \citet{rozitis2024pre}. However, applying this to a super-fast rotator to demonstrate a highly uniform longitudinal surface temperature distribution driven entirely by the extreme spin rate is an original contextual result for this target.

This paper is structured as follows: Sec. \ref{methodology} gives the methodology adopted in our work. In Sec. \ref{sec:topographic}, we investigate the shape model of the 2011~UW$_{158}$ and how its irregular surface differs from a simple spherical one. In Sec. \ref{sec:surface}, we qualitatively discuss how its surface dynamic environment is dominated by its rotation and the absence of loose regolith outside the poles. Cohesive forces are fundamental to the existence of 2011~UW$_{158}$, and in Sec. \ref{sec:particle_stability}, we substantiate the surface results by analyzing how cohesive particles would behave on the surface of this dynamic environment and comparing the behavior of particles with and without cohesion. Section \ref{sec:ZVC} extends the dynamic environment of the surface to its surroundings, demonstrating that its geopotential structure acts in a repulsive, unstable manner. Section \ref{sec:escape_speed} shows that when conventional escape speed analyses are pushed to the SFR regime, a new mathematical definition becomes necessary, one that preserves established results while encompassing the dynamical peculiarities of objects such as 2011~UW$_{158}$. Section \ref{sec:thermophysical} presents a thermophysical model showing that the rapid rotation produces a nearly uniform surface temperature distribution. Section \ref{sec:HarmonicModel} explains the application of a spherical harmonics method to 2011~UW$_{158}$, which has small errors compared to the polyhedral method when far from the surface, justifying its use in computing families of periodic orbits and in assessing the fuel cost of station-keeping maneuvers around the asteroid (Sec. \ref{sec:orbits}). Finally, in Sec. \ref{finalcomm} we give our final comments.

\section{Methodology}
\label{methodology}
Because the shape of 2011~UW$_{158}$ is highly irregular, with asymmetric surfaces and complex structures, a simple geometric representation cannot adequately describe its dynamical environment. To accurately capture its specific morphological features and investigate its gravitational environment, surface material flow, regolith retention, and orbital dynamics, we use a 3-D convex polyhedral shape model and a rotational period derived from photometric observations and light-curve inversion, as presented in \cite{Monteiro_2020}. Based on radar observations \citep{naidu2015radar}, this model is scaled such that its dynamically equivalent triaxial ellipsoid has a semimajor axis of $300\,\text{m}$ and a rotational period of $0.61072\,\text{h}$.

\subsection{Polyhedron Approach}
This approach entails a higher computational cost, but it provides a precise representation of the shape and dynamic properties of irregular celestial bodies. In contrast, conventional spherical harmonic expansions and mass concentration (mascon) models introduce significant inaccuracies when evaluating the gravitational potential and acceleration near the surface \citep{Werner_1997}. The polyhedron method is a well-established, robust approach to orbital dynamics, considered state-of-the-art \citep{Scheeres_1998, Ostro1999Toutatis} and still relevant in the current literature on dynamics, including comparisons aimed at improving other methods \citep{perez2025characterization, braga2026surface, de2026generalized}. Furthermore, because of its extreme rotational characteristics, 2011~UW$_{158}$ lies in a tensile-stress regime where its structural stability depends on internal friction and cohesion \citep{Hu2021}. Therefore, idealized rotational-equilibrium shapes (such as Maclaurin or Jacobi sequences for strengthless, self-gravitating fluids) cannot adequately represent its physical state, reinforcing the need to use the polyhedral approach.

Consequently, this study adopts the polyhedron approach because its primary objective is to evaluate the highly perturbed local dynamic environment governed by the asteroid's irregular geometry, rather than to investigate its long-term secular evolution or internal granular mechanics. While alternative numerical approaches are effective for other purposes, they entail trade-offs regarding near-surface accuracy, structural convergence, and mathematical singularities. These constraints render them unsuitable for the specific scope of this work. By exactly preserving the body's topographic shape and local gravitational field, the polyhedron model avoids these mathematical limitations. Therefore, its use in this manuscript is a modeling choice, consistent with the physical need for precise surface mapping.

The gravitational potential $U(\mathbf{r})$ can be modeled using a polyhedral representation of the asteroid. For the body's 3-D shape, we compute the field generated by its mass distribution along its edges and faces \citep{Werner_1997, tsoulis2001}. Thus, we model particle motion under the asteroid's potential, which justifies using the 3-D polyhedral representation to describe both its shape and local gravitational interactions. Under these assumptions, the expression for $U$ is given by: 
\begin{equation}
U = \frac{1}{2} G \sigma \left[\sum_{f \in \text{faces}} \mathbf{r}_f \cdot \mathbf{F}_f \cdot \mathbf{r}_f \cdot \omega_f - \sum_{e \in \text{edges}} \mathbf{r}_e \cdot \mathbf{E}_e \cdot \mathbf{r}_e \cdot L_e \right].
\label{eq:potencial_poliedro}
\end{equation}
From the Eq. \eqref{eq:potencial_poliedro}, the gradient of the potential can be calculated as:  
\begin{equation}
\nabla U = G \sigma \left[\sum_{e \in \text{edges}} \mathbf{E}_e \cdot \mathbf{r}_e \cdot L_e - \sum_{f \in \text{faces}} \mathbf{F}_f \cdot \mathbf{r}_f \cdot \omega_f \right],
\label{eq:gradiente}
\end{equation}
\noindent where \(\sigma\) is the density (assumed uniform) of the asteroid, $G = 6.67428 \times 10^{-20}$ km kg$^{-1}$ s$^{-2}$ is the gravitational constant \citep{mohr2025codata}, and the summations run over all edges \((e)\) and faces \((f)\) of the shape model. The terms \(\mathbf{E}_e\) and \(\mathbf{F}_f\) are the second-order tensors of the edges and faces (edge dyads). At the same time, \(L_e\) is a dimensionless logarithmic factor associated with the length of each edge, and \(\omega_f\) is the signed solid angle viewed from the field point.

The polyhedron gravitational potential was calculated using the \textsc{Minor-Gravity}\footnote{\url{https://github.com/a-amarante/minor-gravity}} \citep{Amarante2020Arrokoth,minor-gravity} package, which provides an efficient and accurate implementation of the polyhedron formulation to compute the gravitational potential and its first and second derivatives.
The equilibrium points found and subsequently discussed in a later section were computed using the \textsc{Minor-Equilibria}\footnote{\url{https://github.com/a-amarante/minor-equilibria-nr}} \citep{Amarante2020Arrokoth,minor-equilibria} package. The \textsc{Minor-Equilibria} package uses the Newton-Raphson method to compute equilibrium points numerically and automatically under Solar Radiation Pressure perturbations around irregularly shaped minor bodies (such as asteroids and comets) to study their stability, using polyhedron or mascons techniques.

We performed the calculations using a convex 3-D shape model represented as a polyhedral mesh with 2,040 facets and 1,022 vertices, as provided by \citet{Monteiro_2020}. Because the model is dimensionless, the vertex coordinates were rescaled by a factor of $\sim0.44$, consistent with radar observations \citep{naidu2015radar}, so that the dynamically equivalent triaxial ellipsoid has a semimajor axis of $300\,\mathrm{m}$.

\subsection{Equations of Motion}
To understand the dynamics of particles around irregularly shaped celestial bodies, it is essential to properly characterize their geopotential field, which results from the combination of the body's mass distribution and its uniform rotation. Rotations and translations of the vertices are performed so that the axes corresponding to the minimum, intermediate, and maximum moments of inertia are, respectively, (\( \hat{{x}}, \hat{{y}}, \hat{{z}}\)) and the origin of the system is the center of mass of the asteroid \citep{Mirtich1996}. So, the geopotential $\mathcal{V}(\mathbf{r})$ is given by \citep{Scheeres2012book}:
\begin{equation}
\mathcal{V}(\mathbf{r}) = - \frac{1}{2} \omega^2(\mathbf{r}_{\perp\hat{z}})\cdot(\mathbf{r}_{\perp\hat{z}})+ U(\mathbf{r}),
\label{eq:geopotencial}
\end{equation}
\noindent where \(\mathbf{r} = (x\hat{x}, y\hat{y}, z\hat{z})\) represents the particle's position, \(\mathbf{r}_{\perp\hat{z}} = (x\hat{x}, y\hat{y})\) represents the particle's position relative to the \(\hat{z}\) axis, \(\omega\) denotes the uniform angular speed (\(\omega = \| \mathbf{\boldsymbol{\omega}} \| \)), with \(\boldsymbol{\omega} = \omega \hat{z}\) and \(U(\mathbf{r})\) represents the gravitational potential. 

We assume that the asteroid is isolated within its sphere of influence \citep{Hill}, so that the gravitational field is sufficiently dominant in its vicinity to govern the dynamics of nearby particles, while neglecting gravitational perturbations from other bodies \citep{domingos2006stable}. We also do not account for solar radiation pressure effects.

\noindent The Hill sphere radius at perihelion is given by \citep{Hamilton1991,murray1999solar}:
\begin{equation}
    r_H^{\text{perihelion}} = a(1 - e)\left(\frac{m}{3M_{\odot}}\right)^{1/3}.
    \label{eq:hill_radius}
\end{equation}
For 2011~UW$_{158}$: $a\sim1.6204$~au (semimajor axis), $e\sim0.3766$ (eccentricity) \citep{JPL_2011}, $m\sim3.6746\times10^{10}$~kg (total mass of asteroid assuming a bulk density of $2.0~\mathrm{g\,cm^{-3}}$ (a density discussed in Sec. \ref{sec:slope}) applied to the shape model by \citet{Monteiro_2020}) and $M_{\odot}\sim1.9885\times10^{30}$~kg \citep{prvsa2016_sunmass}. This yields a perihelion Hill radius of $r_H^{\text{perihelion}} = 27.7$~km ($\sim1.8517\times10^{-7}$~au) (Eq. \eqref{eq:hill_radius}).

The gravitational acceleration \(\nabla U\) is directly obtained from Eq. \eqref{eq:gradiente}. The gradient of the geopotential $\mathcal{V} (\mathbf{r})$ defines the acceleration that a massless particle will experience relative to a reference frame fixed to the body. This equation of motion is written as \citep{Jiang2014motion}:
\begin{equation}
\ddot{\mathbf{r}} + 2(\mathbf{\omega} \times \dot{\mathbf{r}}) =-\nabla \mathcal{V}(\mathbf{r}) .
\label{eq:acct}
\end{equation}
Integrating the equation of motion (Eq. \eqref{eq:acct}) under the assumption of a uniform rotator ($\dot{\boldsymbol{\omega}} = \mathbf{0}$), where the geopotential (Eq. \eqref{eq:geopotencial}) has no explicit time dependence, yields a constant of motion for the system, namely the Jacobi integral \citep{murray1999solar,Scheeres_2016}. Applying this integral to the dynamics around irregular bodies \citep{Scheeres_1994} forms the basis of our analysis. This constant of motion allows us to define the Zero-Velocity Curves (ZVCs), which constrain the regions of motion, and to identify the equilibrium points that govern the general dynamical structure around the asteroid. These concepts, directly derived from the geopotential presented here, will be explored in detail in the following sections.

\section{Topographic Characteristics}
\label{sec:topographic}
\begin{figure}[pos=h!]
    \centering
    \includegraphics[width=\linewidth]{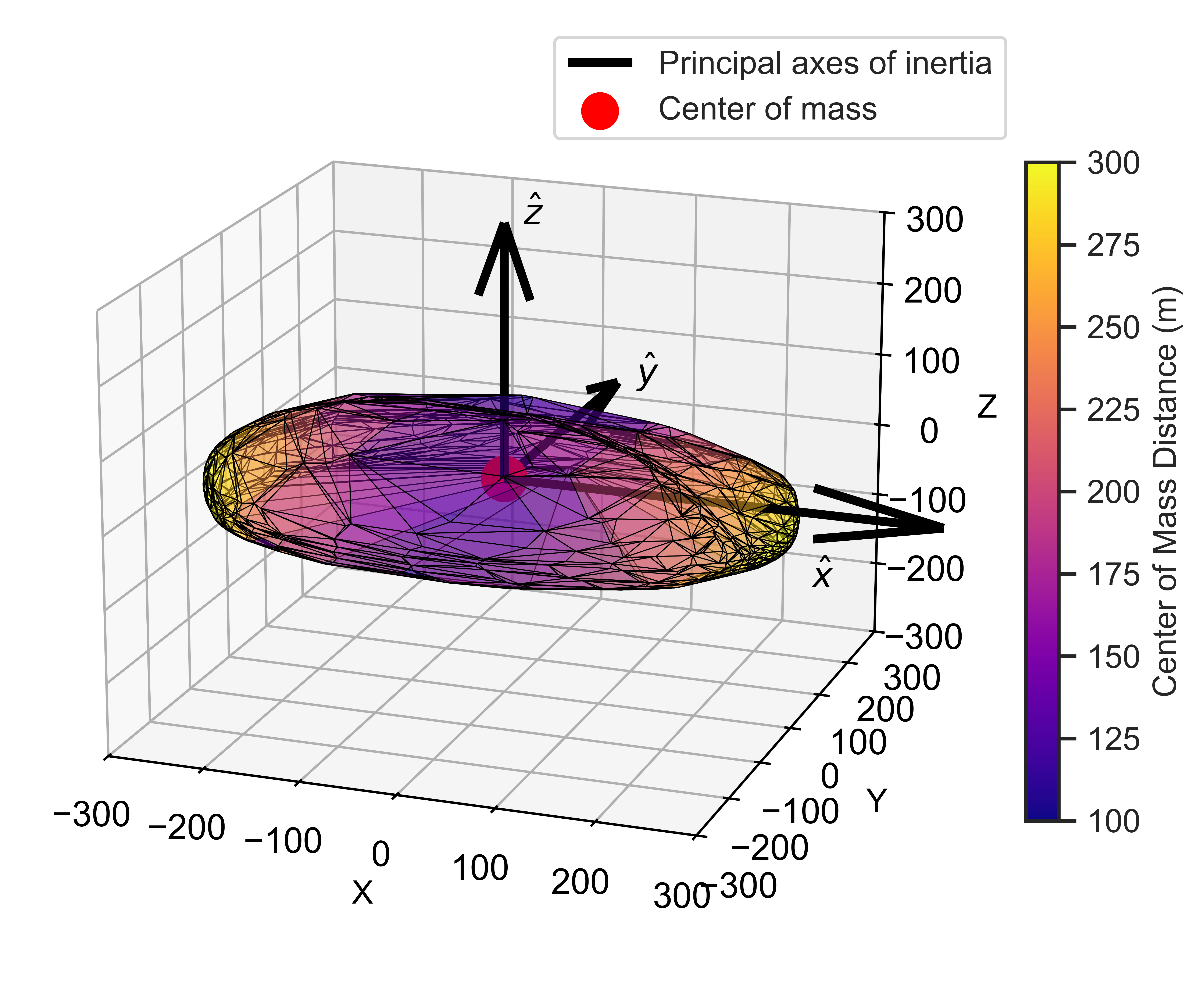}
    \caption{A detailed 3-D representation of the polyhedral shape model of asteroid 2011~UW$_{158}$. The model consists of 1,022 vertices and 2,040 faces. The rectangular color bar indicates each face's distance to the red point marking the asteroid's center of mass, measured in meters. The three vectors represent $\hat{\mathbf{x}}, \hat{\mathbf{y}}$, and $\hat{\mathbf{z}}$, which are aligned with the principal axes of inertia.}
    \label{fig:shapemodel}
\end{figure}

Figure \ref{fig:shapemodel} shows a visualization of the elongated shape of asteroid 2011~UW$_{158}$, with a color map indicating distances from the center of mass in meters. The largest values along the longitudinal regions of the xOy plane present distances $\sim$3\,to\,4$\times$ greater than those measured in the polar regions.

Figure \ref{fig:tilt} shows the presence of a few faces that are significantly flattened by convex modeling, which may suggest that this region across the surface of the asteroid is flatter.

\subsection{Tilt}
\begin{figure*}[t]
    \centering
    \includegraphics[width=\textwidth]{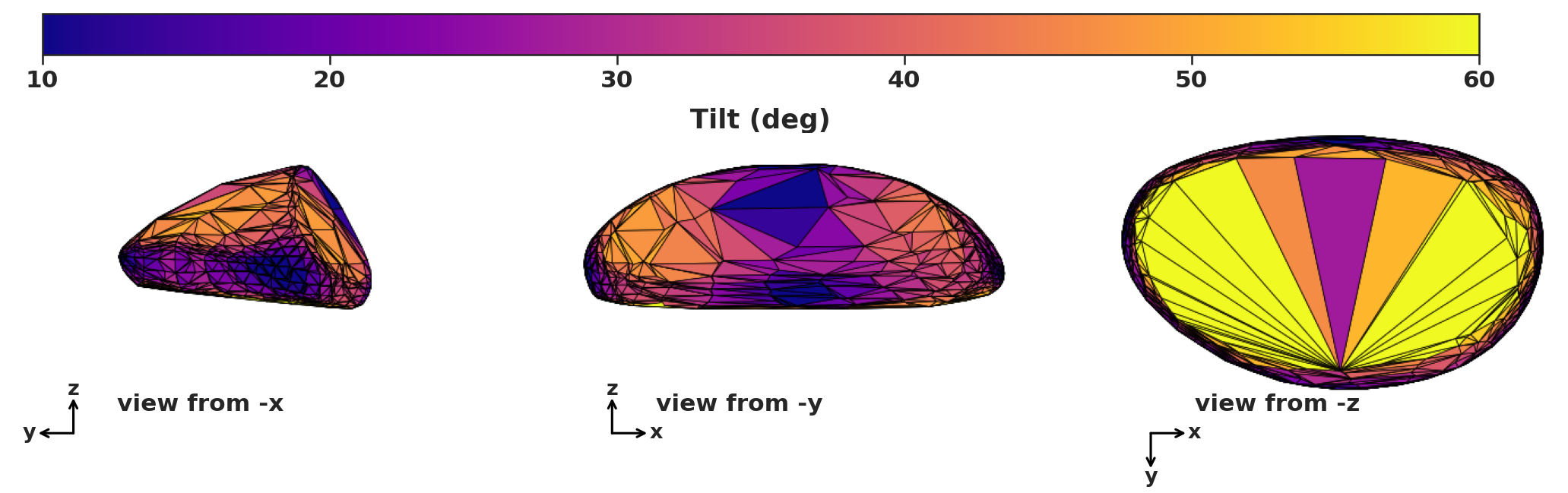}
    \caption{Tilt angle mapped across the surface of 2011~UW$_{158}$, presented from various perspectives, measured in degrees.}
    \label{fig:tilt}
\end{figure*}

The tilt angle in Fig. \ref{fig:tilt} refers to the angle ($\varpi$) measured between the surface normal vector and the position vector extending from the center of mass to the centroid of a given face. 
These angles can directly affect various physical analyses, such as escape speed and slope calculations. Depending on the tilt orientation, it can increase or decrease escape slopes and velocities. The tilt behavior along the asteroid tends toward lower values farther from the center of mass, ranging from $\sim10^\circ$ to $\sim60^\circ$.

\section{Dynamical Surface Features}
\label{sec:surface}
The compositional analysis of 2011~UW$_{158}$ is complex, and the work of \citet{Monteiro_2020} highlights the uncertainty in its composition due to its extremely rapid rotation. To explain how the body withstands such rotation, the authors explored two possible internal mechanisms. The first is a C-type asteroid, analogous to a rubble pile, which would require sufficient internal cohesive forces to avoid disruption. The second is an E-type asteroid, whose structure would be consistent with a monolithic body that is denser and mechanically stronger.  

For this purpose, we analyzed the slopes in the next section by adopting three density values: one representing the mean density for the C-type class ($1.4~\mathrm{g~cm^{-3}}$), one representing the mean density for the E-type class ($2.9~\mathrm{g~cm^{-3}}$), and an intermediate value between these extremes ($2.15~\mathrm{g~cm^{-3}}$) \citep{Carry_2012}.

\subsection{Slope}
\label{sec:slope}
``Slope'' is defined as the angle ($\varphi$) between the effective acceleration vector and the inward-pointing normal surface vector. This means that when \(\varphi>90^\circ\), the acceleration vector points outward from the asteroid, indicating a tendency to eject cohesionless particles. In contrast, when \(\varphi<90^\circ\), the acceleration vector points inward, suggesting the possibility of retaining cohesionless particles \citep{Scheeres_2016}.

\begin{figure}[pos=h!]
    \centering
    \includegraphics[width=\columnwidth]{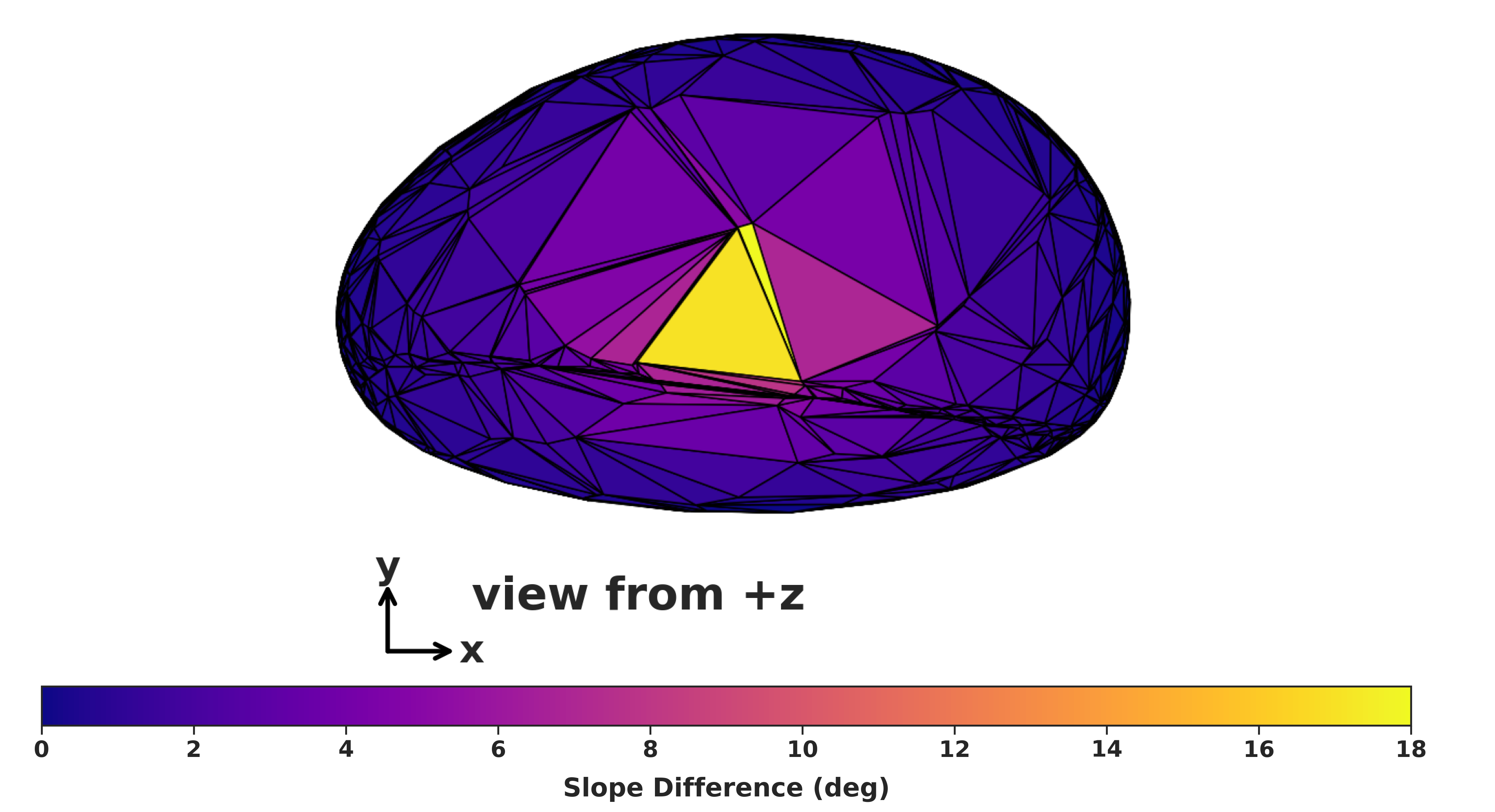}
    \caption{The magnitude of slope variation across each mapped surface region of 2011~UW$_{158}$ when comparing densities of $1.4 \,\text{g\,cm}^{-3}$ and $2.9\,\text{g\,cm}^{-3}$, measured in degrees.}
    \label{fig:slopebeetwen}
\end{figure}

Figure \ref{fig:slopebeetwen} denotes the magnitude of the slope variation for each face when comparing densities of 1.4~g~cm$^{-3}$ and 2.9~g~cm$^{-3}$. These results demonstrate that asteroid 2011~UW$_{158}$, due to its super-fast rotation, exhibits dynamic behavior that is only weakly influenced by its possible mass. Consequently, only the polar regions, where the centrifugal force has progressively weaker effects, show perceptible differences. 
 
\begin{figure*}[t]
    \centering
    \includegraphics[width=\textwidth]{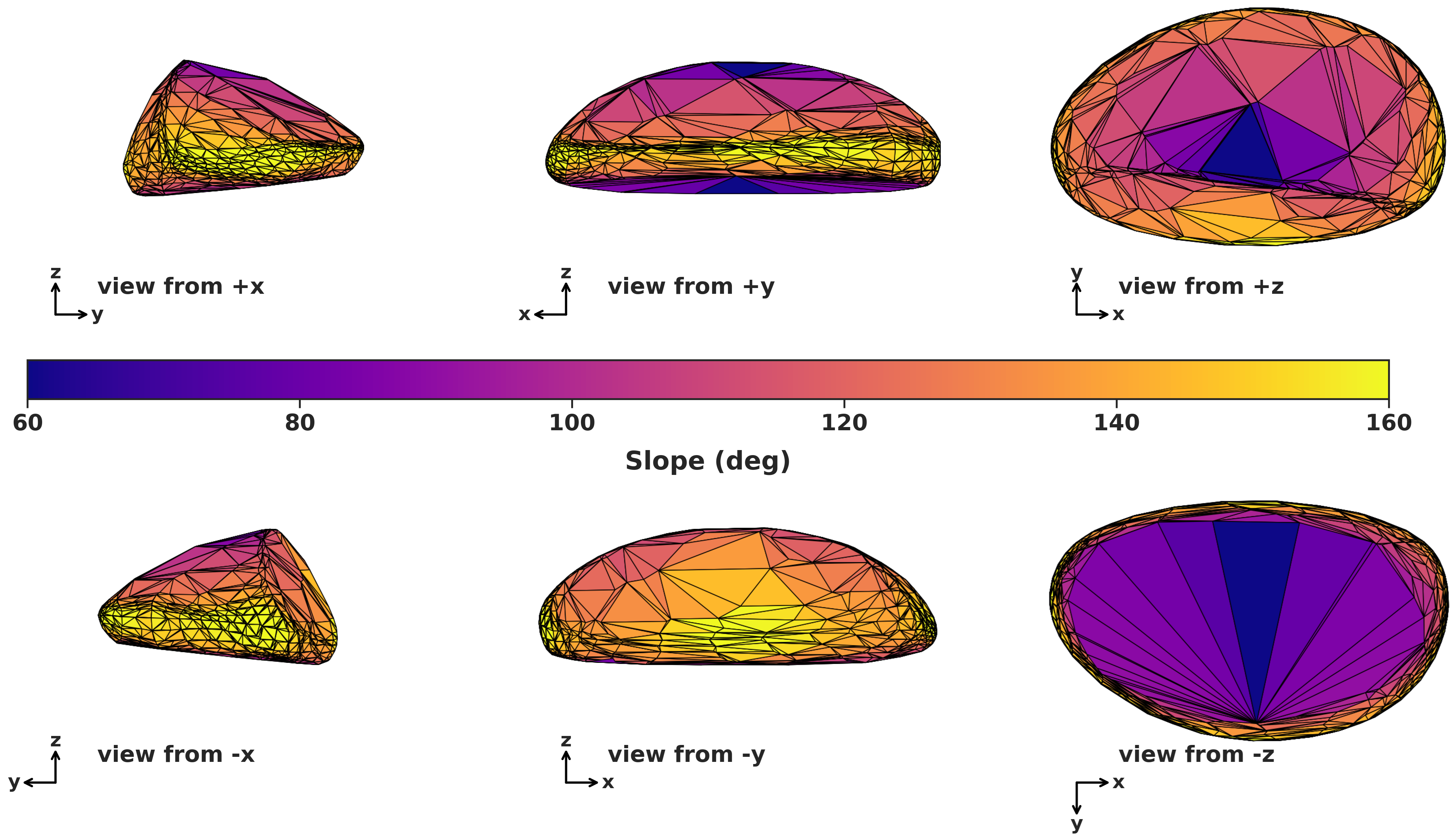}
    \caption{Slope angle mapped across the surface of 2011~UW$_{158}$, considering a density of \(2.9\,\text{g\,cm}^{-3}\), presented from various perspectives and measured in degrees.}
    \label{fig:slope29}
\end{figure*}

Figure \ref{fig:slope29} displays the slopes for 2.9~g~cm$^{-3}$, with a range of values from $\varphi\sim 50^\circ$ to $\varphi \sim170^\circ$, and the data analysis shows that the steepest slopes occur in regions farthest from the rotation axis. In contrast, the gentlest slopes are near the poles, close to the rotation axis. This phenomenon results from the object's super-fast rotation and elongated shape.
This dynamic analysis qualitatively confirms the observational study reported in \citet{Gary_2016}, which states that, when extrapolating the spin barrier, it is essential to consider that the edges of 2011~UW$_{158}$ experience significant centrifugal forces, ejecting loose regolith particles and leaving the edges devoid of rock. In contrast, the rest of the surface closer to the poles appears able to retain regolith.
As mentioned earlier, we used slope as the primary metric to analyze the dynamic differences between extreme densities.

As the slopes do not show significant global changes for the previously adopted density, we assume a uniform bulk density of $d_{\rm ast} = 2.0~\mathrm{g\,cm^{-3}}$ as our primary baseline for the dynamical analyses. This value represents an intermediate estimate between typical C-type and E-type asteroids \citep{Carry_2012}.
The motion of cohesionless particles on $\varphi <90^\circ$ is associated with the angle of repose, generally when $\varphi$ is between $35^\circ$ and $40^\circ$ for geological material \citep{lambe2008_soil, al2018_repose2, muller2021_repose3, valvano2022_repose1}. Assuming a cohesionless surface, the results indicate that the body retains negligible amounts of regolith, mostly confined to the poles.

According to Eq. \eqref{eq:geopotencial}, the geopotential value is the joint effect of the gravitational potential and the centrifugal potential. The centrifugal potential is directly related to the asteroid's rotation rate, which we assume corresponds to its maximum moment of inertia. The gravitational potential is inversely related to the distance from the center of mass.

It is important to note that introducing small perturbations in the rotation about the other principal axes would lead to non-principal-axis rotation, making the angular velocity vector time-dependent, $\boldsymbol{\omega}(t)$. Our methodology focuses on the asteroid's instantaneous surface dynamics. Consequently, we simplify the approach by adopting $\dot{\boldsymbol{\omega}} = \mathbf{0}$, allowing the local dynamical environment to be described by the conventional equations of motion (Eqs. \ref{eq:motion}).
Since the rotational period of asteroid 2011~UW$_{158}$ is very short, the contributions of gravitational potential and acceleration (as discussed in later subsections) are relatively minor. However, we accounted for both contributions in our dynamical analysis. Together, they let us determine the relative potential in any given region on the object's surface. Overall, this analysis provides information for identifying the topographic features of each region.

\section{Surface Particle Stability}
\label{sec:particle_stability}
To retain material particles on the surface of a super-fast-spinning asteroid, cohesive and gravitational forces must counterbalance the high centrifugal force. The equilibrium forces of a simplified one-dimensional ejection condition for a particle at the equator and for a spherical model can be expressed as \citep{hirabayashi2015_boulder, polishook2017_boulder}:
\begin{equation}
    F_{\text{centrifugal}} \le F_{\text{gravity}} + F_{\text{cohesion}}
    \label{eq:force_balance} .
\end{equation}
Expanding these terms for a particle of radius $r$, mass $m$, and rock density $d_{\text{rock}}$, located at a distance $(R+r)$ from the center of mass (with $R$ being the spherical equivalent radius of the asteroid), we obtain:
\begin{equation}
    m(R+r)\omega^2  \le \frac{GMm}{(R+r)^2} + F_{\text{cohesion}} ,
    \label{eq:expanded_forces}
\end{equation}
\noindent where $M$ is the mass of the primary body; assuming a spherical particle, its mass is defined as $m = (4/3)\pi r^3d_{\text{rock}}$. The cohesive force is given by $F_{\text{cohesion}} \sim K A_{\text{contact}}$, where $K$ is the cohesive strength (in $\text{Pa}$).

The effective cohesive approximation contact area for spherical material is approximated as $A_{\text{contact}} \sim 2\pi r^2$ \citep{polishook2017_boulder}. Substituting these relationships into Eq. \eqref{eq:expanded_forces} yields:
\begin{equation}
    \left[ (4/3)\pi r^3 d_{\text{rock}} \right] (R+r)\omega^2  \le \frac{GM \left[ (4/3)\pi r^3 d_{\text{rock}} \right]}{(R+r)^2} + 2\pi r^2 K ,
\end{equation}
\noindent rearranging the terms to isolate $K$, we derive the general condition for surface boulder stability:
\begin{equation}
    (2/3) d_{\text{rock}} r \left[ (R+r)\omega^2  - \frac{GM}{(R+r)^2} \right] \le K
    \label{eq:exact_condition} .
\end{equation}
Unlike the standard $r \ll R$ approximation, Eq. \eqref{eq:exact_condition} explicitly accounts for the outward displacement of the particle's center of mass by $r$. This geometric shift locally increases the centrifugal term and decreases the gravitational pull. Because Eq. \eqref{eq:exact_condition} is non-linear in $r$, the maximum stable radius $r_{\text{max}}$ (where the inequality becomes an equality) is computed using numerical root-finding methods in Python.

\subsection{Maximum Boulder Analysis}
We adopted $R = 0.164$~km (spherical equivalence radius), $M = 3.67 \times 10^{10}$~kg, $\omega \sim 2.86 \times 10^{-3}$~rad~s$^{-1}$ and assumed a rock density typical of enstatite or ordinary chondrites of $d_{\text{rock}} = 3.5~\mathrm{g\,cm^{-3}}$ \citep{Carry_2012}. For $d_{\rm ast}= 2.0~\mathrm{g\,cm^{-3}}$, we performed a linear regression of the results of $K$ of four different bulk densities from \citet{Monteiro_2020}, yielding a cohesive strength of $K \sim 260$~Pa. Under these conditions, the maximum retainable boulder size at the equator is $r_{\text{max}} \sim 60~m$, consistent with boulder sizes found on other asteroids \citep{chapman2002_sizeboulder, michikami2008_sizeboulder}.

\begin{figure}[pos=h!]
    \raggedleft
    \includegraphics[width=\columnwidth]{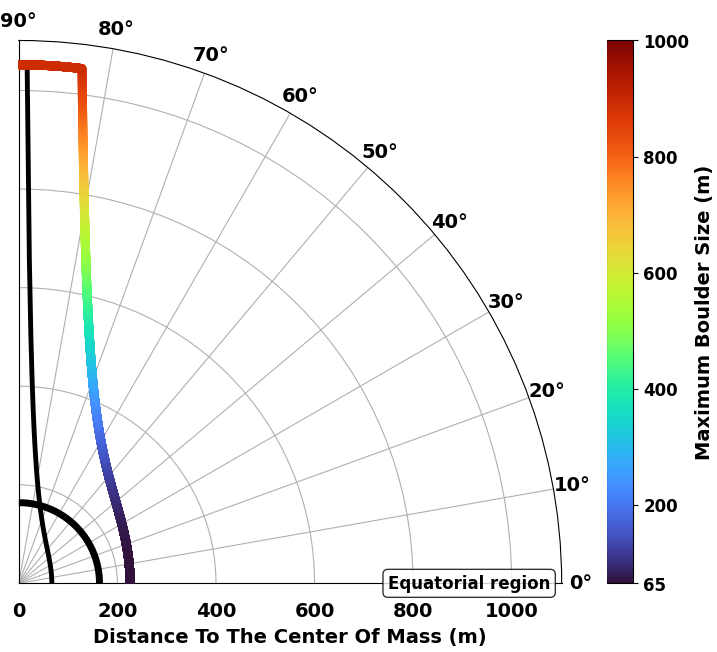}
    \caption{Spherical simplification maximum boulder size, measured in meters, according to latitude. The black circle represents the spherical-equivalence radius for 2011~UW$_{158}$, while the black line represents the maximum values for cohesionless boulders.}
    \label{fig:sphericalbouldersize}
\end{figure}

For a 3-D spherical generalization, not only at the equator, but also:
\begin{equation}
    (2/3) d_{\text{rock}} r \left[ \overbrace{\underbrace{\sin\phi (R+r)}_{\text{Distance from }\hat{z}\text{ axis} } \omega^2  \hat{r}_{\perp \hat{z}}  \cdot \hat{\rho}}^{\text{Projected in } \hat{\rho}}- \frac{GM}{(R+r)^2} \right] \le K ,
    \label{eq:exact_angular0_condition}
\end{equation}
\noindent where $\rho$ is the position unit vector (surface normal for the spherical generalization), $\phi$ is the colatitude (making $(R+r) \sin\phi$ the distance from the rotation axis), $\hat{r}_{\perp \hat{z}} = \cos\theta \hat{i}\text{ +}\sin\theta \hat{j}$ (the unit vector that belongs to the plane of rotation) and $\theta$ is the polar angle with respect to the x-axis. 
Since the $\hat{r}_{\perp \hat{z}}  \cdot \hat{\rho} = \cos(90^\circ-\phi)$, the inequality results in:
\begin{equation}
    (2/3) d_{\text{rock}} r \left[ \sin^2\phi (R+r)\omega^2- \frac{GM}{(R+r)^2} \right] \le K .
    \label{eq:exact_angular_condition}
\end{equation}
From the Eq. \eqref{eq:exact_angular_condition}, each $\phi$ angle, there is a critical $r$ that represents the maximum boulder size within that region of spherical symmetry (Fig. \ref{fig:sphericalbouldersize}) and, $\lim_{r \to \infty} r \implies \sin^2\phi \text{ it can only be }= 0$, meaning that larger boulders are possible the more $\phi \to 0^\circ$ (in the polar regions).
In Fig. \ref{fig:sphericalbouldersize} black line represents the inequality for when $K = 0$ (cohesionless situation), the only possible $r$ is obtained through the condition $\sin^2\phi (R+r)\omega^2 = \frac{GM}{(R+r)^2}$, from Eq. \eqref{eq:exact_angular_condition}. Resulting in:
\begin{equation}
    (R+r) = \left( {\frac{GM}{\omega^2\sin^2\phi}}\right)^{1/3}.
    \label{eq:rsync_caio}
\end{equation}
From Fig. \ref{fig:sphericalbouldersize}, the black line demonstrates that cohesionless materials can be retained only within $\phi <\sim 15^\circ$ of either pole. This approximate limit quantitatively exemplifies the framework proposed by \citet{hirabayashi2015_boulder} and \citet{sanchez2020cohesive}, showing that even under the extreme centrifugal environment of 2011~UW$_{158}$, the poles remain dynamically viable for regolith to exist. The color gradient represents the maximum boulder size possible at each latitude angle (represented by the arc), with the distance to the center of mass plotted on the horizontal axis.

\begin{figure}[pos=h!]
    \centering
    \includegraphics[width=\columnwidth]{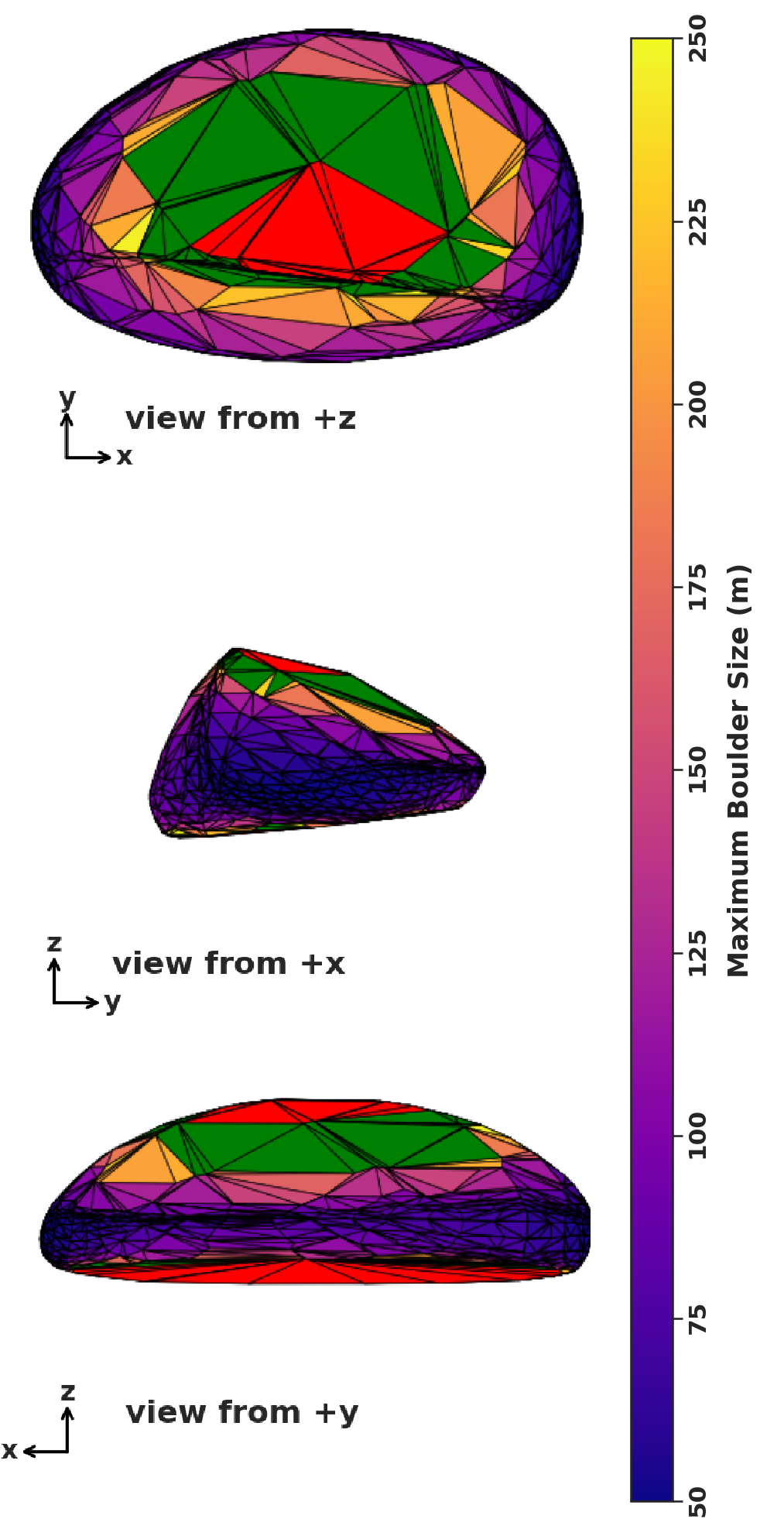}
    \caption{3-D polyhedron analysis of maximum boulder size, measured in meters, for each face. The red faces represent areas where there is no lift-off for boulders, and the green faces represent maximum boulder sizes greater than 300 m.}
  \label{fig:polyhedralbouldersize}
\end{figure}

Furthermore, we expanded the structural analysis of Eq. \eqref{eq:exact_condition} into a simplified analysis of a 3-D polyhedral shape of asteroid 2011~UW$_{158}$ in Fig. \ref{fig:polyhedralbouldersize}. To evaluate the stability across the entire irregular polyhedral mesh, we adopted the $r \ll R$ approximation. The one-dimensional terms within the brackets of Eq. \eqref{eq:exact_condition} were thereby replaced by their respective vector components projected onto the local outward normal ($\hat{n}$) of each face:
\begin{equation}
    (2/3) d_{\text{rock}} r \left[ \underbrace{\vec{a}_{\omega} \cdot \hat{n}}_{\text{Centrifugal projection}} + \underbrace{ \vec{a}_{\text{g}} \cdot \hat{n} }_{\text{Gravitational projection}} \right] \le K
    \label{eq:3d_condition},
\end{equation}
\noindent where $\vec{a}_{\omega}$ is  $(R \sin\phi)\omega^2 \hat{r}_{\perp \hat{z}}$ and $\vec{a}_{\text{g}}$ is the gravitational acceleration derived from the polyhedron analysis. This vector formulation inherently accounts for the object's physical imperfections, recognizing that the local gravity vector need not be directed toward the center of mass.

From Fig. \ref{fig:polyhedralbouldersize}, regions in green have $\mathbb{R}$ values of $r$ that satisfy Eq. \eqref{eq:3d_condition}, but they can reach tens of thousands of meters; regions in red indicate where any value of $r$ satisfies the inequality, because on these faces the gravitational term has a greater magnitude than the centrifugal term.

While the analysis of Eq. \eqref{eq:exact_angular_condition} represented in Fig. \ref{fig:sphericalbouldersize} assumes a spherical radius of $\sim164$ m, here the polyhedral shape model is taken into account. This causes the regions furthest from the center of mass to have a separation of $\sim300$~m. This naturally implies smaller maximum boulders of $\sim50$~m because of greater centrifugal acceleration. Spherical modeling allows larger boulders at the equator but generally produces smaller boulders outside the equator.

\subsection{Surface Acceleration}
\begin{figure}[pos=h!]
    \centering
    \includegraphics[width=\columnwidth]{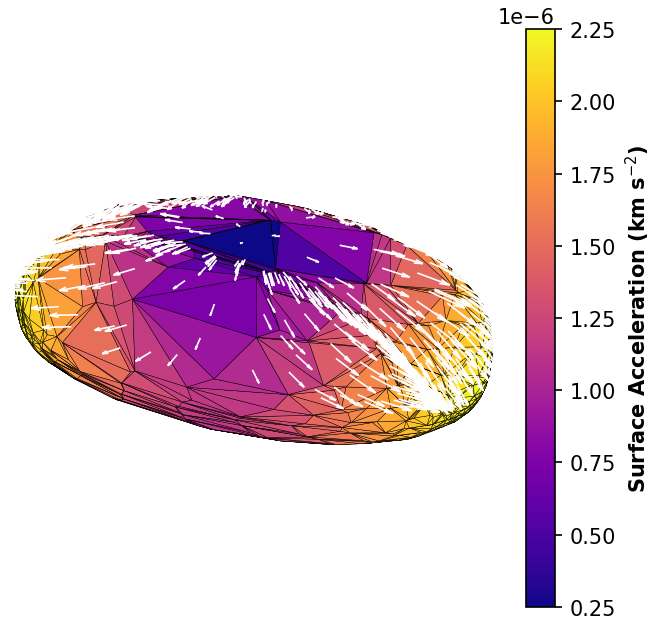}
    \caption{The color bar represents the surface acceleration of 2011~UW$_{158}$, assuming $d_{\rm ast}=2.0~\mathrm{g\,cm^{-3}}$. The vectors represent the tangential component originating in the Northern Hemisphere.}
  \label{fig:surfaceacceleration}
\end{figure}

Figure \ref{fig:surfaceacceleration} illustrates the magnitude of the effective acceleration, defined as the norm of the geopotential gradient ($||\nabla \mathcal{V}(\mathbf{r})||$), derived from Eqs. (\ref{eq:geopotencial}) and (\ref{eq:acct}). Centrifugal acceleration dominates, as the resultant value deviates markedly from that due to gravity alone. The tangential vectors of surface acceleration across the surface of asteroid 2011~UW$ _ {158}$ illustrate the possible flow of material at the surface of asteroid 2011~UW$_{158}$ \citep{scheeres2016orbital}. Due to the high rotation rate, it can be clearly seen that the tendency of loose material to flow is concentrated towards the equatorial region (Fig. \ref{fig:surfaceacceleration}).

This analysis, along with the slope, is fundamental to understanding the orientation of the effective accelerations. In Fig. \ref{fig:surfaceacceleration}, we can see that polar regions have low acceleration intensity, with a medium-high tangential contribution and a normal orientation inwards. Meanwhile, the equatorial regions have high acceleration intensity, with little tangential contribution and normal orientation outwards.

\subsection{Perturbed Slope}
\label{sec:perturbed_slope}
\begin{figure*}[t]
    \centering
    \includegraphics[width=1\textwidth]{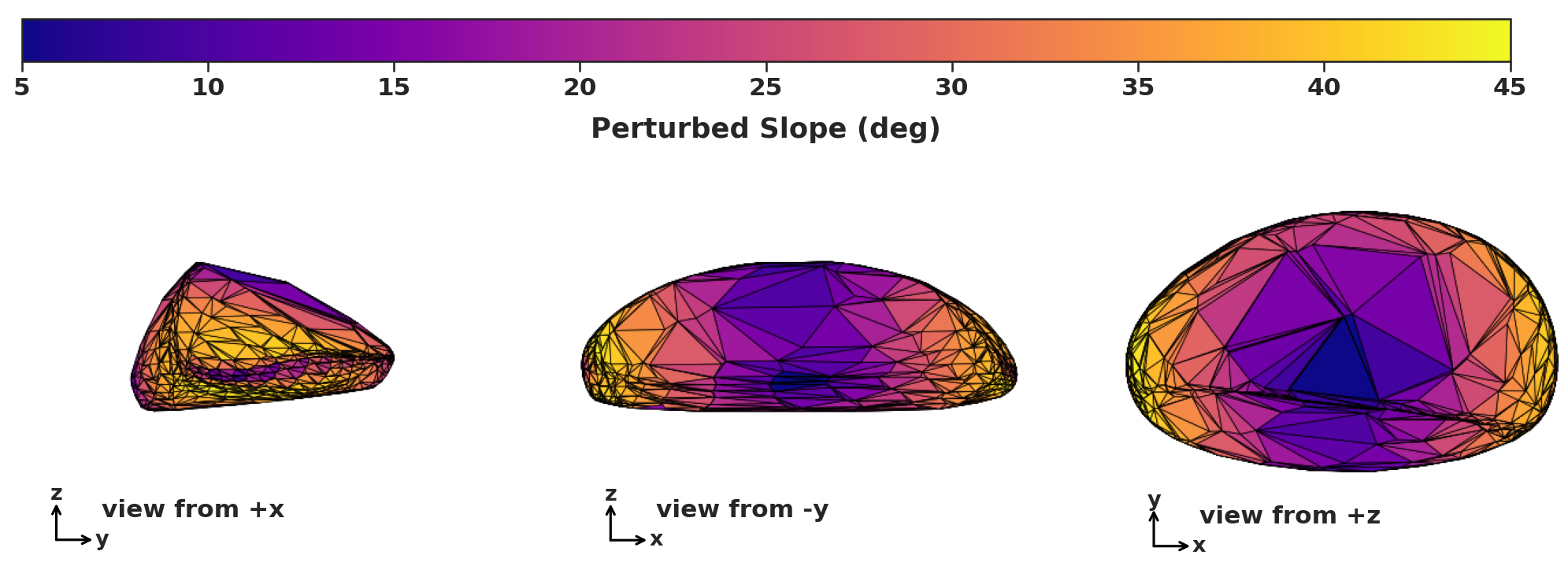}
    \caption{Slope angle perturbed by cohesion acceleration mapped across the surface of 2011~UW$_{158}$, considering $d_{\rm ast}=2.0~\mathrm{g\,cm^{-3}}$, $d_{\rm rock}=2.0~\mathrm{g\,cm^{-3}}$, $r=62~\text{m}$, and $K = 260~\text{Pa}$, presented from various perspectives and measured in degrees.}
    \label{fig:perturbedslope}
\end{figure*}

Using previous insights, we map the effective slopes across the surface of the asteroid for the cohesive material in Fig. \ref{fig:perturbedslope}.
It is important to distinguish between the bulk density of 2011~UW$_{158}$, $d_{\rm ast}=2.0~\mathrm{g\,cm^{-3}}$, which is adopted to compute the asteroid's gravitational field and effective acceleration, and the boulder density, $d_{\rm rock}=2.0~\mathrm{g\,cm^{-3}}$, which is used to estimate the cohesive acceleration acting on the surface material.
In our perturbed slope analysis, we introduced cohesion by adding an inward normal cohesive acceleration to the perturbed surface acceleration. We defined $\mathbf{a}_{\text{eff, pert}} = \mathbf{a}_{\text{eff}} - a_{\text{cohesive}}\hat{n}\text{, with }(\mathbf{a}_{\text{eff}} = \mathbf{a}_{gravitational} + \mathbf{a}_{centrifugal})$, where: 
\begin{equation}
    a_{\text{cohesive}} = KA_{\rm contact}/m_{\rm rock} =  3K / (2 r d_{\text{rock}}).
    \label{eq:a_coh}
\end{equation}
The values used for this parameterization are: $K = 260~\text{Pa}$; $A_{\rm contact}=2\pi r^2$; $m_{\rm rock}=d_{\rm rock}(4/3)\pi r^3$; $d_{\rm rock}=2.0\text~{g~cm}^{-3}$; $r=62$~m. This modification reveals a drastic transformation in the asteroid's surface characteristics. While the unperturbed dynamical environment exhibited extreme slopes reaching $\sim160^\circ$ (Fig. \ref{fig:slope29}), which translates to a widespread tendency toward surface mass shedding, the cohesion-modified slopes are notably reduced, with the global maximum dropping to $\sim45^\circ$ (Fig. \ref{fig:perturbedslope}). This sharp decrease brings the surface slopes within typical angles of repose, quantitatively demonstrating that a cohesive force of 260 Pa can keep material physically bound to the surface. Consequently, this cohesive structure ensures the asteroid's structural stability and allows it to retain the large boulders discussed above, despite the extreme rotational stresses of the Super-Fast regime.

\subsection{Perturbed Surface Acceleration}
\begin{figure}[pos=h!]
    \centering
    \includegraphics[width=0.95\columnwidth]{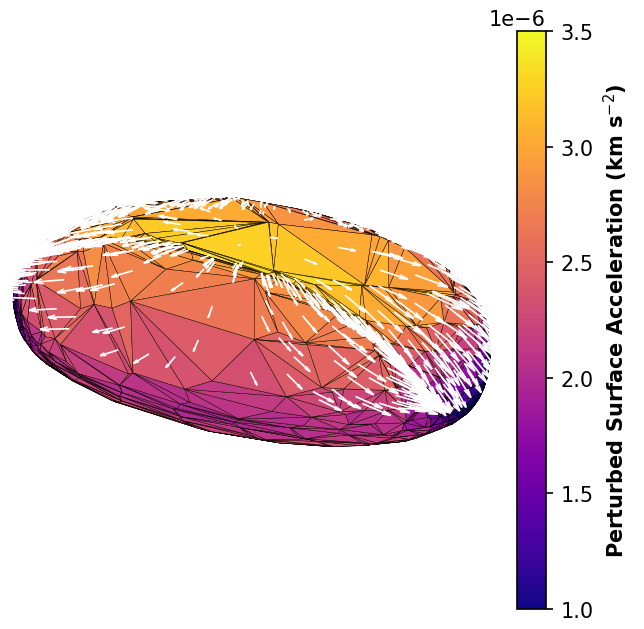}
    \caption{The color bar represents the perturbed surface acceleration (including the cohesion) of 2011~UW$_{158}$, considering $d_{\rm ast}=2.0~\mathrm{g\,cm^{-3}}$, $d_{\rm rock}=2.0~\mathrm{g\,cm^{-3}}$, $r=62~\text{m}$, and $K = 260~\text{Pa}$. The vectors represent the tangential component originating in the Northern Hemisphere.}
  \label{fig:effectivesurfaceacceleration}
\end{figure}

Figure \ref{fig:effectivesurfaceacceleration} indicates the intensity of the effective acceleration perturbed by the cohesive acceleration ($||\mathbf{a_{eff, pert}}||$ from Sec. \ref{sec:perturbed_slope}). Due to the vector direction of cohesion, the tangential components of surface acceleration remain constant.

In contrast to Fig. \ref{fig:surfaceacceleration}, Fig. \ref{fig:effectivesurfaceacceleration} suggests an inversion in how the acceleration intensity behaves on the asteroid's surface.
Polar regions have high acceleration intensity, with very low tangential contribution and normal inward orientation. Meanwhile, equatorial regions exhibit low acceleration intensity, a medium tangential contribution, and a normal inward orientation.

We further assess the sensitivity of $||\mathbf{a}_{\text{coh}}||$ to the assumed contact-area geometry. The adopted $r$ is a reference value based on the maximum boulder size that can exist at the equator (Eq. \eqref{eq:exact_condition}). Thus, changing the assumed contact area also changes this reference $r_{max}$ and, consequently, the resulting cohesive acceleration. Table~\ref{tab:ac} reports the resulting $r_{\max}$ and $a_{\text{coh}}$ for $A_{\text{contact}} = 2\pi r^2$ (as used throughout this work) and for a halved-area case ($A_{\text{contact}} = \pi r^2$), evaluated at both tested rock densities. The relative difference remains below $\sim 16\%$ in all cases, and $||\mathbf{a}_{\text{coh}}||$ stays within the upper range of the unperturbed surface accelerations ($0.25$-$2.25\times10^{-6} \text{ km s}^{-2}$), consistent with cohesion remaining a significant contribution.
\begin{table*}[h!]
\centering
\caption{Parameters and relative errors for cohesive acceleration at different rock densities and contact areas.}
\resizebox{\textwidth}{!}{%
\begin{tabular}{c c c c c c}
\toprule
$d_{\text{rock}}$ (g\,cm$^{-3}$) & $r_{\rm max}(A=\pi r^2)$ (m) & $a_{\rm coh}(A=\pi r^2)$ (km s$^{-2}$) & $r_{\rm max}(A=2\pi r^2)$ (m) & $a_{\rm coh}(A=2\pi r^2)$ (km s$^{-2}$) & $a_{\rm coh}$ Relative Error (\%) \\
\midrule
$3.5$ & $35.583$ & $1.566 \times 10^{-6}$ & $62.0563$ & $1.796 \times 10^{-6}$ & 12.801 \\
$2.0$ & $55.940$ & $1.743 \times 10^{-6}$ & $94.223$ & $2.0696 \times 10^{-6}$ & 15.783 \\
\bottomrule
\end{tabular}%
}
\label{tab:ac}
\end{table*}

\begin{figure*}[t]
    \centering
    \includegraphics[width=\linewidth]{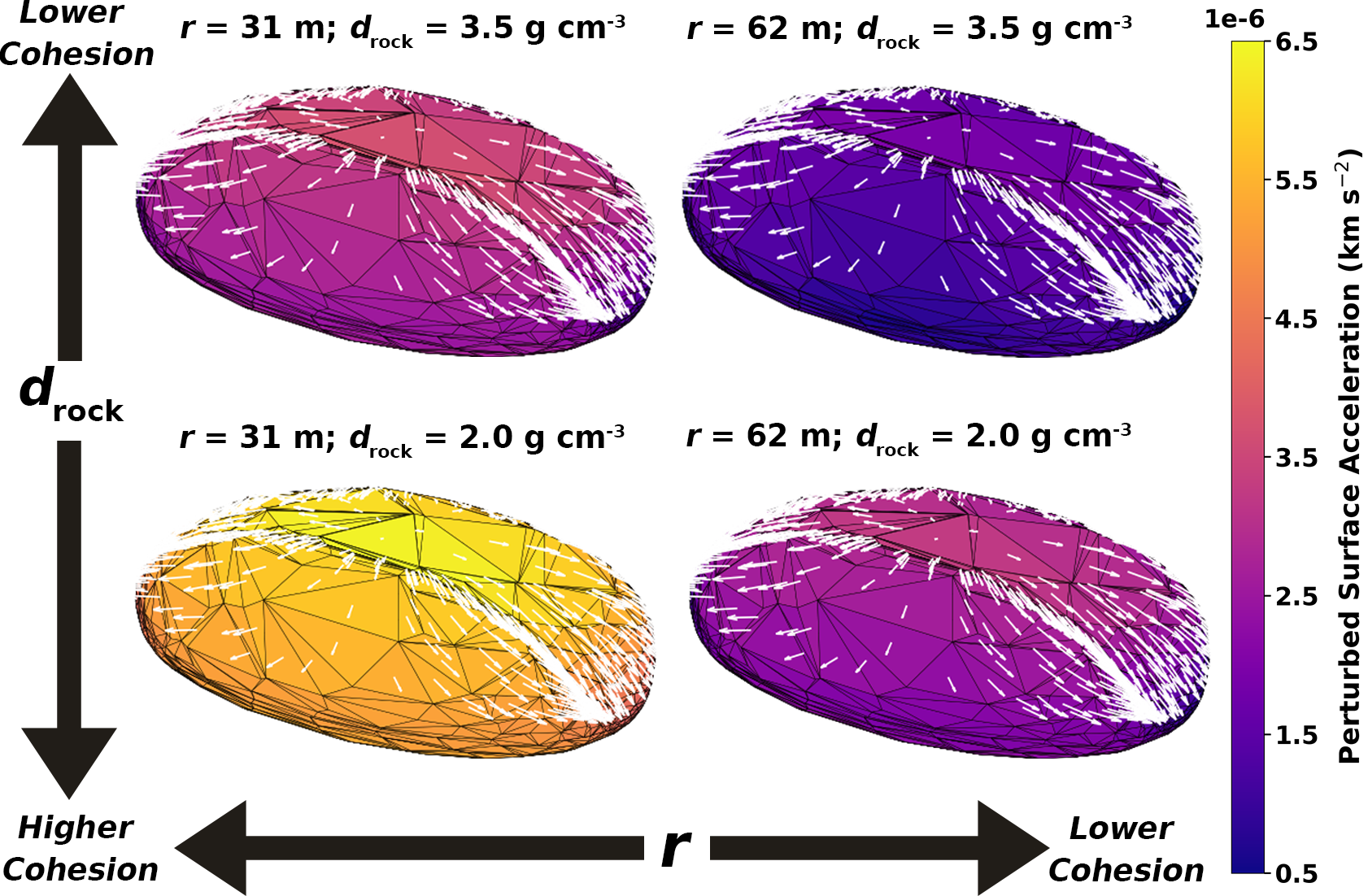}
    \caption{Perturbed surface acceleration with a bulk density of $d_{\rm ast}=2.0\,\text{g\,cm}^{-3}$, for: boulder sizes of 31~m (left) and 62~m (right); boulder densities of $d_{\rm rock}=2.0\,\text{g\,cm}^{-3}$ (bottom) and $d_{\rm rock}=3.5\,\text{g\,cm}^{-3}$ (top). Black arrows indicate the direction of increasing/decreasing cohesive acceleration.}
    \label{fig:variation}
\end{figure*}

Furthermore, to assess the cohesive contribution across a broader parameter space, we analyze the perturbed acceleration field under varying configurations. Figure \ref{fig:variation} presents the perturbed surface acceleration calculated for distinct combinations of boulder sizes ($r=31$ and $62$~m) and rock densities ($d_{\text{rock}}=2.0$ and $3.5~\text{g\,cm}^{-3}$). Across these configurations, the magnitude of the perturbed surface acceleration spans approximately $0.5$-$6.5\times10^{-6}~\text{km s}^{-2}$. This demonstrates that the cohesive contribution remains significant across a broad range of physically plausible parameters.
\section{Zero-velocity curves}
\label{sec:ZVC}
\begin{figure*}[t]
    \centering
    \includegraphics[width=\textwidth]{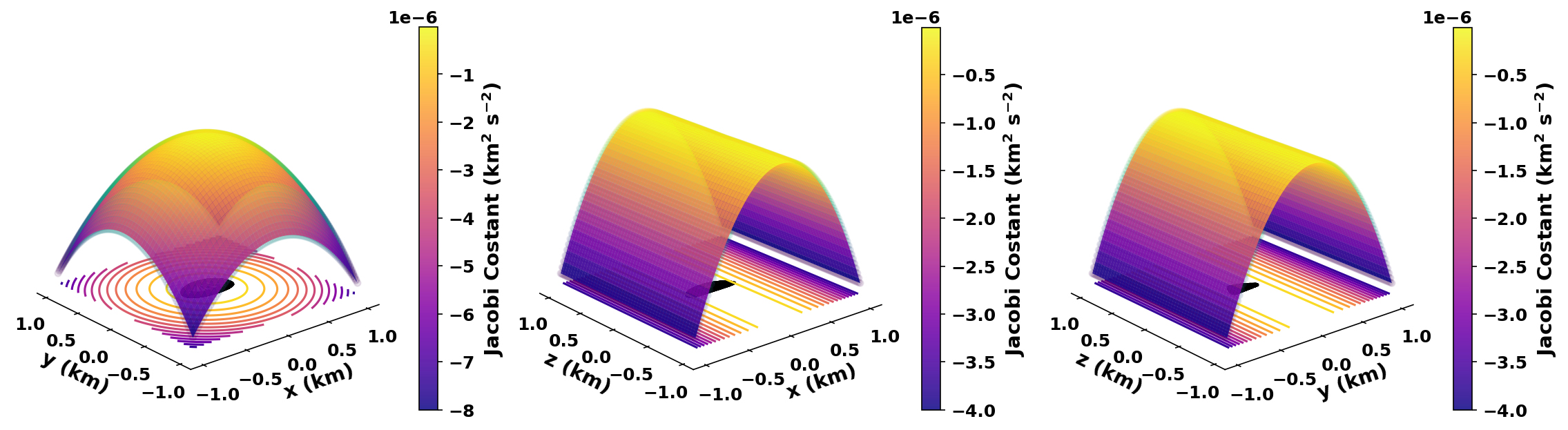}
    \captionof{figure}{The color bar represents the Jacobi Constant of 2011~UW$_{158}$, considering \(d_{\rm ast}=2.0\,\text{g\,cm}^{-3}\) in the three Cartesian planes.}
    \label{fig:ZVC}
\end{figure*}
The equations of motion governing the dynamics of particles near a uniformly rotating asteroid (436724) 2011~UW$_{158}$ can be expressed as \citep{Jiang2014motion}:
\begin{equation}
    \ddot{x} - 2\omega \dot{y} + \frac{\partial \mathcal{V}}{\partial x} = 0, \nonumber
\end{equation}
\begin{equation}
    \ddot{y} + 2\omega \dot{x} + \frac{\partial \mathcal{V}}{\partial y} = 0, \nonumber
\end{equation}
\begin{equation}
    \ddot{z} + \frac{\partial \mathcal{V}}{\partial z} = 0. 
    \label{eq:motion}
\end{equation}
Since these uniformly rotate, the geopotential $\mathcal{V}$ also forms a conserved quantity analogous to energy in the rotating frame, the Jacobi Constant:
\begin{equation}
    \mathcal{J}(\mathbf{r}) = \frac{1}{2} (\mathbf{\dot{r}})\cdot(\mathbf{\dot{r}}) + \mathcal{V}(\mathbf{r}).
\end{equation}
For particles with \( \mathbf{\dot{r}}=0\), the previous equation becomes:
\begin{equation}
    \mathcal{J}(\mathbf{r}) = \mathcal{V}(\mathbf{r}).
    \label{geopotential_v=0}
\end{equation}
The analysis of zero-velocity curves (ZVCs) is given by Eq. \eqref {geopotential_v=0}. In a certain plane, there will be equipotential values of $\mathbf{r}$ that form a path, and continuous intervals of $\mathcal{J}(\mathbf{r})$ form a zero-velocity geopotential surface. Given this surface, there will be local critical points that indicate the locations of the equilibrium points of a given object in that plane.

Figure \ref{fig:ZVC} illustrates the $\mathcal{J}(\mathbf{r})$ of asteroid 2011~UW$_{158}$ in different planes. The color bar represents the values of the Jacobi constant $\mathcal{J}(\mathbf{r})$. The projected lines are contour lines of this zero-velocity geopotential surface, representing different values of $\mathbf{r}$ with the same $\mathcal{J}(\mathbf{r})$. The projections reveal that the asteroid has a single equilibrium point located near its center of mass. Unlike bilobed or highly irregular asteroids, which may have multiple equilibrium points distributed across their surfaces, the rapid rotation of 2011~UW$_{158}$ strongly influences the shape of the zero-velocity curves, leading to the convergence of the outer equilibrium points at a single location (including the Roche lobes) \citep{Fu2024,braga2026surface}.

Unlike bilobed or highly irregular asteroids, which may have multiple equilibrium points distributed across their surfaces, the rapid rotation of 2011~UW$_{158}$ strongly influences the shape of the zero-velocity curves. This extreme centrifugal environment forces the outer equilibrium points to migrate inward and converge at a single location: the center of mass. Consequently, the zero-velocity surfaces associated with these points and the Roche lobes sink below the asteroid's physical surface  \citep{Fu2024,braga2026surface}.

The ZVCs indicate no stable equilibrium regions on the asteroid's surface, as centrifugal forces far outweigh gravity. These results suggest that any cohesionless particle in the system would tend to be attracted to the outermost regions. This configuration reflects a unique dynamical regime, distinct from that of asteroids with conventional rotational periods. It is an energetic analysis of material flow behavior on the surface of an SFR \citep{scheeres2015landslides, Scheeres_2016}.

\subsection{Equilibrium points}
Equilibrium points are locations where the net force acting on a particle in the rotating reference frame of the asteroid is zero. Assume that in Eq. \eqref{eq:motion}, \(\mathbf{\dot{r}}, \mathbf{\ddot{r}}=0\); these points are determined by solving the system of equations:
\begin{equation}
    \frac{\partial \mathcal{V}}{\partial x} = 0, \quad
    \frac{\partial \mathcal{V}}{\partial y} = 0, \quad
    \frac{\partial \mathcal{V}}{\partial z} = 0.
\end{equation}
For irregularly shaped bodies, the distribution and stability of equilibrium points depend on the interplay between gravitational and rotational effects \citep{scheeres2016orbital}. In the case of 2011~UW$_{158}$, its rapid rotation and elongated shape significantly influence the number and location of these points.

The analysis of the equilibrium points for 2011~UW$_{158}$ shows that, unlike bilobed asteroids such as (486958) Arrokoth \citep{Amarante2020Arrokoth} or highly irregular asteroids \citep{braga2025equilibrium}, where multiple equilibrium points can emerge due to complex geopotential distributions. The 2011~UW$_{158}$ exhibits a single equilibrium point located at its center of mass (Fig. \ref{fig:ZVC}). This is primarily due to its fast rotational period of \( \sim 36 \) min, which induces strong centrifugal forces that reshape the geopotential surfaces of zero velocity.

The analysis of the equilibrium points for 2011~UW$_{158}$ shows that, unlike bilobed asteroids such as (486958) Arrokoth \citep{Amarante2020Arrokoth} or highly irregular asteroids \citep{braga2025equilibrium}, where multiple equilibrium points can emerge due to complex geopotential distributions, 2011~UW$_{158}$ exhibits a single equilibrium point located at its center of mass (Fig. \ref{fig:ZVC}). This is primarily due to its super-fast rotational period of $\sim$36 min, which induces strong centrifugal forces that drastically reshape the geopotential surfaces of zero velocity.

\section{Escape Speed}
\label{sec:escape_speed}
A particularly compelling analysis for SFRs concerns the minimum normal velocity required for loose, cohesionless particles to overcome local gravity and escape the asteroid's gravitational influence.

The mathematical development of the escape speed $v_{min}$ for a particle on the surface is given by \citep{Scheeres1996escapespeed}:
\begin{align}
    v_{\text{min}} 
    &= -\hat{n} \cdot (\bm{\omega} \times \mathbf{r}) \notag \\
    &\quad + \sqrt{\bigl[\hat{n} \cdot (\bm{\omega} \times \mathbf{r})\bigr]^2 - 2U_{\text{min}}(\mathbf{r}) - (\bm{\omega} \times \mathbf{r})^2}\nonumber \\
    &= -A + \sqrt{\Delta}, \label{eq:escapespeed}
\end{align}
\noindent where: $U_{min}(\mathbf{r}) = \text{min} \left[ U(\mathbf{r}),-(GM/||r||)\right]$ since the gravitational potential $U$ is defined as negative,

\noindent
$A = \hat{n} \cdot (\bm{\omega} \times \mathbf{r})$; it is the projection of the inertial velocity due to rotation onto the face's normal vector,

\noindent
$B =- (\bm{\omega}\times \mathbf{r})^2,$ which represents$-\omega^2(\mathbf{r}_{\perp\hat{z}})\cdot(\mathbf{r}_{\perp\hat{z}})$ (a proportional term of the centrifugal potential) and repels the particle from the surface, contrasting the term proportional to the gravitational potential ($- 2U_{min}(\mathbf{r})$) and

\noindent
$\Delta = A^2  - 2U_{min}(\mathbf{r}) + B.$

\begin{figure}[pos=h!]
    \centering
    \includegraphics[width=\columnwidth]{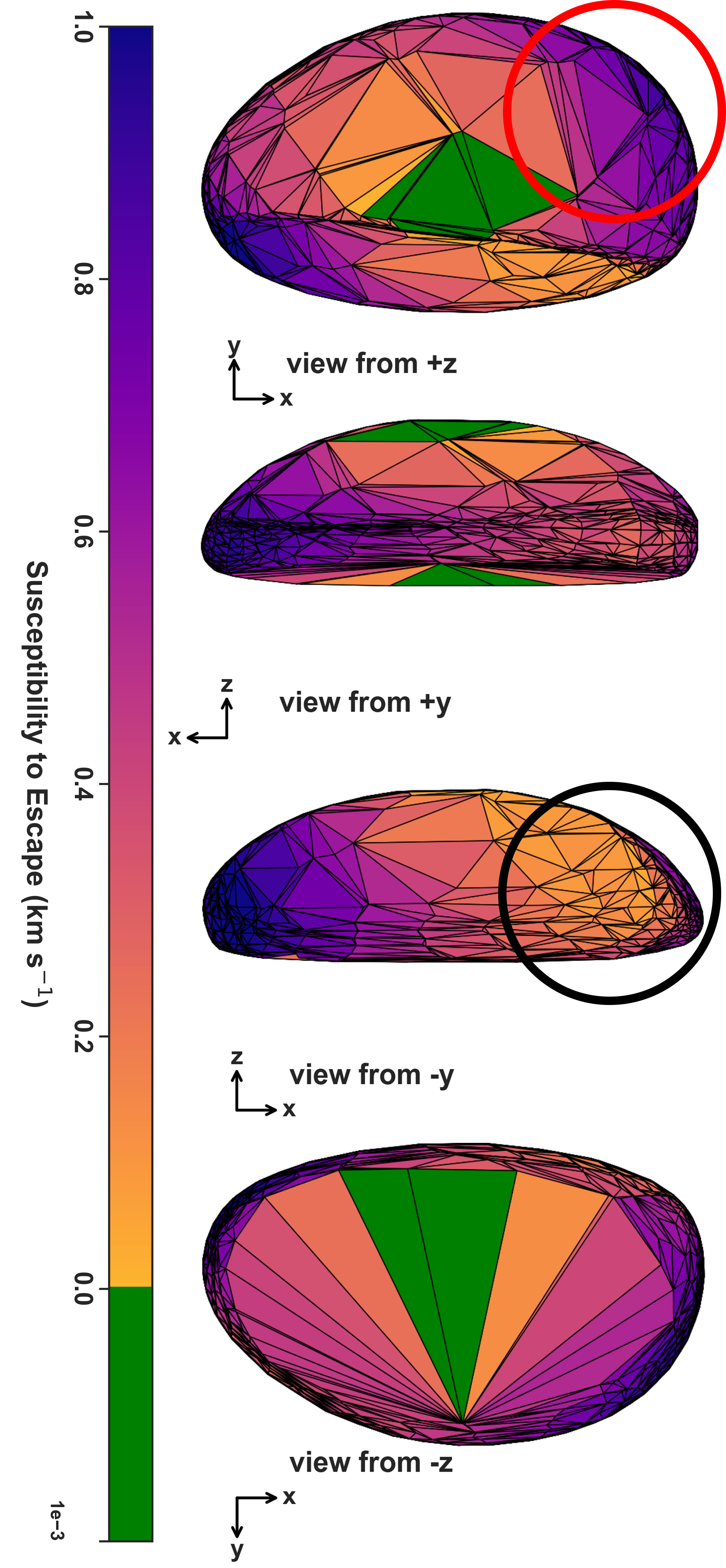}
    \caption{The color bar quantifies the magnitude of the equivalent instability speed ($|v_{ei}|$) for the 2,020 faces that are in the inevitable ejection regime ($\Delta < 0$), considering \(d_{\rm ast}=2.0\,\text{g\,cm}^{-3}\) and represented in different perspectives.}
    \label{fig:escape}
\end{figure}

When $-A + \sqrt{\Delta} \leq 0$, this implies the escape condition has already been reached. For 2011~UW$_{158}$, 2,020 of the 2,040 faces have a $\Delta < 0$, generating values of $v_{min} \in \mathbb{C}$. This suggests that the particle does not need a minimum normal velocity to escape, creating regions of unavoidable escape. For this analysis of loose, cohesionless particles, the result in Fig. \ref{fig:escape} agrees with our previous analyses and discussions, such as those in Figs. \ref{fig:slope29}, \ref{fig:sphericalbouldersize}, and \ref{fig:polyhedralbouldersize}, where we established that due to their high rotation, such particles can only exist in regions close to the poles, such that a real and positive escape speed exists only in these regions.

When $\Delta < 0$, the classical definition of escape speed is found in the complex plane, physically indicating a state where the particle possesses excess mechanical energy. To globally map the intensity of this excess energy onto a continuous scalar metric, one cannot simply use the modulus of the complex root, $|z|=\sqrt{A^2 + |\Delta|}$, because extracting the modulus from the topographic term $A$ would lose the physical distinction between leading and trailing edges.

To preserve both the topographic influence ($-A$) and the magnitude of the excess or deficit of kinetic energy $(|\Delta|)$, we define an equivalent instability speed $(v_{ei})$:
\begin{equation}
    v_{ei} = -A + \operatorname{sgn}(\Delta)\sqrt{|\Delta|},
\end{equation}
when
\[\operatorname{sgn}(\Delta) = \begin{cases}
    1 & \text{if } \Delta > 0, \\
    0 & \text{if } \Delta = 0,\\
    \text{-1} & \text{if } \Delta < 0.
\end{cases}\]

In this formulation, the sign function ($\operatorname{sgn}$) acts as a switch between physical regimes. $(\Delta >0~\text{and}~{-A + \sqrt{\Delta} > 0})$ is a bound state, $\operatorname{sgn}(\Delta) = 1$ and $v_{ei}$ reduces to the classical escape speed $v_{\min}$. For non-bound states ($\Delta < 0$, or $\Delta > 0$ with $v_{ei} \leq 0$), $\operatorname{sgn}(\Delta)$ converts the square root into a signed measure of excess rotational energy, reflecting the centrifugal repulsive nature of the SFR regime and identifying regions from which cohesionless particles cannot escape.
 
The regions highlighted by the circles in Fig. \ref{fig:escape} are examples discussed in \citet{scheeres2016orbital} of how the orientations of $(\bm{\omega} \times \mathbf{r})$ influence the values of $v_{ei}$, where the region in red is a leading edge, and the region in black is a trailing edge. Notably, at trailing edges, where $A < 0$, it is geometrically possible to find $\Delta < 0$ yet $v_{ei} > 0$; in such cases, the positive value reflects only the topography of the instability when the particle remains in the unavoidable escape regime, and the $\Delta < 0$ criterion takes precedence over the sign of $v_{ei}$ in classifying the dynamical state.

Only the 20 faces highlighted in green denote the regions that remain in the conventional escape regime, where the $v_{min}$ values $\in\mathbb{R}$ and gravitational retention is theoretically possible without cohesive forces.
The other 2,020 faces are represented by the color gradient, where all of them are in the unavoidable energetically unbound regime, and it is not necessary to apply a normal velocity to the surface for cohesionless particles to escape (an unbound state).

The regions highlighted in purple indicate areas with a greater propensity for material shedding, where rotational inertial energy and local topographic tilt maximize the susceptibility of non-cohesive material to escape. This analysis considers only the energy required for cohesionless particles to escape. However, it agrees with the previous discussion that its Roche lobes converge within the asteroid due to its super-fast rotation rate, allowing immediate escape from its gravitational influence at the surface \citep{scheeres2015landslides}.

\section{3-D Thermophysical Model for Asteroid 2011~UW$_{158}$}
\label{sec:thermophysical}
\begin{figure*}[t]
    \centering
    \includegraphics[width=1\textwidth]{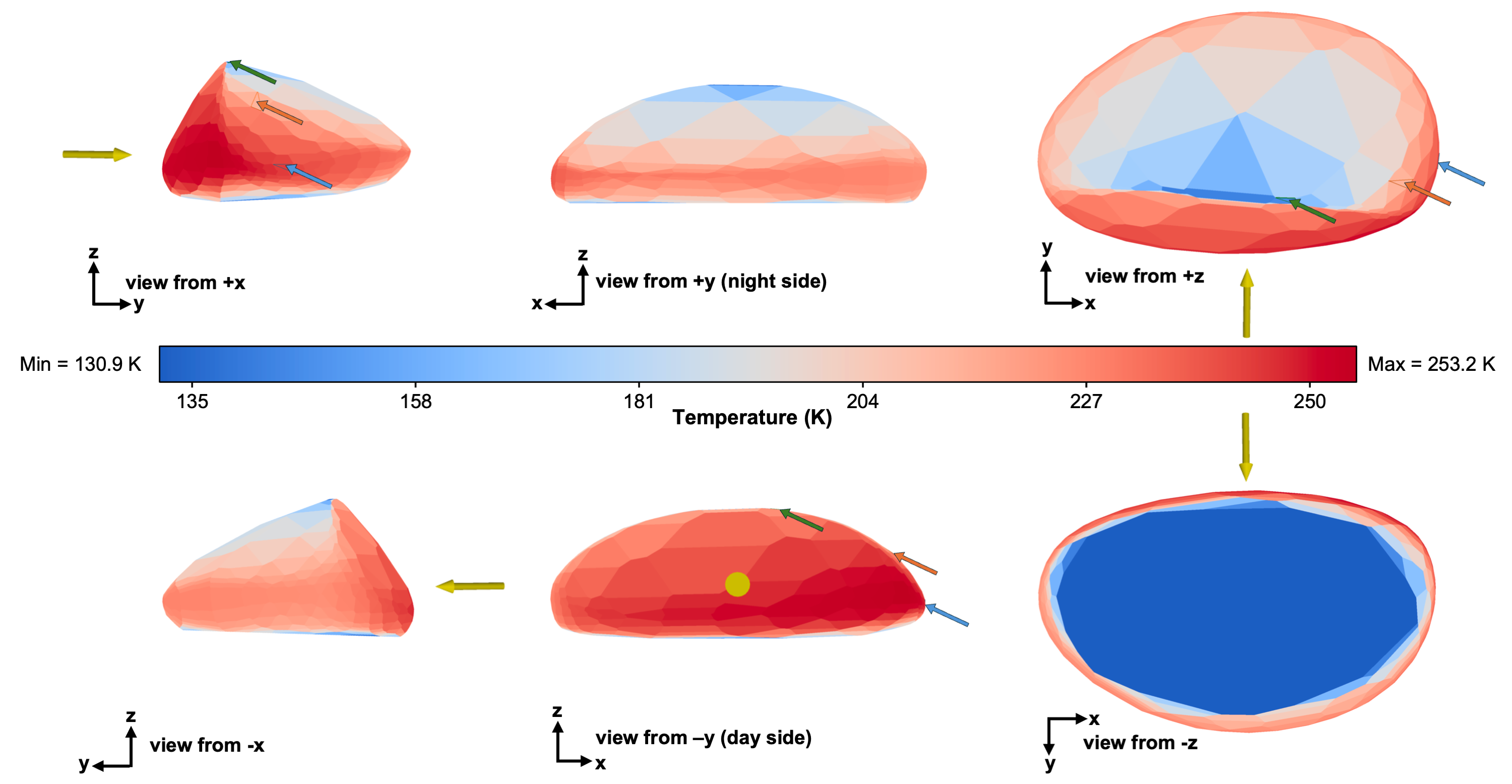}
    \caption{Surface temperature map predicted by TEMPEST for a thermal inertia of 300~J~m$^{-2}$~K$^{-1}$~s$^{-1/2}$, at a heliocentric distance corresponding to the semimajor axis of 1.62~au at a sub-solar latitude of $0^{\circ}$. The facets highlighted by the blue, orange, and green arrows on the surface represent the equator, mid-latitudes, and polar regions, respectively. The yellow arrow shows the direction of incoming sunlight.}
    \label{thermal2011}
\end{figure*}

In this section, we use the TEMPEST software \citep{1989Icar...78..337S,2024EPSC...17.1121L, lyster2025tempest, tempest}, which solves the one-dimensional heat diffusion equation using an explicit finite-difference time-stepping scheme. The model assumes simple rotation about a fixed axis, a constant heliocentric distance, and thermophysical properties such as thermal inertia, albedo, and heat capacity for each facet of the given shape model. This approach lets us estimate the temperature distribution on the asteroid's surface and in its subsurface.

We model the asteroid as rotating simply about a principal axis perpendicular to its orbital plane, with the Sun in the $-y$ direction, corresponding to a sub-solar latitude of $0^{\circ}$. The preferred pole solution of \citet{Monteiro_2020}, $\beta = 36^{\circ}$, instead implies an obliquity of $\sim54^{\circ}$ to the orbit normal; a solstice case at this obliquity is presented at the end of this section.
\begin{table*}[t]
\centering
\caption{Physical and thermophysical properties adopted for asteroid 2011 for different taxonomic classes.}
\label{tabthermal}
\begin{tabular}{lcccc}
\hline
\textbf{Parameter} & \textbf{C-type} & \textbf{E-type} & \textbf{Bare rock} & \textbf{Units} \\
\hline
Thermal inertia ($\Gamma$) & 300 & 200 & 2000 & J m$^{-2}$ K$^{-1}$ s$^{-1/2}$ \\
Albedo & 0.045 & 0.30 & 0.10 & - \\
Semimajor axis & \multicolumn{3}{c}{1.62} & au \\
Rotational period & \multicolumn{3}{c}{0.6107} & h \\
Density & 1.4 & 2.9 & 2.0 &~g~cm$^{-3}$ \\
Specific heat capacity & 500 & 300 & 800 & J kg$^{-1}$ K$^{-1}$ \\
\hline
\end{tabular}
\end{table*}

For our analysis, we consider two possible taxonomic classes for asteroid 2011 (C-type or E-type), which remain uncertain. We adopted a heliocentric distance of $r_a = 1.62$\,au (semimajor axis) and a rotational period of $P_{\mathrm{rot}} = 0.6107$\,h \citep{Monteiro_2020}. We also assume an emissivity of 0.9 and bulk densities appropriate for each taxonomic class. We selected the values of thermal inertia ($\Gamma$) and heat capacity based on the possible taxonomic classes of 2011~UW$_{158}$ \citep{muller2007, Okada2020, Piqueaux2021}. Tab. \ref{tabthermal} summarizes the parameter values used in the thermal analysis.

\begin{figure}[pos=h!]
    \centering
    \includegraphics[width=1\columnwidth]{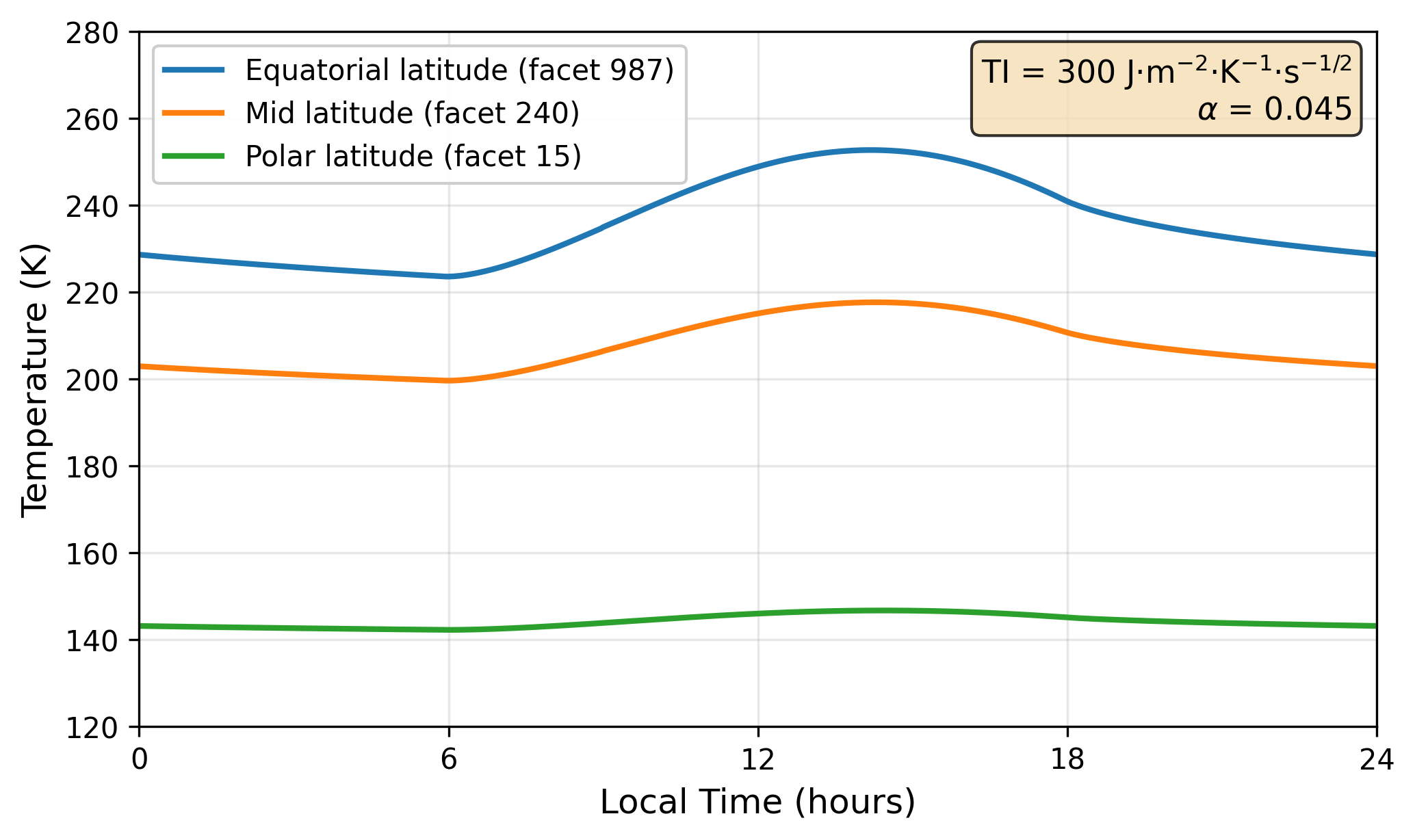}
    \captionof{figure}{Predicted diurnal surface temperatures at three representative surface points corresponding to those shown in Fig. \ref{thermal2011}, considering a thermal inertia of 300~J~m\(^{-2}\)~K\(^{-1}\)~s\(^{-1/2}\).}
    \label{thermal300}
\end{figure}

\begin{figure}[pos=h!]
    \centering
    \includegraphics[width=1\columnwidth]{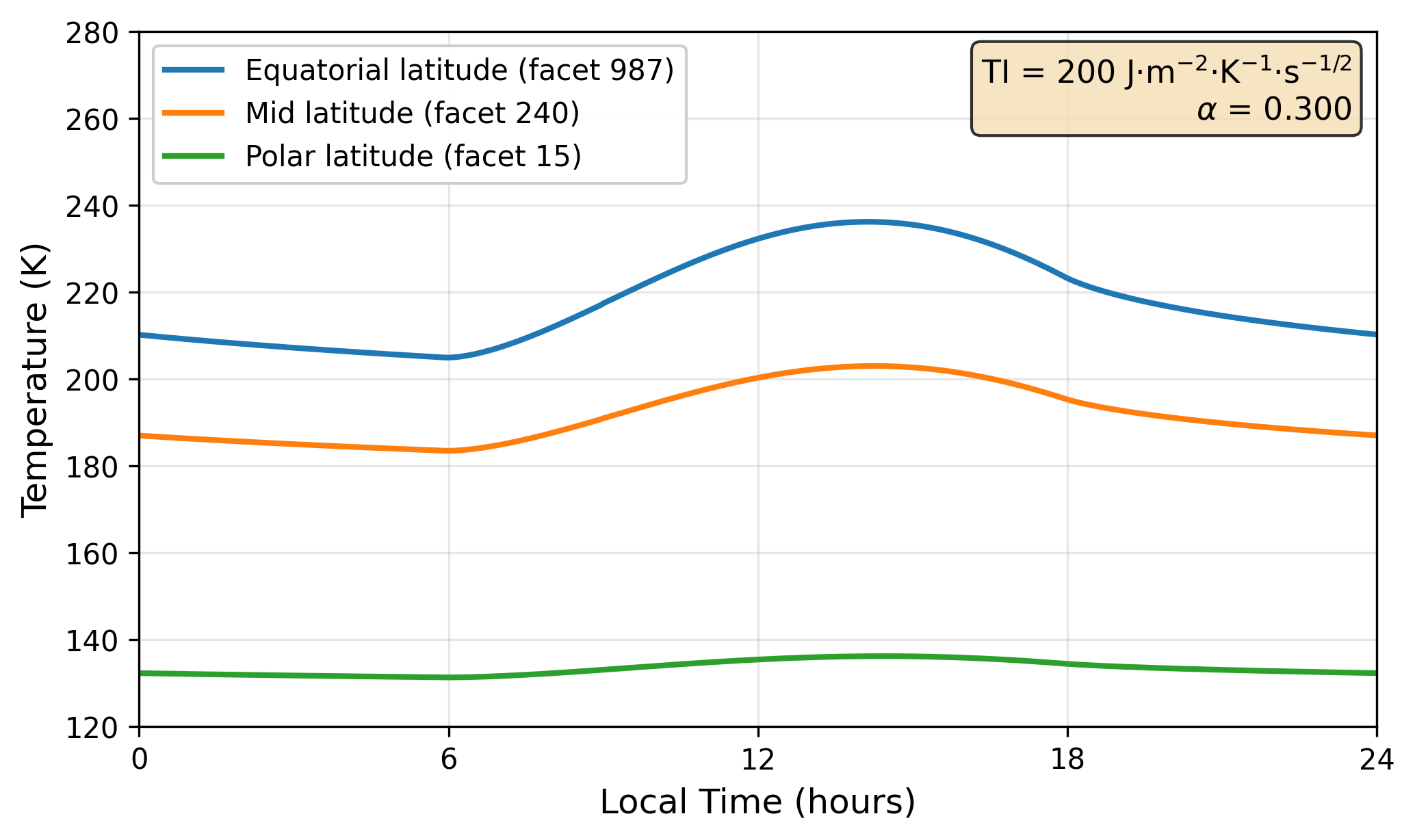}
    \captionof{figure}{Predicted diurnal surface temperatures at three representative surface points corresponding to those shown in Fig. \ref{thermal2011}, considering a thermal inertia of 200~J~m\(^{-2}\)~K\(^{-1}\)~s\(^{-1/2}\).}
    \label{thermal200}
\end{figure}

Figure \ref{thermal2011} shows the surface temperature distribution. For a nominal thermal inertia of $300\,\mathrm{J\,m^{-2}\,K^{-1}}\mathrm{s^{-1/2}}$ and an average density for a C-type asteroid, the diurnal equatorial temperature range is $\sim 222$ - $253$ K. In contrast, the polar temperatures drop to $\sim 130$ K (see Fig. \ref{thermal300}). Considering a typical density of an E-type asteroid, and a thermal inertia of $200\,\mathrm{J\,m^{-2}\,K^{-1}\,s^{-1/2}}$ and a higher albedo of 0.30, diurnal equatorial temperatures decrease to a range of $\sim 205$ - $235$ K. At the same time, the polar regions drop to $\sim 120$ K (see Fig. \ref{thermal200}).

\begin{figure}[pos=h!]
    \centering
    \includegraphics[width=\columnwidth]{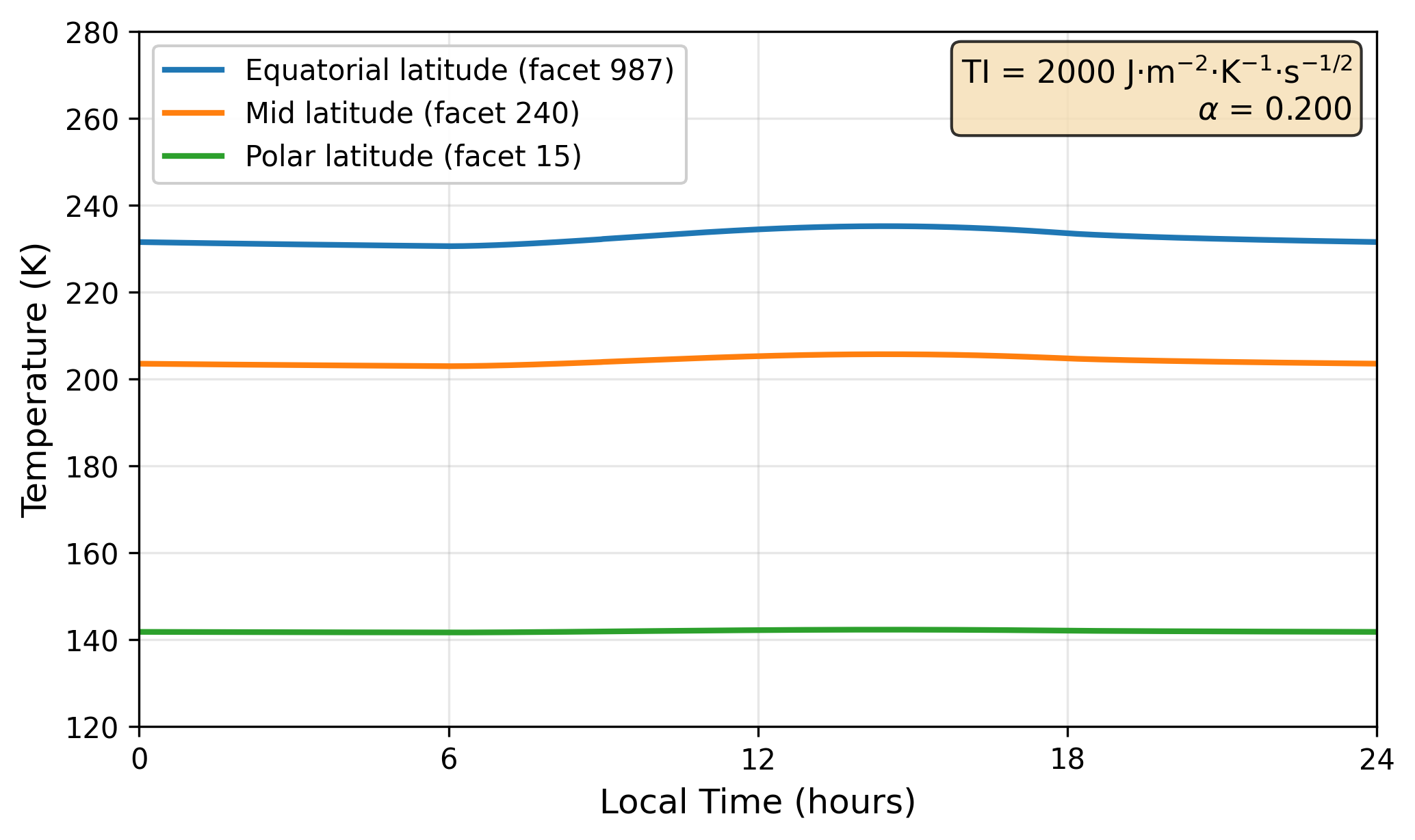}
    \captionof{figure}{Predicted diurnal surface temperatures at three representative surface points corresponding to those shown in Fig. \ref{thermal2011}, considering a thermal inertia of 2000~J~m\(^{-2}\)~K\(^{-1}\)~s\(^{-1/2}\).}
    \label{thermal2000}
\end{figure}

The temperatures strongly depend on the asteroid's super-fast rotation, resulting in small diurnal temperature variations across its regions (from the poles to the equator). This effect is more pronounced when the asteroid lacks regolith or surface rocks, resulting in very high thermal inertia. In such situations, high thermal inertia means that the material responds very slowly to changes in solar flux, producing a very low diurnal variation (as shown in Fig. \ref{thermal2000}). 

Because this asteroid is an SFR, diurnal temperature variation decreases further as thermal inertia increases. Moreover, even with low thermal inertia, rapid rotation prevents the surface from responding quickly, further reducing diurnal variation. We also observed that thermal inertia generally has a relatively small impact on mean temperature, whereas albedo has a very substantial influence.

The obliquity implied by the preferred pole solution, $\sim54^{\circ}$, means that 2011~UW$_{158}$ experiences a seasonal cycle. The sub-solar latitude of $0^{\circ}$ modeled above is the midpoint of this cycle. As a bounding case, we therefore repeated the C-type calculation with the Sun placed $54^{\circ}$ out of the asteroid's equatorial plane, corresponding to solstice at the preferred pole. Latitudes greater than $36^{\circ}$ are then continuously illuminated or in continuous darkness; the equator cools to a mean of $\sim 211$~K, and the summer pole reaches $\sim 298$~K, exceeding the highest temperature found anywhere on the surface at equinox.

At solstice, the diurnal amplitude is $\sim 4$~K at the summer pole and $\sim 20$~K at the equator, comparable to the equinox values, as the small amplitude follows from the rotation period being short compared with the thermal response time of the surface. This condition holds at any spin-axis orientation.

\section{Spherical Harmonic Representation for Asteroid 2011~UW$_{158}$}
\label{sec:HarmonicModel}
Modeling the gravitational field of Small Solar System Bodies (SSSBs), such as asteroid (436724) 2011~UW$_{158}$, is requisite for a detailed assessment of their orbital and surface environments. Due to their highly irregular shapes and potentially non-uniform mass distribution, simplified gravitational models may fail to capture essential dynamical features. Among the available methods, the polyhedron approach stands out for its high fidelity, as it directly incorporates the body's detailed geometry through its numerous faces and vertices \citep{Werner_1997, tsoulis2001}. Nevertheless, the numerical burden of this formulation significantly limits its applicability to large-scale or long-term dynamical investigations.

To balance accuracy and efficiency, we use a truncated spherical harmonic expansion of the gravitational potential. This analytical representation expresses the potential as a series of spherical harmonics, yielding a smooth, differentiable function that can be evaluated at considerably lower computational cost. The harmonic coefficients are determined through an optimization process that reproduces the gravitational potential generated by the polyhedron approach within a prescribed spatial domain surrounding the asteroid. In the present study, the normalized coefficients $C_{20}$, $C_{22}$, $C_{30}$, $C_{31}$, $C_{32}$, $C_{33}$, $C_{40}$, $C_{41}$, $C_{42}$, $C_{43}$, and $C_{44}$, were estimated by minimizing the mean squared deviation between the polyhedron-based potential and its harmonic approximation.

The primary benefit of the spherical harmonic formulation is its substantial computational performance gains, combined with analytical simplicity. While simulations based directly on the polyhedron method may require several days of computation, the harmonic representation enables comparable analyses in minutes. This efficiency gain enables extensive exploration of initial conditions and orbital configurations, facilitating a more comprehensive assessment of dynamical behavior near 2011~UW$_{158}$ while preserving the leading effects of its irregular mass distribution.

\subsection{Mathematical Model}
The external gravitational field of 2011~UW$_{158}$  is modeled using a truncated spherical harmonic expansion of the potential, retaining terms up to degree $n=4$. The set of normalized harmonic coefficients considered in the model is
\[
\{C_{20},\, C_{22},\, C_{30},\, C_{31},\, C_{32},\, C_{33},\, C_{40},\, C_{41},\, C_{42},\, C_{43},\, C_{44}\},
\]
while the coefficient $C_{21}$ is constrained to zero because the origin of the adopted body-fixed reference frame is located at the asteroid's center of mass and its axes are aligned with the principal axes of inertia.

Let $(x,y,z)$ denote the Cartesian coordinates of an arbitrary point expressed in the body-fixed frame. In compact form, the gravitational potential associated with the spherical harmonic approximation can be written as
\begin{equation}
U_{\mathrm{harm}}(x,y,z) \;=\; -\frac{k}{r}\left[\,1 + \sum_{n=2}^{4}\left(\frac{1}{r}\right)^{n} \Phi_n(x,y,z)\,\right],
\label{U_harm_compact}
\end{equation}
where $r=\sqrt{x^2+y^2+z^2}$ and $k$ is a dimensionless scaling parameter. The functions $\Phi_n$ are homogeneous polynomials of degree $n$ that group all spherical harmonic contributions of the corresponding order.

By explicitly expanding the terms retained in the series, the gravitational potential can be expressed in Cartesian coordinates by Eq. \eqref{U_harm}, where $C_{nm}$ denotes the normalized spherical harmonic coefficients up to degree and order $n=4$. The absence of the $C_{21}$ term reflects the symmetry properties imposed by the selected reference frame.

The parameter $k$ appearing in Eq. \eqref{U_harm_compact} and Eq. \eqref{U_harm} represents the \textit{force ratio}, which defines the relative magnitude of the gravitational potential with respect to the centrifugal contribution. It is given by:
\begin{equation}
k = \frac{G M}{\omega^2 L^3},
\label{eq:k}
\end{equation}
where $L$ is a characteristic length scale.

\begin{figure*}[!t]
\centering
\begin{multline}
U_{\mathrm{harm}}(x,y,z) = -\frac{k}{r}\Bigg\{ 1 
- \left(\frac{1}{r}\right)^{2}\Bigg[ 
\frac{C_{20}\,\bigl(x^{2}+y^{2}-2z^{2}\bigr)}{2\,r^{2}}
- \frac{3\,C_{22}\,\bigl(x^{2}-y^{2}\bigr)}{r^{2}}
\Bigg]
+ \left(\frac{1}{r}\right)^{3}\Bigg[
\frac{C_{30}\,z\bigl(-3x^{2}-3y^{2}+2z^{2}\bigr)}{2\,r^{3}}
\\+ \frac{3\,C_{31}\,x\bigl(x^{2}+y^{2}-4z^{2}\bigr)}{2\,r^{3}}
+ \frac{15\,C_{32}\,\bigl(x^{2}-y^{2}\bigr)}{r^{3}}
- \frac{15\,C_{33}\,x\bigl(x^{2}-3y^{2}\bigr)}{r^{3}}
\Bigg] 
\\- \left(\frac{1}{r}\right)^{4}\Bigg[
\frac{C_{40}\,\bigl(3(x^{2}+y^{2})^{2}-24(x^{2}+y^{2})z^{2}+8z^{4}\bigr)}{8\,r^{4}}
+ \frac{5\,C_{41}\,x z\bigl(3x^{2}+3y^{2}-4z^{2}\bigr)}{2\,r^{4}}
\\- \frac{15\,C_{42}\,(x^{2}-y^{2})\bigl(x^{2}+y^{2}-6z^{2}\bigr)}{2\,r^{4}} 
- \frac{105\,C_{43}\,x z\bigl(x^{2}-3y^{2}\bigr)}{r^{4}}
+ \frac{105\,C_{44}\,\bigl(x^{4}-6x^{2}y^{2}+y^{4}\bigr)}{r^{4}}
\Bigg]
\Bigg\},
\label{U_harm}
\end{multline}
\end{figure*}
\vspace{10pt}


\subsection{Determination of the Spherical Harmonic Parameters}
\label{S2}
The spherical harmonic formulation introduced in the previous section depends on the accurate estimation of the normalized gravitational coefficients $C_{nm}$ up to the selected degree and order, together with the force ratio parameter $k$. These quantities must be properly determined in order for the truncated expansion to provide a reliable approximation of the asteroid's gravitational field.

To this end, we adopt a parameter-estimation strategy based on potential matching, in which we adjust the spherical-harmonic model to reproduce the gravitational potential generated by the polyhedron method. The reference potential is computed for asteroid (436724) 2011~UW$_{158}$ at a set of $N=10{,}000$ sampling points distributed at radial distances ranging from 0.4~km to 0.8~km from the asteroid's center of mass. This region ensures that the fitting procedure captures the external gravitational field relevant to the orbital regime investigated in this study.

The calibration process minimizes the discrepancy between the two potential models. The associated objective function is defined as
\begin{equation}
\mathbf{J}(C_{nm}, k) = \sum_{i=1}^{N} \left[ U_{\mathrm{harm}}(x_i,y_i,z_i) - U_{\mathrm{poly}}(x_i,y_i,z_i) \right]^2,
\label{opt_harm}
\end{equation}
where $U_{\mathrm{poly}}$ denotes the gravitational potential computed using the polyhedron approach, and $U_{\mathrm{harm}}$ corresponds to the truncated spherical harmonic expansion given in Eq. \eqref{U_harm}.

The resulting nonlinear least-squares problem is solved using standard nonlinear programming techniques available in MATLAB, with the performance index $\mathbf{J}$ serving as the quantity to be minimized \citep{2021MNRAS.502.4277S, 2022AdSpR..70.3362S, article22}. The optimized parameter set yields a harmonic representation that closely reproduces the gravitational field of 2011~UW$_{158}$ while substantially reducing computational cost compared with direct polyhedron evaluations.

\subsection{Boundary Constraints and Initial Conditions}
The nonlinear optimization procedure requires appropriate parameter bounds and initial estimates to ensure numerical stability and physically meaningful solutions. These constraints were based on the expected magnitudes of the asteroid (436724)~2011~UW$_{158}$'s gravitational harmonics, as well as preliminary numerical tests to ensure robust convergence.
The vector of optimization parameters is defined as
\begin{equation}
\begin{split}
\mathbf{x} = [\,C_{20},~C_{22},~C_{30},~C_{31},~C_{32}, \\
~C_{33},~C_{40},~C_{41},~C_{42},~C_{43},~C_{44},~k\,],
\end{split}
\end{equation}
where $C_{nm}$ denote the normalized spherical harmonic coefficients and $k$ is the force ratio parameter introduced previously.

The initial guess $\mathbf{x}_0$, together with the lower and upper bounds, $\mathbf{lb}$ and $\mathbf{ub}$, was selected to restrict the search space to physically admissible regions while preserving sufficient flexibility for the optimizer. These values are summarized in Eqs. \eqref{x0}, \eqref{lb}, and \eqref{ub}, respectively.
\begin{figure*}[!b]
\centering
\begin{multline}
\mathbf{x_0} =
\begin{bmatrix}
-8.30\times10^{-3},~
2.62\times10^{-3},~
1.09\times10^{-4},~
-2.64\times10^{-5},~
-2.90\times10^{-6},~
-6.18\times10^{-6},\\
7.41\times10^{-5},~
-1.26\times10^{-6},~
-3.04\times10^{-6},~
-3.78\times10^{-7},~
2.85\times10^{-7},~
3.01\times10^{-4}
\end{bmatrix},
\label{x0}
\end{multline}
\vspace{5pt}
\begin{multline}
\mathbf{lb} =
\begin{bmatrix}
-5.06\times10^{-2},~
1.31\times10^{-3},~
0,~
-1.69\times10^{-4},~
-1.53\times10^{-2},~
-1.02\times10^{-2},\\
-2.01\times10^{-2},~
-4.03\times10^{-4},~
-4.13\times10^{-2},~
-1.28\times10^{-4},~
-1.57\times10^{-2},~
1.0\times10^{-9}
\end{bmatrix},
\label{lb}
\end{multline}
\vspace{5pt}
\begin{multline}
\mathbf{ub} =
\begin{bmatrix}
5.06\times10^{-2},~
1.82\times10^{-2},~
7.26\times10^{-3},~
1.87\times10^{-4},~
7.27\times10^{-3},~
1.22\times10^{-3},\\
4.01\times10^{-2},~
1.20\times10^{-3},~
2.13\times10^{-2},~
1.08\times10^{-4},~
1.17\times10^{-2},~
15
\end{bmatrix}.
\label{ub}
\end{multline}
\end{figure*}

The selected bounds prevent the optimizer from exploring non-physical regions of the parameter space while ensuring numerical robustness throughout the iterative process. The initial parameter vector $\mathbf{x}_0$ was derived from exploratory simulations and sensitivity analyses, which indicated stable convergence behavior and reduced residual errors between the spherical harmonic approximation and the polyhedral reference model. We also tested several initial guesses during the exploratory stage to assess the optimization procedure's convergence behavior and verify the robustness of the resulting solution.

The optimized values of the spherical harmonic coefficients and the force ratio parameter obtained from the procedure described above are summarized in Tab. \ref{tab:coefotimizados}. These parameters define the final gravitational model adopted for 2011~UW$_{158}$.
\begin{table*}[t]
\centering
\renewcommand{\arraystretch}{1.1}
\resizebox{1.0\textwidth}{!}{
\begin{tabular}{|l||c|c|c|c|c|c|c|c|c|c|c|c|}
\hline
\textbf{Coefficients} &
$C_{20}$ & $C_{22}$ & $C_{30}$ & $C_{31}$ & $C_{32}$ & $C_{33}$ &
$C_{40}$ & $C_{41}$ & $C_{42}$ & $C_{43}$ & $C_{44}$ & $k$ \\
\hline
Values &
$-1.06\times10^{-2}$ &
$2.76\times10^{-3}$ &
$1.59\times10^{-4}$ &
$-4.50\times10^{-5}$ &
$-3.38\times10^{-6}$ &
$-8.73\times10^{-6}$ &
$1.17\times10^{-4}$ &
$-2.77\times10^{-6}$ &
$-3.72\times10^{-6}$ &
$-6.94\times10^{-7}$ &
$2.31\times10^{-7}$ &
$3.01\times10^{-4}$ \\
\hline
\end{tabular}}
\caption{Optimized spherical harmonic coefficients and force ratio parameter for asteroid (436724)~2011~UW$_{158}$.}
\label{tab:coefotimizados}
\end{table*}

\subsection{Analysis of the Optimized Gravitational Coefficients}
The optimized spherical harmonic coefficients for asteroid (436724)~2011~UW$_{158}$ reveal a clear hierarchy in the relative importance of the retained terms, reflecting the body's global shape and mass distribution. The dominant contribution arises from the second-degree harmonics, particularly the zonal coefficient $C_{20}$ and the tesseral coefficient $C_{22}$, which exhibit magnitudes one to two orders of magnitude larger than most higher-degree terms.

The negative value of $C_{20}$ indicates a deviation from spherical symmetry associated with the asteroid's overall elongation or flattening along its principal axis of rotation. This behavior is consistent with the irregular, elongated shapes commonly inferred for small near-Earth asteroids, in which the mass distribution departs significantly from that of a uniform sphere. The relatively large magnitude of $C_{20}$ confirms that the quadrupole component is the leading correction to the central potential and dominates the long-range gravitational field.

The coefficient $C_{22}$, although smaller than $C_{20}$, remains a significant contributor to the gravitational potential. Its non-negligible value reflects equatorial asymmetry, indicating that the mass distribution in the equatorial plane is not axisymmetric. This tesseral term plays a fundamental role in shaping the non-axisymmetric component of the gravitational field. It is expected to strongly influence the dynamics of particles and spacecraft near the asteroid.

In contrast, the third- and fourth-degree coefficients exhibit substantially smaller magnitudes, with several terms reaching values of order $10^{-6}$ to $10^{-7}$. These coefficients correspond to finer-scale variations in the gravitational field associated with local shape irregularities rather than the body's global geometry. Their reduced magnitude suggests that such high-order features contribute only weak corrections to the external potential within the region considered for the optimization.

Among the higher-degree terms, coefficients such as $C_{30}$ and $C_{40}$ display slightly larger values compared to the remaining harmonics of the same degree, indicating that some degree of axial asymmetry persists beyond the quadrupole level. Nevertheless, their contribution remains secondary when compared to the dominant second-degree harmonics. The remaining tesseral and sectoral terms, particularly those with magnitudes below $10^{-6}$, are effectively negligible for many practical applications and primarily serve to refine the local accuracy of the potential.

Overall, the observed distribution of coefficient magnitudes supports the physical consistency of the optimization procedure. The dominance of low-degree harmonics reflects the fact that the gravitational field at distances comparable to several characteristic radii of the asteroid is primarily governed by its global shape. At the same time, higher-degree terms decay rapidly with distance. This behavior justifies using truncated spherical harmonic models for dynamical studies around 2011~UW$_{158}$, since a limited number of coefficients captures the essential gravitational features.

From a dynamical perspective, these results indicate that models that retain only second-degree harmonics provide a robust first-order approximation of the gravitational environment. At the same time, including higher-degree terms offers incremental improvements at the cost of increased complexity. Consequently, the optimized coefficient set obtained here establishes a physically meaningful and computationally efficient representation of the gravitational field.

\subsection{Numerical comparison between the Spherical Harmonic and Polyhedron Method}
\label{Section5}
In this section, we validate the spherical harmonic representation by direct comparison with the polyhedron method, which we adopt as the reference solution. The comparison uses the complete optimized harmonic expansion, including all coefficients up to degree 4 and order 4, as determined in the previous sections.

To quantify the agreement between the two models, the relative error of the gravitational potential is evaluated at each sampled spatial point according to
\begin{equation}
\varepsilon_i = \left| \frac{U^{\mathrm{harm}}_i - U^{\mathrm{poly}}_i}{U^{\mathrm{poly}}_i} \right| \times 100\%,
\label{relative_harmonics}
\end{equation}
where $U^{\mathrm{harm}}_i$ and $U^{\mathrm{poly}}_i$ denote the gravitational potentials computed using the spherical harmonic Eq. \eqref{U_harm} and polyhedron Eq. \eqref{eq:potencial_poliedro} models, respectively, at the same location.

\begin{figure}[pos=h!]
\centering
\includegraphics[width=\columnwidth]{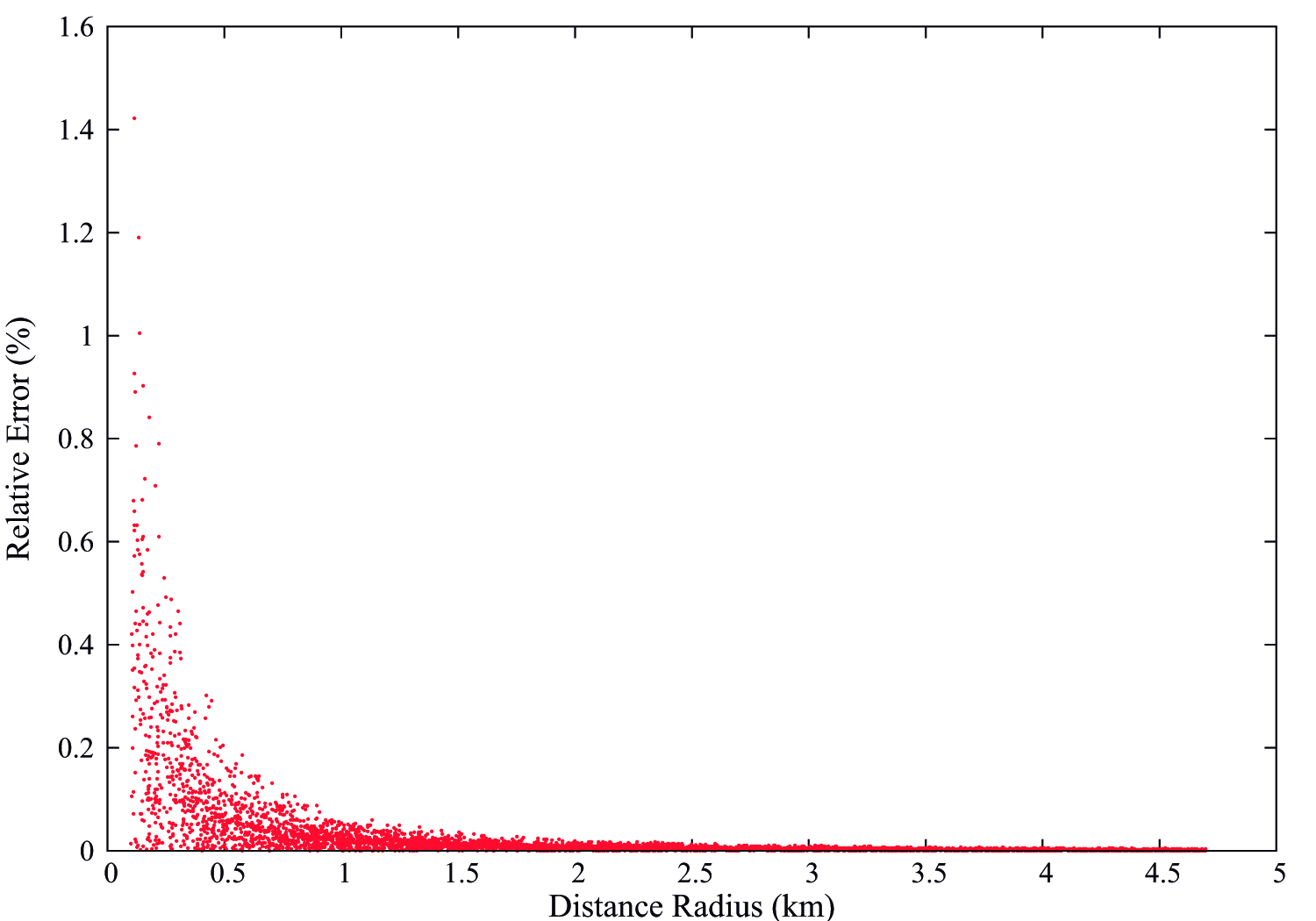}
\caption{Relative error of the gravitational potential between the optimized spherical harmonic expansion and the polyhedron method for 2011~UW$_{158}$.}
\label{errorUW158}
\end{figure}

Figure \ref{errorUW158} presents the spatial distribution of the relative error as a function of the distance from the surface of asteroid (436724)~2011~UW$_{158}$. 

The results show excellent agreement between the spherical harmonic approximation and the reference polyhedron solution across the entire analyzed region. As shown in Fig. \ref{errorUW158}, the relative error reaches its maximum values in the immediate vicinity of the asteroid surface, where the gravitational field is more sensitive to small-scale geometric features. Even in this region, however, the relative error remains limited to $\sim$1.4\%, indicating that the truncated harmonic expansion provides a reliable representation of the near-field gravitational potential.

As distance from the surface increases, the relative error decreases rapidly, falling below 0.05\% at distances on the order of 1~km. This behavior reflects the diminishing influence of local shape irregularities and the increasing dominance of the lower-degree harmonic terms, which efficiently capture the global mass distribution.

Including the full set of coefficients up to degree and order four enables the spherical harmonic model to reproduce both the large-scale asymmetry and the dominant non-spherical contributions of the gravitational field. The low error levels observed throughout the domain confirm that the optimized harmonic expansion closely matches the polyhedron potential, not only in the far-field region but also near the asteroid surface.

Overall, these results validate the optimized spherical harmonic model as a robust and computationally efficient representation of the gravitational field. A key aspect of the present work is the systematic determination of the spherical harmonic coefficients via an optimization procedure that directly references the high-fidelity polyhedron potential. By minimizing the discrepancy between the harmonic expansion and the polyhedron method, the optimized coefficients provide a more consistent representation of the gravitational environment than simpler analytical fits or heuristic assumptions.

This optimization-based calibration not only closely matches the reference potential across the region of interest but also provides a principled way to truncate the harmonic series with minimal fidelity loss.
As a result, the derived harmonic model achieves high accuracy while retaining the analytical and computational advantages of spherical harmonic representations. Such a calibrated harmonic field thus represents a powerful and versatile tool for dynamical investigations, including near-surface orbital motion, long-term propagation, and extensive parameter-space explorations where computational efficiency is essential.
In particular, a consistent, smooth harmonic representation of the gravitational potential facilitates identifying and analyzing periodic orbits around irregular bodies, enabling systematic searches, stability assessments, and investigation of dynamical structures near the asteroid.

\section{Orbits around the Asteroid 2011~UW$_{158}$} \label{sec:orbits}
The literature shows that near small bodies, the body's shape and rotation strongly perturb the dynamics, and that small changes in initial conditions can lead to escape or collision, indicating extreme sensitivity around the body \citep{scheeres2012orbital}.
Since asteroid 2011~UW$_{158}$ is classified as an SFR, its high angular speed strongly influences its dynamic environment.
In these cases, the synchronous radius (see Eq. \eqref{eq:rsincrono}), which defines the critical distance where the orbital frequency equals the rotation of the central body, marking the transition between regimes dominated by gravity and centrifugal force in the rotating frame of reference, tends to be very close to or inside the body. 
\begin{equation}
    r_{sync}=\left( \frac{G M}{\omega^2}\right)^{1/3}.
    \label{eq:rsincrono}
\end{equation}
The $r_{sync} = 66.965$\,m is placed inside the body, causing the centrifugal force to dominate in the rotating frame of reference, eliminating stable direct orbits and favoring the existence of naturally stable retrograde orbits \citep{Hu_Scheeres_2004,Amarante2021bennu,AmaranteWinter2022}.

\subsection{Planar Symmetric Periodic Orbits}
\label{sec:PO}
From this, we used the Grid Search Method \citep{Markellosetal1974} to find planar symmetric periodic orbits around the asteroid and then classify their directions and stabilities.

It is a numerical technique used to identify periodic orbits, particularly effective in non-integrable problems, such as those associated with irregular gravitational fields or simplified models of rotating small bodies. The method is based on the exhaustive scanning of a discrete set of previously defined initial conditions, in this case a grid $(x_0,C_J)$ in phase space, where $x_0$ represents the position $x$ at the initial instant ($t=0$) and $C_J=(x^2 + y^2)-2U_{harm} -(\dot{x}^2+\dot{y}^2)$ is the Jacobi constant. A planar symmetrical periodic orbit crosses the $x$-axis perpendicularly twice, at times $T$ and $T/2$ (where $T$ is the orbital period), so that $y=\dot{x}=0$, and then $C_J(x_0,0,0,\dot{y}_0)$ are the initial conditions. The motion is restricted to the equatorial plane of the body, and the gravitational potential is symmetrical with respect to the axis perpendicular to the plane.

We also approximated the body's irregular gravitational potential with an equivalent ellipsoid, expanded in spherical harmonics up to fourth order. Despite the planarity of the problem, we included all the optimized coefficients and the value of $k$ from Tab. \ref{tab:coefotimizados} in the equation of the effective potential $V=-\frac{1}{2}(x^2+y^2) + U_{harm}$, for $z=0$ ($U_{harm}$ is in Eq. \eqref{U_harm}).

Regarding the linear stability of the orbit, we calculate it from the stability index $SI=\mid \lambda_1+\frac{1}{\lambda_1}\mid$, since $\lambda_i$ for $i=1,2,3,4$ represents the four eigenvalues of the monodromy matrix $M = \Phi(t_0+T,t_0)$, obtained by integrating the state transition matrix of the problem ($\dot \Phi$) over a complete orbital period ($T$) \citep{parker2014low}. Due to the symplectic and planar structure of the problem, the eigenvalues appear in reciprocal pairs and, as a consequence of the periodicity of the solution, two of them are equal to one, so that $\lambda_1, \lambda_2=\frac{1}{\lambda_1}, \lambda_3=1, \lambda_4=1$. Thus:

$SI < 2$, the orbit is linearly stable, 

$SI = 2$, the orbit is neutrally stable,

$SI > 2$, the orbit is unstable.

\begin{figure}[pos=h!]
    \centering
    \includegraphics[width=\columnwidth]{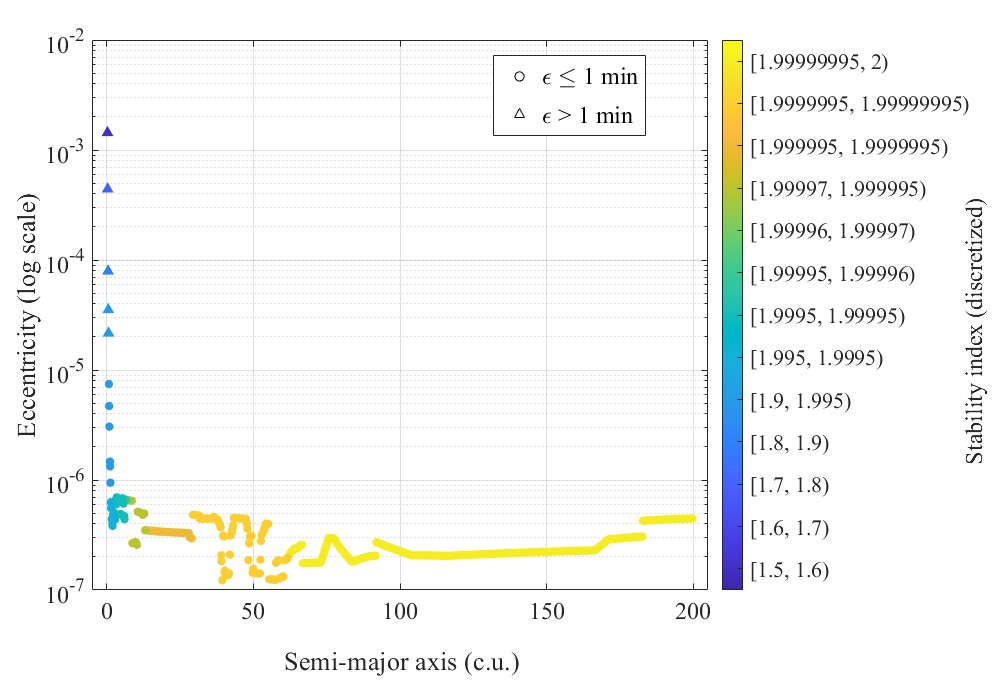}
    \caption{Map of the initial conditions $a \times e$, stability, and absolute error of periods of the orbits of Family 1.}
    \label{fig:Fam1}
\end{figure}

\begin{figure}[pos=h!]
    \centering
    \includegraphics[width=\columnwidth]{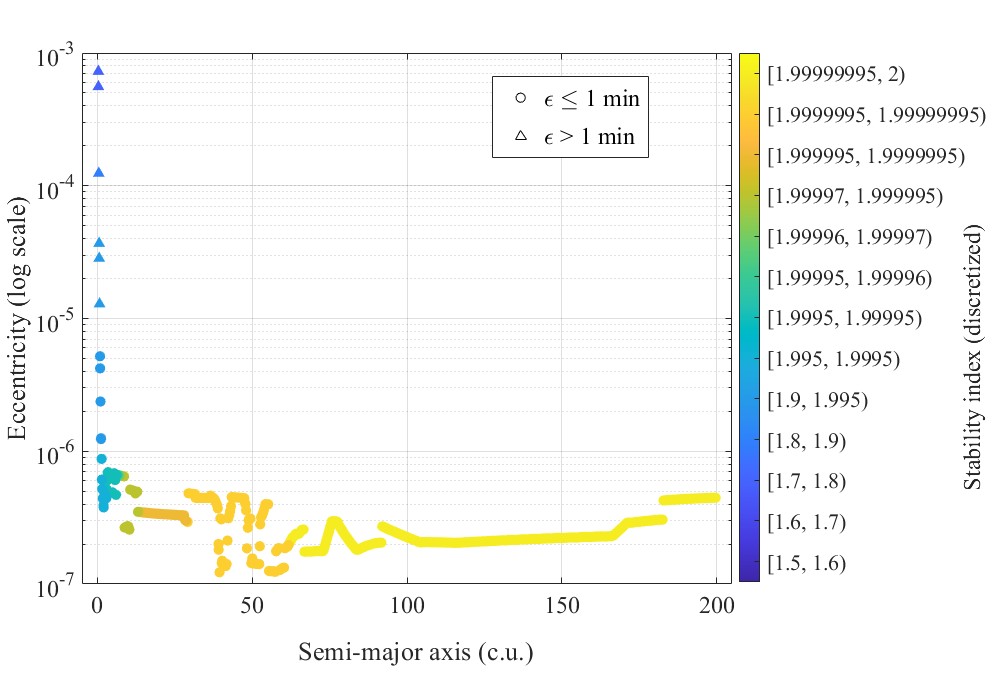}
    \caption{Map of the initial conditions $a \times e$, stability, and absolute error of periods of the orbits of Family 2.}
    \label{fig:Fam2}
\end{figure}

\begin{figure}[pos=h!]
    \centering
    \includegraphics[width=\columnwidth]{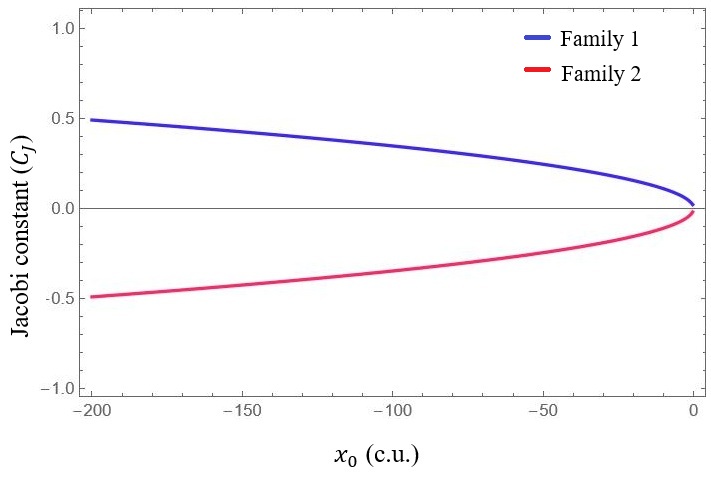}
    \caption{Jacobi Constant as a function of $x_0$, i.e., $x$ at initial time zero, of Families 1 and 2.}
    \label{fig:CJ}
\end{figure}

We found two families of stable, retrograde quasi-circular orbits. Figures \ref{fig:Fam1} and \ref{fig:Fam2} show the initial conditions $a \times e$ of these families, respectively. Given that $a$ and $e$ are calculated from Eq. \eqref{eq:a} and Eq. \eqref{eq:e}, being $r_{min}$ and $r_{max}$ the minimum and maximum distance of the particle from the asteroid, respectively, calculated via numerical integration (checking the greatest and smallest distances at each step). The symbols $\circ$ and $\triangle$ differentiate the absolute error $\epsilon$ (Eq. \eqref{eq:errorabsolute}) of the orbital period compared to the rotational period of the asteroid ($P$), since the orbital period of the particle in the inertial system is $T_{i}=T \frac{1}{\omega}$, where $1/\omega$ is a time unit. This calculation was done to identify doubly synchronous orbits, that is, orbits in which the orbital period of the particle/spacecraft is equal to the rotational period of the orbited body. For cases represented by $\triangle$ with $\epsilon > 1$ min, the maximum value achieved is $\sim$4\,min. The colors represent the orbits' stability indices. 
\begin{equation}
    a = \frac{r_{min}+r_{max}}{2},
    \label{eq:a}
\end{equation}
\begin{equation}
    e = 1-\frac{r_{min}}{a},
    \label{eq:e}
\end{equation}
\begin{equation}
     \epsilon = T_{i} - P.
    \label{eq:errorabsolute}
\end{equation}
Figures \ref{fig:Fam1} and \ref{fig:Fam2} have very similar initial conditions, with small variations in the values of $a$ and $e$, and mainly variations in the Jacobi constant, which for Family 1 (Fig. \ref{fig:Fam1}) is positive and for Family 2 (Fig. \ref{fig:Fam2}) is negative (see Fig. \ref{fig:CJ}). 

Orbits with very similar initial conditions may have a positive Jacobi constant in one and a negative one in the other because, in nonlinear dynamical systems described in the rotating frame of reference, $C_J$ results from the delicate balance between the pseudo-potential $((x^2 + y^2)-$$2U_{harm})$ and the quadratic kinetic term ($\dot{x}^2+\dot{y}^2$), so that small variations in the position components or, mainly, in the velocity, especially when the system is close to a critical energy surface, are sufficient to change this balance and reverse the sign of the constant, reflecting a qualitative change in the energy level and topology of the dynamically accessible regions.

The Jacobi constant shows the energy level of the orbit in the rotating frame of reference. If $C_J > 0$, it means that the potential term dominates, and the rotational energy is negative. In this case, the particle is at a lower energy level. In general, it is more confined to the allowed regions of space. If $C_J < 0$, kinetic energy dominates, and rotational energy is positive. The orbit is at a higher energy level. The forbidden regions diminish or disappear.

The eccentricity in both cases is on the order of $10^{-3}$ or less, with the vast majority varying between $10^{-6}$ and $10^{-7}$. The semimajor axis extends up to 200 canonical units (where 1 c.u. $\sim$ 0.998 km). In Family 1, there are 5 orbits with an absolute error greater than 1 minute, and in Family 2, there are 6 orbits; all others are characterized as synchronous double orbits with an error less than 1 minute. Regarding stability, all cases are stable, ranging from 1.5 to near the $SI$ limit of 2.0, but not exactly 2.0. All orbits in both families are retrograde, meaning they have a direction opposite to the motion of the orbited body.

\subsection{Station-Keeping around Asteroid 2011~UW$_{158}$}
Station-keeping is a maneuver performed to maintain a spacecraft's orbit at its nominal value, correcting for perturbations. In this section, we evaluate the costs associated with maintaining the spacecraft's periodicity relative to the asteroid. The objective is to quantify the influence of the non-sphericity of the body on the fuel-equivalent $\Delta V$ costs.

To quantify this influence on costs, compare the perturbed orbit with a Keplerian one, which naturally arises from dynamics around a spherical body. A Keplerian orbit (with respect to the inertial frame) in the $x$-$y$ plane of motion is periodic in the rotating frame of reference when its period is given by
\begin{equation}
    T=\frac{2 \pi}{\omega - \sqrt{\frac{GM}{a^3}}},
    \label{eq:periodic}
\end{equation}
where $a$ is the semimajor axis of this reference orbit.

We adopt a circular Keplerian orbit as reference. Thus, the semimajor axis $a$ is the initial distance of the spacecraft from the barycenter of 2011~UW$_{158}$. The spacecraft is initially (at time 0) located at the coordinates $[\,a,0,0\,]$ and performs a complete orbit around 2011~UW$_{158}$, returning to the same point $[\,a,0,0\,]$ after a time of flight $T$. If we neglect the perturbation, assuming a spherical shape by making every one of the harmonic coefficients \( \{C_{20},\, C_{22},\, C_{30},\, C_{31},\, C_{32},\, C_{33},\, C_{40},\, C_{41},\, C_{42},\, C_{43},\, C_{44}\} \) 
\noindent \\equal to zero, then the initial velocity equals the final velocity after one relative revolution (after a time of flight $T$) in the rotating frame. This result holds because the reference orbit is circular. On the other hand, when the coefficients are taken into account, the orbit is perturbed, and a difference between the final and initial velocities arises. We define $\Delta V$ as the magnitude of this difference. It is considered the fuel-equivalent index because the spacecraft can perform an impulsive maneuver at $[\,a,0,0\,]$ every period $T$ to maintain the orbit's periodicity with respect to the rotating body.

When we consider the harmonic coefficients, the orbit starting at the point $[\,a,0,0\,]$ is perturbed; then, after a time $T$, a velocity associated with a Keplerian orbit does not ensure that the orbit will return to the same starting point $[\,a,0,0\,]$. To enforce this situation, we adopted the recently developed overdetermined constraints technique based on the Theory of Functional Connections (TFC) shown in \citep{tfc_segmentation}. TFC has been successfully applied to solve important astrodynamics problems, from evaluating Earth-Moon transfer costs \citep{fastTFC} to maneuvering a spacecraft with no fuel propellant consumption \citep{almeida24solar}.

We adopt the constrained functional derived for the non-segmented case shown in Eq.~(7) of \citet{tfc_segmentation} and analytically enforce the periodicity through the constraint $\mathbf{r}_i=\mathbf{r}_f$ as defined in the same reference. After applying the transformation to the constrained functional, we discretize the resulting differential equation of motion in time using the collocation method. We then use a nonlinear least-squares optimization procedure to minimize a loss function based on the magnitude of the acceleration minus the specific forces in the rotating frame of reference. We implement the numerical technique in Python, using automatic differentiation \citep{10.1145/355586.364791} and the TFC module \citep{tfc2021github}.

\begin{figure}[pos=h!]
\centering
\includegraphics[width=\columnwidth]{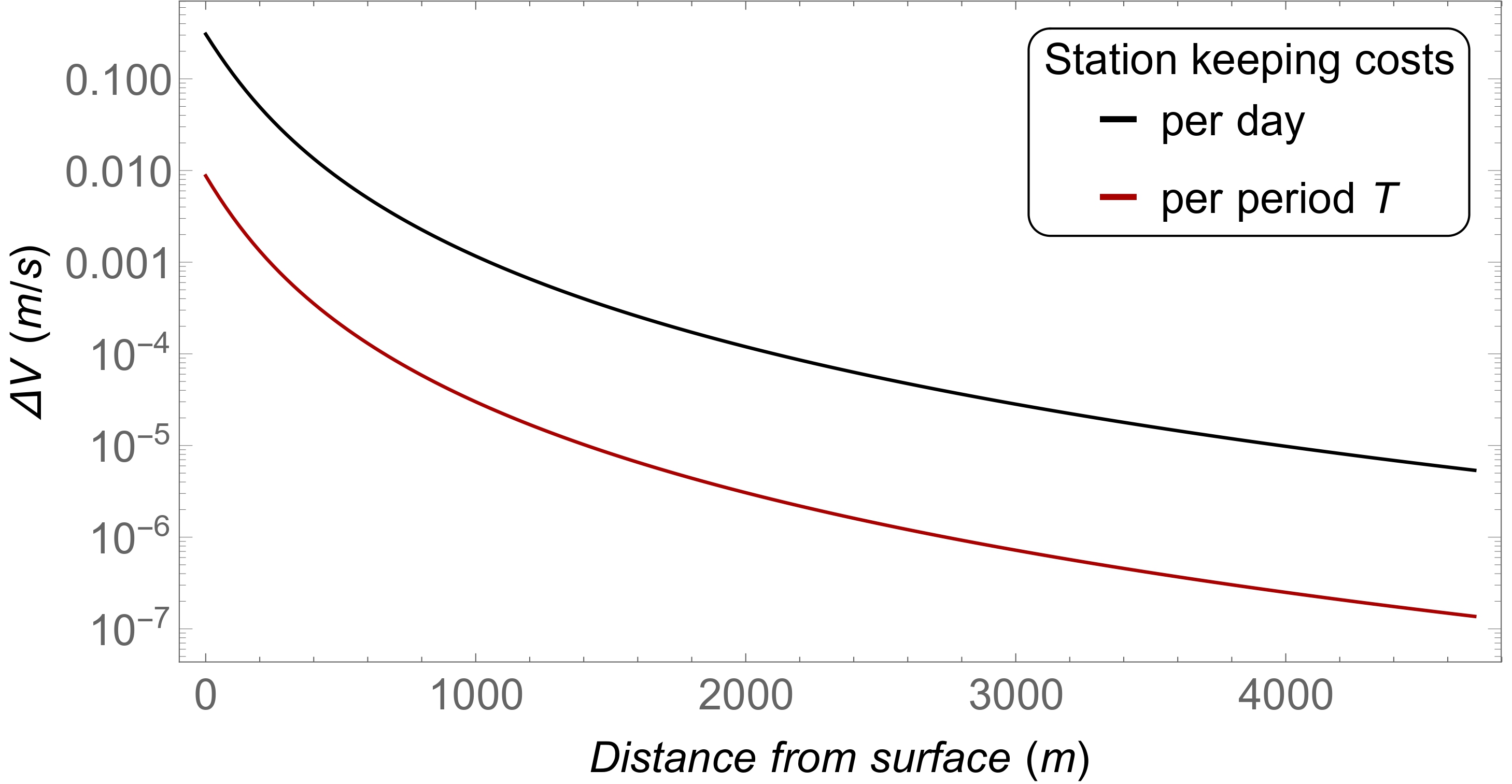}
\caption{Fuel equivalent $\Delta V$ to correct the deviation of the perturbed orbit with respect to the Keplerian one for every period of time $T$ shown in Eq. \eqref{eq:periodic} and for one day as functions of the distance from the surface.}
\label{fig:deltav}
\end{figure}

The fuel-equivalent cost as a function of the distance from the barycenter of the asteroid 2011~UW$_{158}$ is shown in Fig. \ref{fig:deltav}. This value represents the fuel equivalent $\Delta V$ required to maintain the periodicity of the orbit with respect to the asteroid 2011~UW$_{158}$, considering its spherical harmonic coefficients shown in Tab. \ref{tab:coefotimizados}.

From Fig. \ref{fig:deltav}, it can be observed that the fuel equivalent cost decreases monotonically with the distance from the surface. Within the analyzed interval, $\Delta V$ varies by approximately five orders of magnitude. Close to the surface, the required correction is on the order of $10^{-2}~\mathrm{m~s^{-1}}$ per orbital period, whereas at distances of about 4.5 km it decreases to approximately $10^{-7}~\mathrm{m~s^{-1}}$.

This strong attenuation matches the spatial decay of the higher-order spherical harmonic terms, whose contribution to the gravitational potential scales with higher inverse powers of distance. In the near-field region, the asteroid's irregular mass distribution strongly influences the spacecraft.
In practical terms, when flying close to the surface, the spacecraft effectively ``perceives'' the non-spherical shape of 2011~UW$_{158}$, and the resulting gravitational field departs significantly from the ideal central-field assumption. These deviations accumulate over one orbital period $T$, producing a noticeable velocity mismatch.

As the orbital radius increases, local mass asymmetries have progressively less influence. The gravitational field becomes increasingly dominated by its monopole component, and from the spacecraft's perspective, the body behaves more like an equivalent point mass. Consequently, the motion approaches the Keplerian solution, and the station-keeping requirement becomes negligible.

The curve's smooth behavior indicates that no strong resonant effects arise within the analyzed distance range. Therefore, the dominant mechanism driving the $\Delta V$ demand is the gradual reduction of the non-spherical gravitational perturbations rather than discrete dynamical transitions. It should be noted that this result holds for an orbit whose plane is perpendicular to the axis of rotation of 2011~UW$_{158}$. Orbits in other directions may not strategically take advantage of this symmetry, whereas the cost $\Delta V$ may behave differently.

From an operational standpoint, the results suggest that proximity operations around 2011~UW$_{158}$ may require non-negligible periodic corrections, especially for low-altitude orbits. The period to update the orbit varies from 2,425 seconds at a distance of 325~m from the barycenter to 2,202 seconds at a distance of 5,000~m, according to Eq. \eqref{eq:periodic}. Maintaining a spacecraft within a few hundred meters of the surface may require cumulative corrections that become necessary over extended mission durations. On the other hand, at kilometer-scale distances, station-keeping costs become practically negligible, which may favor higher operational altitudes when propulsion resources are limited.

\section{Final Comments}
\label{finalcomm}
We investigated the strongly dynamical environment of the sub-kilometer Super-Fast Rotator (436724) 2011~UW$_{158}$. Disregarding external disturbances, the dynamical structure of this asteroid is completely dominated by its exceptionally short rotational period of $0.61072$~h. Using a 3-D polyhedral shape model, we mapped regions of material flow, retention, and structural stability across the surface.

Owing to its super-fast rotational regime and elongated shape, the centrifugal potential strongly dominates the effective acceleration. Our topographic mapping shows that the tangential vectors of surface acceleration drive the flow of loose material outwards towards the equatorial regions. Assuming a completely cohesionless surface, the body retains negligible regolith, with stability confined solely to regions near the poles, where centrifugal force exerts a weaker effect. 

To retain material particles on the surface of such a dynamic environment, cohesive and gravitational forces must counteract the high rotational stresses. By incorporating a cohesive strength of $K \sim 260$~Pa into the polyhedral analysis, the maximum effective slopes decrease drastically to $\sim 45^\circ$. This cohesion allows the asteroid to retain large boulders up to $r_{max} \sim 50$~m near the equator, neutralizing the generalized tendency toward material ejection.

Our escape speed analysis further supports this instability. Because conventional equations yield complex roots across most of the asteroid's surface, we introduced the equivalent instability speed ($v_{ei}$). This metric confirms that a cohesionless particle needs no minimum normal velocity to escape the surface, creating regions of unavoidable ejection and highlighting the repulsive topology of the geopotential environment.

Beyond surface dynamics, our thermal modeling shows that the asteroid's rapid rotation strongly suppresses diurnal temperature variations across its surface. This effect is evident at both high thermal inertia, where the material responds slowly to changes in solar flux, and low thermal inertia, where the rapid spin still prevents significant temperature fluctuations. We also found that thermal inertia has only a minor influence on the mean temperature, whereas albedo exerts a substantial control. These results highlight that the extreme rotation rate is the dominant factor driving the uniform longitudinal temperature distribution.

Expanding our analysis to the orbital environment, we showed that station-keeping costs drop substantially with distance. Fuel expenditure drops by up to five orders of magnitude, from $10^{-2}~\mathrm{m~s^{-1}}$ per orbit close to the surface down to minimal corrections for an orbit with an altitude close to 5~km and a period of $T=2200$~s. 

Overall, the fact that 2011~UW$_{158}$ maintains its structural integrity with no signs of disintegration reinforces the hypothesis that it possesses significant internal cohesion. This allows the body to withstand rotational stresses far exceeding the classical cohesionless spin barrier, classifying it as either a coherent monolithic fragment or a highly cohesive aggregate.

Ultimately, the dynamical mapping and stability constraints in this study provide critical baseline data for designing future close-proximity observation missions to this Potentially Hazardous Asteroid.

\section*{Acknowledgements}
We thank the two anonymous Reviewers for their constructive comments and suggestions, which significantly improved the discussion and interpretation of the results presented in this study.
A.~Amarante and A.~Ferreira thank the financial support of the São Paulo Research Foundation (FAPESP) [grants \#2023/11781-5 \& \#2025/15438-9].
Santos, L. B. T. thanks the Polytechnic School of the University of Pernambuco (POLI-UPE) and the PostGrad Program in Systems Engineering (PPGES) for their support.
AKAJ acknowledges support from the SPACE project, ref. COMPETE2030-FEDER-00860300, funded by COMPETE 2030 and FCT, Portugal.
The authors also acknowledge the financial support of the Coordination for the Improvement of Higher Education Personnel (CAPES) - Brazil (finance code 001). This research was also supported by computational resources supplied by the Center for Scientific Computing (NCC/GridUNESP) of the São Paulo State University (UNESP) and the Center for Mathematical Sciences Applied to Industry (CeMEAI), funded by FAPESP [grant \#2013/07375-0].

\printcredits

\bibliographystyle{cas-model2-names}

\bibliography{cas-refs}

@article{cruikshank2005high,
  title={The high-albedo Kuiper belt object (55565) 2002 AW197},
  author={Cruikshank, Dale P and Stansberry, John A and Emery, Joshua P and Fern{\'a}ndez, Yanga R and Werner, Michael W and Trilling, David E and Rieke, George H},
  journal={Astrophys. J. Lett.},
  volume={624},
  number={1},
  pages={L53--L56},
  year={2005},
  doi={10.1086/430420}
}

@article{rozitis2024pre,
  title={Pre-impact thermophysical properties and the Yarkovsky effect of NASA DART target (65803) Didymos},
  author={Rozitis, Benjamin and Green, Simon F and Jackson, Samuel L and Snodgrass, Colin and Opitom, Cyrielle and M{\"u}ller, Thomas G and Kolb, Ulrich C and Chesley, Steven R and Daly, R Terik and Thomas, Cristina A and others},
  journal={Planet. Sci. J.},
  volume={5},
  number={3},
  pages={66},
  year={2024},
  publisher={The American Astronomical Society},
  doi={10.3847/PSJ/ad23eb}
}

@Article{ershkov2025particular,
AUTHOR = {Ershkov, Sergey},
TITLE = {On the Particular Dynamics of Rubble-Pile Asteroid Rotation Following Projectile Impact on the Surface During Planetary Approach},
JOURNAL = {Mathematics},
VOLUME = {13},
YEAR = {2025},
NUMBER = {21},
ARTICLE-NUMBER = {3412},
ISSN = {2227-7390},
DOI = {10.3390/math13213412}
}

@article {ershkov2026semi,
author = {Ershkov, Sergey},
title = {Semi-analytical Ansatz for Approximating Dynamics of Asteroid Rotation Approaching Planet},
journal = {J. Appl. Comput. Mech.},
volume = {12},
number = {2},
pages = {483-490},
year  = {2026},
publisher = {Shahid Chamran University of Ahvaz},
issn = {2383-4536}, 
eissn = {2383-4536}, 
doi = {10.22055/jacm.2025.48493.5273},	

}

@article{walsh2012spin,
  title={Spin-up of rubble-pile asteroids: Disruption, satellite formation, and equilibrium shapes},
  author={Walsh, Kevin J and Richardson, Derek C and Michel, Patrick},
  journal={Icarus},
  volume={220},
  number={2},
  pages={514--529},
  year={2012},
  publisher={Elsevier},
  doi={10.1016/j.icarus.2012.04.029}
}

@article{holsapple2006tidal,
  title={Tidal disruptions: A continuum theory for solid bodies},
  author={Holsapple, Keith A and Michel, Patrick},
  journal={Icarus},
  volume={183},
  number={2},
  pages={331--348},
  year={2006},
  publisher={Elsevier},
  doi={10.1016/j.icarus.2006.03.013}
}

@article{luo2024taxonomic,
  title={Taxonomic analysis of asteroids with artificial neural networks},
  author={Luo, Nanping and Wang, Xiaobin and Gu, Shenghong and Penttil{\"a}, Antti and Muinonen, Karri and Liu, Yisi},
  journal={Astron. J.},
  volume={167},
  number={1},
  pages={13},
  year={2024},
  publisher={The American Astronomical Society},
  doi={10.3847/1538-3881/ad0b7a}
}

@article{roh2022new,
  title={A new approach to feature-based asteroid taxonomy in 3D color space: I. SDSS photometric system},
  author={Roh, Dong-Goo and Moon, Hong-Kyu and Shin, Min-Su and DeMeo, Francesca E},
  journal={Astron. Astrophys.},
  volume={664},
  pages={A51},
  year={2022},
  publisher={EDP Sciences},
  doi={10.1051/0004-6361/202039551}
  
}

@article{chapman1975surface,
  title={Surface properties of asteroids: A synthesis of polarimetry, radiometry, and spectrophotometry},
  author={Chapman, Clark R and Morrison, David and Zellner, Ben},
  journal={Icarus},
  volume={25},
  number={1},
  pages={104--130},
  year={1975},
  publisher={Elsevier},
  doi= {10.1016/0019-1035(75)90191-8}
}

@article{Hu2021,
    author = {Hu, Shoucun and Richardson, Derek C and Zhang, Yun and Ji, Jianghui},
    title = {Critical spin periods of sub-km-sized cohesive rubble-pile asteroids: dependences on material parameters},
    journal = {Mon. Not. Roy. Astron. Soc.},
    volume = {502},
    number = {4},
    pages = {5277-5291},
    year = {2021},
    month = {04},
    issn = {0035-8711},
    doi = {10.1093/mnras/stab412}

}

@article{prvsa2016_sunmass,
  title={Nominal values for selected solar and planetary quantities: IAU 2015 resolution B3},
  author={Pr{\v{s}}a, Andrej and Harmanec, Petr and Torres, Guillermo and Mamajek, Eric and Asplund, Martin and Capitaine, Nicole and Christensen-Dalsgaard, J{\o}rgen and Depagne, {\'E}ric and Haberreiter, Margit and Hekker, Saskia and others},
  journal={Astron. J.},
  volume={152},
  number={2},
  pages={41},
  year={2016},
  publisher={The American Astronomical Society},
  doi={10.3847/0004-6256/152/2/41}
}

@article{jiang2026emerging,
  title={Emerging Chemical and Biological Materials Technologies in the Extraplanetary Environment},
  author={Jiang, Qingyao and Wang, Bin and Cheng, Yifan and Wang, Yiming and Zhao, Hongxin and Lu, Yuan},
  journal={Nano-Micro Lett.},
  volume={18},
  number={1},
  pages={151},
  year={2026},
  publisher={Springer},
  doi={10.1007/s40820-025-01979-8}
}

@article{jedicke2018availability,
  title={Availability and delta-v requirements for delivering water extracted from near-Earth objects to cis-lunar space},
  author={Jedicke, Robert and Sercel, Joel and Gillis-Davis, Jeffrey and Morenz, Karen J and Gertsch, Leslie},
  journal={Planet. Space Sci.},
  volume={159},
  pages={28--42},
  year={2018},
  publisher={Elsevier},
  doi={10.1016/j.pss.2018.04.005}
}

@article{braga2026surface,
  title={Surface Properties, Orbital Dynamics, and Thermophysical Modeling of the Primitive Asteroid (269) Justitia},
  author={Braga, Leonardo and Amarante, Andre and Ferreira, Alessandra and Monteiro, Filipe and Martins, Maria},
  journal={Planet. Sci. J.},
  volume={7},
  number={2},
  pages={46},
  year={2026},
  publisher={The American Astronomical Society},
  doi= {10.3847/PSJ/ae3d93}
}

@article{mohr2025codata,
  title={CODATA recommended values of the fundamental physical constants: 2022},
  author={Mohr, Peter J and Newell, David B and Taylor, Barry N and Tiesinga, Eite},
  journal={J. Phys. Chem. Ref. Data},
  volume={54},
  number={3},
  year={2025},
  publisher={AIP Publishing},
  doi={10.1063/5.0279860}
}

@article{domingos2006stable,
  title={Stable satellites around extrasolar giant planets},
  author={Domingos, RC and Winter, OC and Yokoyama, T},
  journal={Mon. Not. Roy. Astron. Soc.},
  volume={373},
  number={3},
  pages={1227--1234},
  year={2006},
  publisher={The Royal Astronomical Society},
  doi={10.1111/j.1365-2966.2006.11104.x}
}

@article{scheeres2015landslides,
  title={Landslides and mass shedding on spinning spheroidal asteroids},
  author={Scheeres, Daniel J},
  journal={Icarus},
  volume={247},
  pages={1--17},
  year={2015},
  publisher={Elsevier},
  doi={10.1016/j.icarus.2014.09.017}
}

@article{perez2025characterization,
  title={Characterization of asteroid shapes and stability on their surface using super-ellipsoids},
  author={P{\'e}rez Molina, Manuel and Campo Bagatin, Adriano},
  journal={Celest. Mech. Dyn. Astron.},
  volume={137},
  number={5},
  pages={31},
  year={2025},
  publisher={Springer},
  doi={10.1007/s10569-025-10261-3}
}

@article{braga2025equilibrium,
  title={Equilibrium points and surface dynamics about comet 67P/Churyumov--Gerasimenko},
  author={Braga, Leonardo and Amarante, Andre and Ferreira, Alessandra and Gomes, Caio and Ceranto, Luis},
  journal={Eur. Phys. J.-Spec. Top.},
  pages={1--21},
  year={2025},
  publisher={Springer},
  doi={10.1140/epjs/s11734-025-01986-1}
}

@article{de2026generalized,
  title={A generalized dipole-segment model for the gravitational field of elongated bodies},
  author={de Almeida Jr, AK and Ferreira, AFS and Santos, LBT and Monteiro, F and Amarante, A and Tresaco, E and Sanchez, DM and Gomes, C and Prado, AFBA},
  journal={Astron. Astrophys.},
  volume={706},
  pages={A355},
  year={2026},
  publisher={EDP Sciences},
  doi={10.1051/0004-6361/202557978}
}

@article{sanchez2020cohesive,
  title={Cohesive regolith on fast rotating asteroids},
  author={S{\'a}nchez, Paul and Scheeres, Daniel J},
  journal={Icarus},
  volume={338},
  pages={113443},
  year={2020},
  publisher={Elsevier},
  doi={10.1016/j.icarus.2019.113443}
}

@inproceedings{naidu2015radar,
  title={Radar observations of near-Earth asteroid (436724) 2011 UW158 using the Arecibo, Goldstone, and Green Bank Telescopes},
  author={Naidu, Shantanu P and Benner, Lance AM and Brozovic, Marina and Giorgini, Jon D and Jao, Joseph S and Busch, Michael W and Taylor, Patrick A and Richardson, James E and Rivera-Valentin, Edgard G and Ford, Linda A and others},
  booktitle={AAS/Division for Planetary Sciences Meeting Abstracts\# 47},
  volume={47},
  pages={204--08},
  year={2015}
}

@article{Scheeres1996escapespeed,
title = {Orbits Close to Asteroid 4769 Castalia},
journal = {Icarus},
volume = {121},
number = {1},
pages = {67-87},
year = {1996},
issn = {0019-1035},
doi = {10.1006/icar.1996.0072},
author = {D.J. Scheeres and S.J. Ostro and R.S. Hudson and R.A. Werner}
}

@article{Jiang2014motion,
  title={Orbital mechanics near a rotating asteroid},
  author={Jiang, Yu and Baoyin, Hexi},
  journal={J. Astrophys. Astron.},
  volume={35},
  number={1},
  pages={17--38},
  year={2014},
  publisher={Springer},
  doi={10.1007/s12036-014-9259-z}
}

@article{Hill,
 ISSN = {00029327, 10806377},
 author = {G. W. Hill},
 journal = {Am. J. Math.},
 number = {1},
 pages = {5--26},
 publisher = {Johns Hopkins University Press},
 title = {Researches in the Lunar Theory},
 urldate = {2026-03-14},
 volume = {1},
 year = {1878}
}

@INPROCEEDINGS{Pravec2002b,
       author = {{Pravec}, Petr and {Ku{\v{s}}nir{\'a}k}, Peter and {{\v{S}}arounov{\'a}}, Lenka and {Harris}, Alan W. and {Binzel}, Richard P. and {Rivkin}, Andrew S.},
        title = "{Large coherent asteroid 2001 OE$_{84}$}",
    booktitle = {Asteroids, Comets, and Meteors: ACM 2002},
         year = 2002,
       editor = {{Warmbein}, Barbara},
       series = {ESA Special Publication},
       volume = {500},
        month = nov,
        pages = {743-745},
       adsurl = {https://ui.adsabs.harvard.edu/abs/2002ESASP.500..743P}
}

@article{Pravec2007,
title = {Binary asteroid population: 1. Angular momentum content},
journal = {Icarus},
volume = {190},
number = {1},
pages = {250-259},
year = {2007},
issn = {0019-1035},
doi = {10.1016/j.icarus.2007.02.023},
author = {P. Pravec and A.W. Harris}
}

@ARTICLE{Kevin2018,
       author = {{Walsh}, Kevin J.},
        title = "{Rubble Pile Asteroids}",
      journal = {Annu. Rev. Astron. Astrophys.},
         year = 2018,
        month = sep,
       volume = {56},
        pages = {593-624},
          doi = {10.1146/annurev-astro-081817-052013},
 primaryClass = {astro-ph.EP},
       adsurl = {https://ui.adsabs.harvard.edu/abs/2018ARA&A..56..593W}
}

@ARTICLE{Ipatov2016,
       author = {{Ipatov}, A.~V. and {Bondarenko}, Yu. S. and {Medvedev}, Yu. D. and {Mishina}, N.~A. and {Marshalov}, D.~A. and {Benner}, L.~A.},
        title = "{Radar observations of the asteroid 2011 UW158}",
      journal = {Astron. Lett.},
         year = 2016,
        month = dec,
       volume = {42},
       number = {12},
        pages = {850-855},
          doi = {10.1134/S1063773716120021},
       adsurl = {https://ui.adsabs.harvard.edu/abs/2016AstL...42..850I}
}

@article{lsst_2026,
doi = {10.3847/2041-8213/ae2a30},
year = {2026},
month = {jan},
publisher = {The American Astronomical Society},
volume = {996},
number = {2},
pages = {L33},
author = {Greenstreet, Sarah and Li, Zhuofu (Chester) and Vavilov, Dmitrii E. and Singh, Devanshi and Jurić, Mario and Ivezic, Zeljko and Eggl, Siegfried and Koumjian, Alec and Moeyens, Joachim and Carruba, Valerio and Womack, Maria and Granvik, Mikael and Alexov, Anastasia and Antilogus, Pierre and Baumanć, B̌rian J. and Bellm, Eric C. and Boucaud, Alexandre and Bradshaw, Andrew and Carlin, Jeffrey L. and Chiang, Hsin-Fang and Daly, Philip N. and Daruich, Felipe and Daubard, Guillaume and Dennihy, Erik and Deppe, Stephanie JH and Drass, Holger and Gangler, Emmanuel and Le Guillou, Laurent and Guy, Leanne P. and Hascall, Patrick A. and Ingraham, Patrick and Jee, M. James and Jenness, Tim and Kahn, Steven M. and Kannawadi, Arun and Kelvin, Lee S. and Kurlander, Jacob A. and Laporte, Didier and Lust, Nate B. and Lutfi, Mostafa and MacArthur, Lauren A. and Mainetti, Gabriele and Marc, Moniez and Plazas Malagón, Andrés A. and Mejías, David Jiménez and Menanteau, F̌elipe and Mills, David J. and O’Mullane, William and Neto, Angelo Fausti and Neveu, Jeremy and Nourbakhsh, Erfan and Park, HyeYun and Patterson, Maria T and Peterson, John R. and Quint, Bruno C. and Ribeiro, Tiago and Ridgway, Stephen T. and van Reeven, Wouter and Sebag, Jacques and Sedaghat, Nima and Shaw, Richard A. and Strauss, Alan L. and Suberlak, Krzysztof and Sullivan, Ian S. and Swinbank, John D. and Thomas, Sandrine and Thornton, Adam and Wood-Vasey, W. M. and Walter, Christopher W. and Ward, Charlotte and Willman, Beth},
title = {Lightcurves, Rotation Periods, and Colors for Vera C. Rubin Observatory’s First Asteroid Discoveries},
journal = {Astrophys. J. Lett.}
}

@BOOK{asteroidsiv,
       author = {{Michel}, Patrick and {DeMeo}, Francesca E. and {Bottke}, William F.},
        title = "{Asteroids IV}",
         year = 2015,
          doi = {10.2458/azu_uapress_9780816532131},
       adsurl = {https://ui.adsabs.harvard.edu/abs/2015aste.book.....M},
      publisher ={University of Arizona Press}
}

@book{scheeres2016orbital,
  title={Orbital motion in strongly perturbed environments: applications to asteroid, comet and planetary satellite orbiters},
  author={Scheeres, Daniel J},
  year={2016},
  publisher={Springer}
}

@article{chapman2002_sizeboulder,
  title={Impact history of Eros: Craters and boulders},
  author={Chapman, Clark R and Merline, William J and Thomas, Peter C and Joseph, Jonathan and Cheng, Andrew F and Izenberg, Noam},
  journal={Icarus},
  volume={155},
  number={1},
  pages={104--118},
  year={2002},
  publisher={Elsevier},
  doi={10.1006/icar.2001.6744}
}

@article{michikami2008_sizeboulder,
  title={Size-frequency statistics of boulders on global surface of asteroid 25143 Itokawa},
  author={Michikami, Tatsuhiro and Nakamura, Akiko M and Hirata, Naru and Gaskell, Robert W and Nakamura, Ryosuke and Honda, Takayuki and Honda, Chikatoshi and Hiraoka, Kensuke and Saito, Jun and Demura, Hirohide and others},
  journal={Earth Planets Space},
  volume={60},
  number={1},
  pages={13--20},
  year={2008},
  publisher={Springer},
  doi={10.1186/BF03352757}
}

@book{lambe2008_soil,
  title={Soil mechanics SI version},
  author={Lambe, T William and Whitman, Robert V},
  year={2008},
  publisher={John Wiley \& Sons}
}

@article{muller2021_repose3,
  title={Algorithm for the determination of the angle of repose in bulk material analysis},
  author={M{\"u}ller, Dominik and Fimbinger, Eric and Brand, Clemens},
  journal={Powder Technol.},
  volume={383},
  pages={598--605},
  year={2021},
  publisher={Elsevier},
  doi={10.1016/j.powtec.2021.01.010}
}

@article{al2018_repose2,
  title={A review on the angle of repose of granular materials},
  author={Al-Hashemi, Hamzah M Beakawi and Al-Amoudi, Omar S Baghabra},
  journal={Powder Technol.},
  volume={330},
  pages={397--417},
  year={2018},
  publisher={Elsevier},
  doi={10.1016/j.powtec.2018.02.003}
}

@article{valvano2022_repose1,
  title={APOPHIS--effects of the 2029 Earth’s encounter on the surface and nearby dynamics},
  author={Valvano, Giulia and Winter, Othon Cabo and Sfair, Rafael and Machado Oliveira, R and Borderes-Motta, G and Moura, TS},
  journal={Mon. Not. Roy. Astron. Soc.},
  volume={510},
  number={1},
  pages={95--109},
  year={2022},
  publisher={Oxford University Press},
  doi={10.1093/mnras/stab3299}
}

@article{hirabayashi2015_boulder,
  title={Stress and failure analysis of rapidly rotating asteroid (29075) 1950 DA},
  author={Hirabayashi, Masatoshi and Scheeres, Daniel J},
  journal={Astrophys. J. Lett.},
  volume={798},
  number={1},
  pages={L8},
  year={2015},
  publisher={The American Astronomical Society},
  doi={10.1088/2041-8205/798/1/L8}
}

@article{polishook2017_boulder,
  title={The fast spin of near-Earth asteroid (455213) 2001 OE84, revisited after 14 years: constraints on internal structure},
  author={Polishook, David and Moskovitz, Nicholas and Thirouin, Audrey and Bosh, Amanda and Levine, Stephen and Zuluaga, Carlos and Tegler, SC and Aharonson, Oded},
  journal={Icarus},
  volume={297},
  pages={126--133},
  year={2017},
  publisher={Elsevier},
  doi={10.1016/j.icarus.2017.06.036}
}

@article{fastTFC,
   author = {Allan K. {de Almeida Jr} and Hunter Johnston and Carl Leake and Daniele Mortari},
   doi = {10.1140/epjp/s13360-021-01151-2},
   issn = {21905444},
   issue = {2},
   journal = {Eur. Phys. J. Plus},
   title = {Fast 2-impulse non-Keplerian orbit transfer using the Theory of Functional Connections},
   volume = {136},
   year = {2021},
}

@article{almeida24solar,
author = {Allan Kardec de Almeida Jr and Timothée Vaillant and Leonardo Santos and Dalmiro Maia},
title = {Low-thrust transfer with Theory of Functional Connections: application to 243 Ida with a solar sail},
journal = {Adv. Space Res.},
volume = {75},
number = {2},
pages = {2108-2125},
year = {2025},
issn = {0273-1177},
doi = {10.1016/j.asr.2024.09.069}
}

@article{10.1145/355586.364791,
	title        = {A Simple Automatic Derivative Evaluation Program},
	author       = {Wengert, R. E.},
	year         = 1964,
	month        = {aug},
	journal      = {Commun. ACM},
	publisher    = {Association for Computing Machinery},
	address      = {New York, NY, USA},
	volume       = 7,
	number       = 8,
	pages        = {463–464},
	doi          = {10.1145/355586.364791},
	issn         = {0001-0782},
	issue_date   = {Aug. 1964},
	numpages     = 2
}

@article{tfc_segmentation,
title = {Segmentation of the spacecraft transfer problem based on overdetermined and continuity constraints using the Theory of Functional Connections},
journal = {Acta Astronaut.},
volume = {},
pages = {},
doi = {10.1016/j.actaastro.2026.01.028},
year = {2026},
author = {Allan Kardec {de Almeida Jr}}
}

@ARTICLE{Carry_2012,
       author = {{Carry}, B.},
        title = "{Density of asteroids}",
      journal = {Planet. Space Sci.},
         year = 2012,
        month = dec,
       volume = {73},
       number = {1},
        pages = {98-118},
          doi = {10.1016/j.pss.2012.03.009},
 primaryClass = {astro-ph.EP},
       adsurl = {https://ui.adsabs.harvard.edu/abs/2012P&SS...73...98C}
}

@ARTICLE{2022AdSpR..70.3362S,
       author = {{Santos}, L.~B.~T. and others},
        title = "{Optimal transfers from Moon to L$_{2}$ halo orbit of the Earth-Moon system}",
      journal = {Adv. Space Res.},
         year = 2022,
        month = dec,
       volume = {70},
       number = {11},
        pages = {3362-3372},
          doi = {10.1016/j.asr.2022.08.035},
 primaryClass = {astro-ph.EP},
       adsurl = {https://ui.adsabs.harvard.edu/abs/2022AdSpR..70.3362S}
}

@ARTICLE{article22,
author = {De Almeida Junior, Allan and Santos, Leonardo and Gomes, C.E.S. and Andrade, E.V.M. and Barros, A.L.S. and Santos, K.G.F. and Fernandes, G.M. and Monteiro, Filipe and Amarante, Andre and Bastos, R.I.S. and Lima, Nathalia and Nascimento, H.C.B. and Lima, Nathan and Prado, Antonio},
year = {2026},
month = {01},
pages = {},
title = {Equilibrium Points and Stability Analysis in Binary Asteroid Systems Using a Double Mass Dipole Model},
journal = {Adv. Space Res.},
doi = {10.1016/j.asr.2026.01.009}
}

@ARTICLE{2021MNRAS.502.4277S,
       author = {{Santos}, L.~B.~T. and {Marchi}, L.~O. and {Aljbaae}, S. and {Sousa-Silva}, P.~A. and {Sanchez}, D.~M. and {Prado}, A.~F.~B.~A.},
        title = "{A particle-linkage model for elongated asteroids with three-dimensional mass distribution}",
      journal = {Mon. Not. Roy. Astron. Soc.},
         year = 2021,
        month = apr,
       volume = {502},
       number = {3},
        pages = {4277-4289},
          doi = {10.1093/mnras/stab198},
 primaryClass = {astro-ph.EP},
       adsurl = {https://ui.adsabs.harvard.edu/abs/2021MNRAS.502.4277S}
}

@article{Scheeres_1994,
title = {Dynamics about Uniformly Rotating Triaxial Ellipsoids: Applications to Asteroids},
journal = {Icarus},
volume = {110},
number = {2},
pages = {225-238},
year = {1994},
issn = {0019-1035},
doi = {10.1006/icar.1994.1118},
author = {D.J. Scheeres}
}

@misc{tfc2021github,
	title        = {{A Functional Interpolation Framework TFC v0.1.2 [software]}},
	author       = {Carl Leake and Hunter Johnston},
	year         = 2021,
	url         = {https://github.com/leakec/tfc},
	version      = {0.1.2}
}

@misc{JPL_2011,
  author = {{Jet Propulsion Laboratory}},
  title = {NASA Small-Body Database},
  year = {2026},
  url = {https://ssd.jpl.nasa.gov/tools/sbdb_lookup.html#/?sstr=2011%20UW158},
  note = {(accessed 25 August 2026)}
}

@misc{UAT,
  author = {{Unified Astronomy Thesaurus}},
  title = {Unified Astronomy Thesaurus},
  year = {2025},
  url = {https://astrothesaurus.org/},
  note = {(accessed 25 August 2026)}
}

@misc{nasa_sbdb,
  author = {{Jet Propulsion Laboratory}},
  title = {NASA Small-Body Database},
  year = {2026},
  url = {https://ssd.jpl.nasa.gov/tools/sbdb_query.html},
  note = {(accessed 13 March 2026)}
}

@misc{tempest,
	title        = {{A Modular Thermophysical Model for Airless Bodies with Support for Surface Roughness and Non-Periodic Heating TEMPEST v1.0.0 [software]}},
	author       = {Duncan Lyster},
	year         = 2025,
	note         = {\url{https://github.com/duncanLyster/TEMPEST}},
	version      = {1.0.0}
}

@misc{minor-gravity,
  author       = {{Amarante}, A.},
  title        = {Minor-Gravity},
  month        = jan,
  year         = 2026,
  publisher    = {Zenodo},
  version      = {v2.1},
  doi          = {10.5281/zenodo.18408101},
  url          = {https://doi.org/10.5281/zenodo.18408101},
}

@misc{minor-equilibria,
  author       = {{Amarante}, A.},
  title        = {Minor-Equilibria-NR Package},
  month        = jan,
  year         = 2026,
  publisher    = {Zenodo},
  version      = {v1.2},
  doi          = {10.5281/zenodo.18408030},
  url          = {https://doi.org/10.5281/zenodo.18408030},
}

@ARTICLE{tsoulis2001,
       author = {{Tsoulis}, Dimitrios and {Petrovi}, Sveto},
        title = "{On the singularities of the gravity field of a homogeneous polyhedral body}",
      journal = {Geophysics},
         year = 2001,
        month = mar,
       volume = {66},
       number = {2},
        pages = {535-539},
          doi = {10.1190/1.1444944},
       adsurl = {https://ui.adsabs.harvard.edu/abs/2001Geop...66..535T}
}

@book{murray1999solar,
  title={Solar system dynamics},
  author={Murray, Carl D and Dermott, Stanley F},
  year={1999},
  publisher={Cambridge university press}
}

@article{Scheeres_1998,
  title={Dynamics of orbits close to asteroid 4179 Toutatis},
  author={Scheeres, Daniel J and others},
  journal={Icarus},
  volume={132},
  number={1},
  pages={53--79},
  year={1998},
  publisher={Elsevier},
  doi={10.1006/icar.1997.5870}
}

@article{Ostro1999Toutatis,
  title={Asteroid 4179 Toutatis: 1996 radar observations},
  author={Ostro, Steven J and others},
  journal={Icarus},
  volume={137},
  number={1},
  pages={122--139},
  year={1999},
  publisher={Elsevier},
  doi={10.1006/icar.1998.6031}
}

@article{Muller_2017,
  title={Hayabusa-2 mission target asteroid 162173 Ryugu (1999 JU3): Searching for the object’s spin-axis orientation},
  author={M{\"u}ller, TG and others},
  journal={Astron. Astrophys.},
  volume={599},
  pages={A103},
  year={2017},
  publisher={EDP Sciences},
  doi={10.1051/0004-6361/201629134}
}

@article{Ostro1999,
  title={Radar and optical observations of asteroid 1998 KY26},
  author={Ostro, Steven J and others},
  journal={Science},
  volume={285},
  number={5427},
  pages={557--559},
  year={1999},
  publisher={American Association for the Advancement of Science},
  doi={10.1126/science.285.5427.557}
}

@article{Amarante2020Arrokoth,
  title={Surface dynamics, equilibrium points and individual lobes of the Kuiper Belt object (486958) Arrokoth},
  author={Amarante, A and Winter, OC},
  journal={Mon. Not. Roy. Astron. Soc.},
  volume={496},
  number={4},
  pages={4154--4173},
  year={2020},
  publisher={Oxford University Press},
  doi={10.1093/mnras/staa1732}
}

@article{Amarante2021bennu,
   title={Stability and Evolution of Fallen Particles Around the Surface of Asteroid (101955) Bennu},
   volume={126},
   ISSN={2169-9100},
   url={http://dx.doi.org/10.1029/2019JE006272},
   DOI={10.1029/2019je006272},
   number={1},
   journal={Journal of Geophysical Research: Planets},
   publisher={American Geophysical Union (AGU)},
   author={Amarante, A. and Winter, O. C. and Sfair, R.},
   year={2021},
   month=jan }

@article{AmaranteWinter2022,
  author  = {Amarante, A. and Winter, O. C.},
  title   = {The Fate of Particles in the Dynamical Environment around
             Kuiper-Belt Object (486958) Arrokoth},
  journal = {Astrophysics and Space Science},
  volume  = {367},
  number  = {4},
  pages   = {38},
  year    = {2022},
  doi     = {10.1007/s10509-022-04065-2}
}

@article{Werner_1997,
  title={Exterior gravitation of a polyhedron derived and compared with harmonic and mascon gravitation representations of asteroid 4769 Castalia},
  author={Werner, Robert A and Scheeres, Daniel J},
  journal={Celest. Mech. Dyn. Astron.},
  volume={65},
  pages={313--344},
  year={1996},
  publisher={Springer},
  doi={10.1007/BF00053511}
}

@ARTICLE{Icarus_2008,
       author = {{Szab{\'o}}, Gyula M. and {Kiss}, L{\'a}szl{\'o} L.},
        title = "{The shape distribution of asteroid families: Evidence for evolution driven by small impacts}",
      journal = {Icarus},
         year = 2008,
        month = jul,
       volume = {196},
       number = {1},
        pages = {135-143},
          doi = {10.1016/j.icarus.2008.01.019},
 primaryClass = {astro-ph},
       adsurl = {https://ui.adsabs.harvard.edu/abs/2008Icar..196..135S}
}

@BOOK{Lewis_1996,
       author = {{Lewis}, John S.},
        title = "{Mining the sky : untold riches from the asteroids, comets, and planets}",
         year = 1996,
       adsurl = {https://ui.adsabs.harvard.edu/abs/1996msur.book.....L},
      publisher={Reading, Mass. : Addison-Wesley Pub. Co.}
}

@ARTICLE{Gary_2016,
       author = {{Gary}, Bruce L.},
        title = "{Unusual Properties for the NEA (436724) 2011 UW158}",
      journal = {Minor Planet Bulletin},
         year = 2016,
        month = jan,
       volume = {43},
       number = {1},
        pages = {33-38},
       adsurl = {https://ui.adsabs.harvard.edu/abs/2016MPBu...43...33G}
}

@ARTICLE{Pan_2022,
       author = {{Pan}, Kang-Shian and others},
        title = "{Is (3599) Basov a large C-type super-fast rotator?}",
      journal = {Planet Space Sci.},
         year = 2022,
        month = oct,
       volume = {220},
          eid = {105520},
        pages = {105520},
          doi = {10.1016/j.pss.2022.105520},
       adsurl = {https://ui.adsabs.harvard.edu/abs/2022P&SS..22005520P}
}

@article{Polishook_2016,
title = {A 2km-size asteroid challenging the rubble-pile spin barrier – A case for cohesion},
journal = {Icarus},
volume = {267},
pages = {243-254},
year = {2016},
issn = {0019-1035},
doi = {10.1016/j.icarus.2015.12.031},
author = {D. Polishook and others}
}

@ARTICLE{Chang_2022,
       author = {{Chang}, Chan-Kao and others},
        title = "{The Large Superfast Rotators Discovered by the Zwicky Transient Facility}",
      journal = {Astrophys. J. Lett.},
         year = 2022,
        month = jun,
       volume = {932},
       number = {1},
          eid = {L5},
        pages = {L5},
          doi = {10.3847/2041-8213/ac6e5e},
       adsurl = {https://ui.adsabs.harvard.edu/abs/2022ApJ...932L...5C}
}

@ARTICLE{Chang_2014,
       author = {{Chang}, Chan-Kao and others},
        title = "{A New Large Super-fast Rotator: (335433) 2005 UW163}",
      journal = {Astrophys. J. Lett.},
         year = 2014,
        month = aug,
       volume = {791},
       number = {2},
          eid = {L35},
        pages = {L35},
          doi = {10.1088/2041-8205/791/2/L35},
 primaryClass = {astro-ph.EP},
       adsurl = {https://ui.adsabs.harvard.edu/abs/2014ApJ...791L..35C}
}

@ARTICLE{Holsapple_2007,
       author = {{Holsapple}, Keith A.},
        title = "{Spin limits of Solar System bodies: From the small fast-rotators to 2003 EL61}",
      journal = {Icarus},
         year = 2007,
        month = apr,
       volume = {187},
       number = {2},
        pages = {500-509},
          doi = {10.1016/j.icarus.2006.08.012},
       adsurl = {https://ui.adsabs.harvard.edu/abs/2007Icar..187..500H}
}

@ARTICLE{Sanchez_2014,
       author = {{S{\'a}nchez}, P. and {Scheeres}, D.~J.},
        title = "{The strength of regolith and rubble pile asteroids}",
      journal = {Meteorit. Planet. Sci.},
         year = 2014,
        month = may,
       volume = {49},
       number = {5},
        pages = {788-811},
          doi = {10.1111/maps.12293},
 primaryClass = {astro-ph.EP},
       adsurl = {https://ui.adsabs.harvard.edu/abs/2014M&PS...49..788S}
}

@article{Scheeres_2010,
title = {Scaling forces to asteroid surfaces: The role of cohesion},
journal = {Icarus},
volume = {210},
number = {2},
pages = {968-984},
year = {2010},
issn = {0019-1035},
doi = {10.1016/j.icarus.2010.07.009},
author = {D.J. Scheeres and others}
}

@article{richardson_2002rubblepiles,
  title={Gravitational aggregates: Evidence and evolution},
  author={Richardson, Derek C and others},
  journal={Asteroids III},
  volume={1},
  pages={501--515},
  year={2002},
  publisher={University of Arizona Press}
}

@article{Pravec_Harris_2000,
title = {Fast and Slow Rotation of Asteroids},
journal = {Icarus},
volume = {148},
number = {1},
pages = {12-20},
year = {2000},
issn = {0019-1035},
doi = {10.1006/icar.2000.6482},
author = {Petr Pravec and Alan W. Harris}
}

@inproceedings{harris1996rotation,
  title={The rotation rates of very small asteroids: Evidence for'rubble pile'structure},
  author={Harris, Alan W},
  booktitle={Lunar and Planetary Science, volume 27, page 493},
  volume={27},
  year={1996}
}

@article{pravec_2002,
  title={Asteroid rotations},
  author={Pravec, PETR and Harris, Alan W and Michalowski, Tadeusz},
  journal={Asteroids III},
  volume={113},
  pages={35},
  year={2002},
  publisher={University of Arizona Press}
}

@article{Monteiro_2020,
    author = {Monteiro, Filipe and others},
    title = {Shape model and spin direction analysis of PHA (436724) 2011 UW158: a large superfast rotator},
    journal = {Mon. Not. Roy. Astron. Soc.},
    volume = {495},
    number = {4},
    pages = {3990-4005},
    year = {2020},
    month = {05},
    issn = {0035-8711},
    doi = {10.1093/mnras/staa1401},

}

@article{Hu_Scheeres_2004,
title = {Numerical determination of stability regions for orbital motion in uniformly rotating second degree and order gravity fields},
journal = {Planet Space Sci.},
volume = {52},
number = {8},
pages = {685-692},
year = {2004},
issn = {0032-0633},
doi = {10.1016/j.pss.2004.01.003},
author = {W. Hu and D.J. Scheeres}
}

@article{Scheeres_2016,
title = {The geophysical environment of Bennu},
journal = {Icarus},
volume = {276},
pages = {116-140},
year = {2016},
issn = {0019-1035},
doi = {10.1016/j.icarus.2016.04.013},
author = {D.J. Scheeres and others}
}

@article{scheeres2012orbital,
  title={Orbital mechanics about small bodies},
  author={Scheeres, DJ},
  journal={Acta Astronaut.},
  volume={72},
  pages={1--14},
  year={2012},
  publisher={Elsevier},
  doi={10.1016/j.actaastro.2011.10.021}
}

@article{Markellosetal1974,
  title={A grid search for families of periodic orbits in the restricted problem of three bodies},
  author={Markellos, VV and Black, W and Moran, PE},
  journal={Celest. Mech.},
  volume={9},
  pages={507--512},
  year={1974},
  publisher={Springer},
  doi={10.1007/BF01329331}
}

@book{parker2014low,
  title={Low-energy lunar trajectory design},
  author={Parker, Jeffrey S and Anderson, Rodney L},
  year={2014},
  publisher={John Wiley \& Sons}
}

@ARTICLE{1989Icar...78..337S,
       author = {{Spencer}, J.~R. and {Lebofsky}, L.~A. and {Sykes}, M.~V.},
        title = "{Systematic biases in radiometric diameter determinations}",
      journal = {Icarus},
         year = 1989,
        month = apr,
       volume = {78},
       number = {2},
        pages = {337-354},
          doi = {10.1016/0019-1035(89)90182-6},
       adsurl = {https://ui.adsabs.harvard.edu/abs/1989Icar...78..337S}
}

@inproceedings{lyster2025tempest,
  author    = {Lyster, D. and Howett, C. and Penn, J.},
  title     = {TEMPEST: A Modular Thermophysical Model for Airless Bodies with Support for Surface Roughness and Non-Periodic Heating},
  booktitle = {EPSC-DPS Joint Meeting 2025, Helsinki, Finland, 7--12 Sep 2025},
  pages     = {EPSC-DPS2025-1479},
  year      = {2025},
  doi       = {10.5194/epsc-dps2025-1479},
}

@INPROCEEDINGS{2024EPSC...17.1121L,
       author = {{Lyster}, Duncan and {Howett}, Carly and {Penn}, Joseph},
        title = "{Predicting Surface Temperatures on Airless Bodies: An Open-Source Python Tool}",
    booktitle = {European Planetary Science Congress},
         year = 2024,
        month = sep,
          eid = {EPSC2024-1121},
        pages = {EPSC2024-1121},
          doi = {10.5194/epsc2024-1121},
       adsurl = {https://ui.adsabs.harvard.edu/abs/2024EPSC...17.1121L}
}

@ARTICLE{Piqueaux2021,
       author = {{Piqueux}, Sylvain and {Vu}, Tuan H. and {Bapst}, Jonathan and others},
        title = "{Specific Heat Capacity Measurements of Selected Meteorites for Planetary Surface Temperature Modeling}",
      journal = {J. Geophys. Res. Planets},
         year = 2021,
        month = nov,
       volume = {126},
       number = {11},
          eid = {e07003},
        pages = {e07003},
          doi = {10.1029/2021JE007003},
       adsurl = {https://ui.adsabs.harvard.edu/abs/2021JGRE..12607003P}
}

@PHDTHESIS{muller2007,
       author = {{M{\"u}ller}, Michael Migo},
        title = "{Surface properties of asteroids from mid-infrared observations and thermophysical modeling}",
       school = {Free University of Berlin, Germany},
         year = 2007,
        month = jan,
       adsurl = {https://ui.adsabs.harvard.edu/abs/2007PhDT.......401M}
}

@article{Okada2020,
  author = {Okada, Tatsuaki and Fukuhara, Tetsuya and Tanaka, Satoshi and others},
  title = {Highly porous nature of a primitive asteroid revealed by thermal imaging},
  journal = {Nature},
  year = {2020},
  volume = {579},
  number = {7800},
  pages = {518--522},
  doi = {10.1038/s41586-020-2102-6},

}

@ARTICLE{2023Natur.616..443D,
       author = {{Daly}, R. Terik and {Ernst}, Carolyn M. and {Barnouin}, Olivier S. and {Chabot}, Nancy L. and {Rivkin}, Andrew S. and {Cheng}, Andrew F. and {Adams}, Elena Y. and {Agrusa}, Harrison F. and {Abel}, Elisabeth D. and {Alford}, Amy L. and {Asphaug}, Erik I. and {Atchison}, Justin A. and {Badger}, Andrew R. and {Baki}, Paul and {Ballouz}, Ronald-L. and {Bekker}, Dmitriy L. and {Bellerose}, Julie and {Bhaskaran}, Shyam and {Buratti}, Bonnie J. and {Cambioni}, Saverio and {Chen}, Michelle H. and {Chesley}, Steven R. and {Chiu}, George and {Collins}, Gareth S. and {Cox}, Matthew W. and {DeCoster}, Mallory E. and {Ericksen}, Peter S. and {Espiritu}, Raymond C. and {Faber}, Alan S. and {Farnham}, Tony L. and {Ferrari}, Fabio and {Fletcher}, Zachary J. and {Gaskell}, Robert W. and {Graninger}, Dawn M. and {Haque}, Musad A. and {Harrington-Duff}, Patricia A. and {Hefter}, Sarah and {Herreros}, Isabel and {Hirabayashi}, Masatoshi and {Huang}, Philip M. and {Hsieh}, Syau-Yun W. and {Jacobson}, Seth A. and {Jenkins}, Stephen N. and {Jensenius}, Mark A. and {John}, Jeremy W. and {Jutzi}, Martin and {Kohout}, Tomas and {Krueger}, Timothy O. and {Laipert}, Frank E. and {Lopez}, Norberto R. and {Luther}, Robert and {Lucchetti}, Alice and {Mages}, Declan M. and {Marchi}, Simone and {Martin}, Anna C. and {McQuaide}, Maria E. and {Michel}, Patrick and {Moskovitz}, Nicholas A. and {Murphy}, Ian W. and {Murdoch}, Naomi and {Naidu}, Shantanu P. and {Nair}, Hari and {Nolan}, Michael C. and {Orm{\"o}}, Jens and {Pajola}, Maurizio and {Palmer}, Eric E. and {Peachey}, James M. and {Pravec}, Petr and {Raducan}, Sabina D. and {Ramesh}, K.~T. and {Ramirez}, Joshua R. and {Reynolds}, Edward L. and {Richman}, Joshua E. and {Robin}, Colas Q. and {Rodriguez}, Luis M. and {Roufberg}, Lew M. and {Rush}, Brian P. and {Sawyer}, Carolyn A. and {Scheeres}, Daniel J. and {Scheirich}, Petr and {Schwartz}, Stephen R. and {Shannon}, Matthew P. and {Shapiro}, Brett N. and {Shearer}, Caitlin E. and {Smith}, Evan J. and {Steele}, R. Joshua and {Steckloff}, Jordan K. and {Stickle}, Angela M. and {Sunshine}, Jessica M. and {Superfin}, Emil A. and {Tarzi}, Zahi B. and {Thomas}, Cristina A. and {Thomas}, Justin R. and {Trigo-Rodr{\'\i}guez}, Josep M. and {Tropf}, B. Teresa and {Vaughan}, Andrew T. and {Velez}, Dianna and {Waller}, C. Dany and {Wilson}, Daniel S. and {Wortman}, Kristin A. and {Zhang}, Yun},
        title = "{Successful kinetic impact into an asteroid for planetary defence}",
      journal = {Nature},
         year = 2023,
        month = apr,
       volume = {616},
       number = {7957},
        pages = {443-447},
          doi = {10.1038/s41586-023-05810-5},
 primaryClass = {astro-ph.EP},
       adsurl = {https://ui.adsabs.harvard.edu/abs/2023Natur.616..443D}
}

@article{Mirtich1996,
 author = {Mirtich, Brian},
 title = {Fast and Accurate Computation of Polyhedral Mass Properties},
 journal = {J. Graph. Tools},
 issue_date = {February 1996},
 volume = {1},
 number = {2},
 month = feb,
 year = {1996},
 issn = {1086-7651},
 pages = {31--50},
 numpages = {20},
 doi = {10.1080/10867651.1996.10487458},
 acmid = {643322},
 publisher = {A. K. Peters, Ltd.},
 address = {Natick, MA, USA},
}

@BOOK{Scheeres2012book,
       author = {{Scheeres}, Daniel J.},
        title = "{Orbital Motion in Strongly Perturbed Environments}",
         year = 2012,
       adsurl = {https://ui.adsabs.harvard.edu/abs/2012omsp.book.....S}
}

@ARTICLE{Fu2024,
       author = {{Fu}, Xiaoyu and {Soldini}, Stefania and {Ikeda}, Hitoshi and {Scheeres}, Daniel J. and {Tsuda}, Yuichi},
        title = "{The dynamics about asteroid (162173) Ryugu}",
      journal = {Celestial Mechanics and Dynamical Astronomy},
         year = 2024,
        month = aug,
       volume = {136},
       number = {4},
          eid = {29},
        pages = {29},
          doi = {10.1007/s10569-024-10199-y},
       adsurl = {https://ui.adsabs.harvard.edu/abs/2024CeMDA.136...29F}
}

@ARTICLE{Hamilton1991,
       author = {{Hamilton}, Douglas P. and {Burns}, Joseph A.},
        title = "{Orbital stability zones about asteroids}",
      journal = {Icarus},
         year = 1991,
        month = jul,
       volume = {92},
       number = {1},
        pages = {118-131},
          doi = {10.1016/0019-1035(91)90039-V},
       adsurl = {https://ui.adsabs.harvard.edu/abs/1991Icar...92..118H}
}



\end{document}